\documentclass[letter]{sbuthesis}
\usepackage{amssymb}
\usepackage{lscape}
\usepackage{color}
\usepackage{mathrsfs}
\usepackage{graphics}
\usepackage{graphicx}
\usepackage{hyperref}
\usepackage{bm}
\usepackage{pdfsync}

\def\Tr{\mbox{Tr}\,}

\def\sgn{\,\mbox{sgn}\,}
\newcommand{\R} {\mbox{Re}\,}

\newcommand{\la}{\label}
\newcommand{\be}{\begin{equation}}
\newcommand{\ee}{\end{equation}}
\newcommand{\bea}{\begin{eqnarray}}
\newcommand{\eea}{\end{eqnarray}}
\newcommand{\eu}{{\rm e}}
\newcommand{\ii}{{\rm i}}
\newcommand{\de}{{\displaystyle\rm\mathstrut d}}
\newcommand{\Ord}{{\rm O}}
\newcommand{\Pf}{{\rm Pf}}
\newcommand{\mod}{{\rm \: mod \:}}

\def\Xint#1{\mathchoice
   {\XXint\displaystyle\textstyle{#1}}%
   {\XXint\textstyle\scriptstyle{#1}}%
   {\XXint\scriptstyle\scriptscriptstyle{#1}}%
   {\XXint\scriptscriptstyle\scriptscriptstyle{#1}}%
   \!\int}
\def\XXint#1#2#3{{\setbox0=\hbox{$#1{#2#3}{\int}$}
     \vcenter{\hbox{$#2#3$}}\kern-.5\wd0}}

\def\dashint{\Xint-}

\author{Fabio Franchini}

\title{On Hydrodynamic Correlations in Low-Dimensional Interacting Systems}

\month{December} \year{2006}

\advisor{Alexander G. Abanov}{Physics \& Astronomy Department, Stony Brook University}
\chairman{Jacobus Verbaarschot}{Physics \& Astronomy Department, Stony Brook University}
\fstmember{Dominik Schneble}{Physics \& Astronomy Department, Stony Brook University}
\outmember{Robert Konik}{Condensed Matter Theory Group, BNL}

\begin{document}

\pagenumbering{roman}

\begin{abstract}

Low-dimensional systems are an important field of current
theoretical and experimental research. Theoretically, the role of
dimensionality has been recognized for many years and dramatic
\hbox{predictions} have been made that still await experimental
confirmation or are currently under study. Recent technological
developments provide many possible realizations of effectively
one-dimensional systems. These devices promise to give us access
to a new range of phenomena. It is therefore very interesting to
develop theoretical methods specific for such systems to model
their behavior and calculate the correlators of the resulting
theory. Incidentally, one such method exists and is known as
Bosonization. It can be applied to one-dimensional systems and
effectively describes low energy excitations in a universal way.
It was developed in the 1970's when one-dimensional physics was
viewed as a toy model for higher dimensional physics. We use the
example of a correlator known as the Emptiness Formation
Probability to show that Bosonization fails to describe some long
range correlators corresponding to large disturbances (the EFP
measures the probability for the ground state of the system to
develop a region without particles). We trace this failure to the
fact that Bosonization is constructed as a linear approximation of
the full theory and we set up to develop a collective description
with the required non-linearity. The resulting scheme is
essentially a Hydrodynamic paradigm for quantum systems. We show
how to construct such a hydrodynamic description for a variety of
exactly integrable models and illustrate how it can be used to
make new predictions. For the special case of the spin-1/2 XY
model we take advantage of the structure of the model to express
the EFP as a determinant of a very special type of matrix, known
as Toeplitz Matrix. We use the theory of Toeplitz determinants to
calculate the asymptotic behavior of the EFP in the XY model and
discuss its relation with the criticality of the theory. Finally,
we analyze the behavior of a charged particle in a two-dimensional
medium filled with point-like magnetic vortices.

\end{abstract}

\singlespace

\tableofcontents 

\listoffigures

\listoftables


\begin{preface}

This thesis is based on the work I performed during my graduate
studies. Chapters \ref{XYModel} and \ref{EFPinXY} are devoted to
the calculation of a correlator known as Emptiness Formation
Probability for the anisotropic XY model. This analysis is based
on the work published in [\dag] and [\ddag]. In this chapter we
also consider the effect on the block entanglement of the
factorized ground state wavefunction on a line of the
phase-diagram of the model [\S]. Chapter \ref{Spin-ChargeHydro}
describes how to develop a two-fluid description for a system of
spin-1/2 electrons in one dimension. This is still a work in
progress [$\star$] the ultimate goal of which is to address the
problem of calculating the corrections to the exact spin-charge
separation of the Luttinger Liquid model. Chapter
\ref{ManyVortices} describes a two-dimensional problem I studied
with the help and the advice of Prof.\~ A.S. Goldhaber
[$\diamond$]. We consider the Aharonov-Bohm effect for a scalar
electron entering a medium filled with point-like magnetic
vortices pinned to the sites of a square lattice and we consider
the effect of such a configuration in the electron wavefunction.
In Appendix \ref{Integrability} we explicitly construct the
conserved currents of gradient-less hydrodynamic theories. While
the existence of two infinite series of conserved quantities is
known for these systems, we have not found such an explicit
construction in the literature. This work is still unpublished, as
the fundamental origin of these conserved quantities remains
unclear (we address this issue in Appendix \ref{Integrability} but
our results are so far inconclusive).

\subsection*{Publications}
\begin{enumerate}

\item[\dag]
 A.G. Abanov and F. Franchini; Phys. Lett. {\bf A 316} (2003) 342-349. \\
 {\it ``Emptiness Formation Probability for the Anisotropic XY Spin Chain in a Magnetic Field''} \\
 (also available on arXiv:cond-mat/0307001).

\item[\ddag]
 F. Franchini and A.G. Abanov; J. Phys. {\bf A 38} (2005) 5069-5096. \\
 {\it ``Asymptotics of Toeplitz Determinants and the Emptiness Formation Probability for the XY Spin Chain''}\\
 (also available on arXiv:cond-mat/0502015).

\item[\S]
 F. Franchini, A. R. Its, B.-Q. Jin and V. E. Korepin; to appear in  ``Proceedings of the
 {\it 26th International Colloquium on Group Theoretical Methods in Physics}''. \\
 {\it ``Analysis of entropy of XY Spin Chain''} \\
 (also available on arXiv:quant-ph/0606240). \\
\\
 F. Franchini, A. R. Its, B.-Q. Jin, V. E. Korepin, quant-ph/0609098. \\
 {\it ``Ellipses of Constant Entropy in the XY Spin Chain''}.

\item[$\star$]
 F. Franchini and A.G. Abanov; In progress. \\
 {\it ``Coupling of Spin and Charge Degrees of Freedom in a Hydrodynamic Two-Fluid Approach''}.

\item[$\diamond$]
 F. Franchini and A.S. Goldhaber; In preparation. \\
 {\it ``Aharonov-Bohm effect with many vortices''}.

\end{enumerate}

\addcontentsline{toc}{section}{Publications}

\end{preface}

\doublespace

\pagenumbering{arabic}



\chapter{Introduction}
\label{Introduction}
Dimensionality has an extremely important role in determining the
physics of a system. Since the 1960's, one-dimensional physics
stopped being just a theoretical toy, when excitations with
unusual quantum numbers were observed in polyacetylene structures
\cite{solitonsAc}. Later on, in the 1980's, the discovery of novel
effects in two-dimensional systems such as the Quantum Hall
Effect, high-temperature superconductivity, and others, brought
increasing attention to low-dimensional physics.

In recent years, zero-dimensional ({\it ``Quantum Dots''}) and
one-dimensional ({\it ``Quantum Wires''}) systems have been
implemented in laboratories as well, by effectively confining the
electrons in fewer dimensions, with the motion in the transverse
directions quantized so that at sufficiently low temperatures they
are made energetically prohibited.

From a theoretical point of view, one-dimensional models are
particularly interesting, because, in some sense, all 1-D theories
are strongly interacting, since particles cannot avoid each other
in their motion (all collisions are \hbox{`head-on').} Moreover,
in one dimension the distinction between a real scattering event
and the effect of quantum statistics is somewhat arbitrary, since
it is impossible to exchange two particles without having them
interacting.

In addition to these remarkable aspects, there is something
special about one-dimensional theories. The special symmetry of
having just one spatial dimension together with the temporal one
renders one-dimensional models particularly appealing and `easy'
to address, while preserving, and sometimes generating new,
interesting non-trivial situations. For example, a number of
theories are known to be exactly solvable in one dimension,
quantum field theory methods are especially powerful and direct,
and even the study of gravity is so simple in $1+1$ dimensions
that many attempts toward quantum gravity are performed in this
lower dimensionality.

In the 1970's, progress in quantum field theory was always
parallel to the study of the corresponding 1-D theory, since the
latter was often easier to tackle and provided important clues on
the general structure of the methods. These efforts resulted in
the development of the field theory method that we call {\it
``bosonization''} and in the concept of the {\it ``Luttinger
Liquid''}.

\section{Bosonization, linearization of the spectrum, and Luttinger Liquid concept}

One of the most successful methods to study one-dimensional
systems is known as {\it ``bosonization''} \cite{stone} which
effectively describes low energy excitations in 1-D. One of its
biggest advantages is its universality, i.e. the fact that the
theory has only two free parameters to be determined
microscopically, when possible, and then the structure of the
model is always the same for every system.

This universality originates in the fact that the model assumes a
linear spectrum, hence its applicability only at energy close to
the Fermi points, and this assumption results in a quadratic
Lagrangian, even including the interactions.

Physically, the bosonization method provides a very powerful
description of the concept of {\it ``Luttinger Liquid''} (LL)
\cite{Haldane-1981} that is commonly employed to describe
electrons in one dimension. This is the one-dimensional equivalent
of the Fermi Liquid concept, but the fact that in one dimension
the Fermi surface collapses to just two points has profound
consequences on the physics of the system. Bosonization captures
this physics accurately and allows for an easy access to the
calculation of many correlation functions, often on very general
grounds.

The main limitation of the Luttinger Liquid description and of
bosonization is the aforementioned assumption of a linear
spectrum, which guarantees its universality. This approximation is
extremely reasonable when one is interested in low temperature
physics, when only low energy excitations, close to the Fermi
points, can contribute. But we would like to argue that there
exist some important problems in one-dimensional physics where the
non-linearity of the spectrum is essential and non-avoidable. In
these cases one needs to go beyond the LL model.

For instance, one of the problems we are going to address is the
aforementioned prediction of spin-charge separation. It is a
standard result of the Luttinger Liquid description that electrons
in a one-dimensional system exist in the form of spin and charge
density waves whose dynamics is decoupled at low energies. As the
LL model relies on the assumption of a linear spectrum, one can
consider the effect of non-linearity of the spectrum on this
prediction and this would couple spin and charge degrees of
freedom. Any curvature in the spectrum would mix spin and charge
degrees of freedom and if one looks at sufficiently high energy
excitations the effect of this coupling would be measurable.
Unfortunately, a perturbative treatment of the curvature of the
spectrum in the bosonized theory generates divergences in the
calculation of physical quantities and is notoriously difficult to
perform.

The bosonization description of a system is a collective
description in which density fluctuations are the effective
degrees of freedom considered. To overcome the assumption of
linearity, we develop a collective description which retains all
the characteristics of the spectrum, while still relying on just a
few macroscopical fields.

\section{The hydrodynamic approach, integrable models, and Bethe Ansatz}

In fact, the collective description of quantum systems was
suggested long ago by Landau \cite{landau1941}. His method is
essentially a hydrodynamic description in which the system is
described just by its local density $\rho$ and velocity $v$. These
two fields obey a continuity equation and an Euler equation that
depends on the original spectrum of the theory and encodes it into
the dynamics of the hydrodynamic parameters.

We are going to apply this hydrodynamic approach to
one-dimensional systems at zero-temperatures and show how to
derive the hydrodynamic description of different one-dimensional
systems \cite{abanovHydro}.

We will pay special attention to integrable systems. These
theories are exactly solvable and have an infinite series of
conserved quantities. Integrable models are very often good
approximations for physical systems and an important test field
for theoretical methods, since their high degree of symmetry
allows for very controlled calculations.

Results for integrable models are usually derived within the
framework of the Bethe Ansatz technique
\cite{bethe,yangyang,korepin93}. The Bethe Ansatz provides a way
to construct the wavefunctions of the system and to derive many
thermodynamic quantities. The knowledge of the wavefunction,
though, is implicit and this makes it very hard to calculate the
correlation functions of the model. Exact expressions for the
correlators are available in terms of determinants of operator
valued matrices \cite{korepin93}, but these expressions are quite
convoluted and are often not of practical use for the evaluation
of physical quantities.

We are going to show how the hydrodynamic approach derives
directly from the Bethe Ansatz formalism. Therefore, physically
relevant correlators involving the density and velocity of the
fluid can be more easily evaluated within our approach.

The ability of bosonization to calculate correlators for
integrable models has been one of the great advantage of this
method in conjunction with the Bethe Ansatz solution which
provides the parameters of the theory. Clearly, these correlators
are correct up to the linear spectrum approximation. The
hydrodynamic approach overcomes this limitation at the cost of
dealing with a more complicated collective description, since the
hydrodynamic equations are intrinsically non-linear.

\section{The Emptiness Formation Probability}

Throughout most of this work we will concentrate on a particular
correlation function known as {\it ``Emptiness Formation
Probability''} (EFP). This correlator measures the probability
that a region of the system is depleted of particles
\cite{korepin93}. It is clear that a bosonization approach is not
suitable for the calculation of the EFP, since particles from
everywhere in the spectrum are needed to empty a region of space
and the curvature of the spectrum cannot be neglected.

The EFP will be a useful example to show the advantage of the
hydrodynamic approach over bosonization when contributions far
from the Fermi points are involved. But the importance of the EFP
goes beyond its role as an example of such correlators.

The Emptiness Formation Probability was introduced in the
development of the determinant representation for correlation
functions of the integrable models we mentioned before. A good
account of this technique can be found in \cite{korepin93}, where
is it shown how the Bethe Ansatz solution can be manipulated to
produce exact expressions for correlators. A careful analysis of
these expressions revealed that the simplest correlator one can
construct for integrable models is precisely the EFP.

As we mentioned, the practical calculation of correlators for
integrable systems is still an open challenge. Therefore, it is
conceivable that the study of the simplest correlator will bring
insights helpful to carry on the investigation of other
correlators. Probably also for this reason, in recent years a
considerable effort has been devoted to the study of the EFP in
different systems \cite{korepin93} -- \cite{abanovfran}, but a
general method for its calculation is still missing.

We propose the hydrodynamic approach as such a general method for
the calculation of the EFP, at least to leading order.

\section{EFP in different systems}
\label{EFPSystems}

In this section we introduce the EFP for several simple integrable
models.

\paragraph{Emptiness Formation Probability:}

Let us first consider a one-dimensional quantum system of $N$
particles at zero temperature. The wavefunction of the ground
state of the system $\Psi_G (x_1, x_2,\ldots, x_N)$ gives the
probability distribution $|\Psi_G|^2$ of having all $N$ particles
at given positions $x_j$, where $j=1,\ldots, N$.

The Emptiness Formation Probability $P(R)$ is defined as the
probability of having no particles with coordinates $-R<x_j<R$ for
every $j$. Formally we write this as
\be
    P(R) = {1 \over \langle \Psi_G | \Psi_G \rangle } \int_{|x_{j}|>R}
    \de x_1 \ldots \de x_N \; |\Psi_G (x_1,\ldots,x_N)|^2,
    \label{efpdef}
\ee or following \cite{korepin93}
\be
    P(R) = \lim_{\alpha\to +\infty} \left \langle \Psi_G \right|
    \eu^{-\alpha \int_{-R}^{R} \rho(x) \, \de x }
    \left| \Psi_G \right \rangle,
\ee where $\rho(x)$ is the particle density operator
\be
    \rho(x) \equiv \sum_{j=1}^{N} \delta (x-x_{j}).
\ee

\paragraph{Spin chains and lattice fermions:}

The EFP can also be defined for spin chains. In these systems we
are interested in what is known as the {\it ``Probability of
Formation of Ferromagnetic Strings''} (PFFS), where we are looking
for strings of length $n$ in the ground state of the spin chain:
\be
   P(n) \equiv
   \langle 0 | \prod_{i=1}^n {1 - \sigma_i^z \over 2} | 0 \rangle ,
   \label{EFPDefInt}
\ee where $\sigma_i^z$ is the $z$-component of the Pauli matrices
on the $i$-th lattice site.

The Jordan-Wigner transformation (\ref{JW1}-\ref{JW3}) maps a
spin-$1/2$ chain to a one-dimensional lattice gas of spinless
fermions. Under this mapping the ferromagnetic string corresponds
to a string of empty lattice sites and one can write the EFP \be
    P(R) =\left\langle \prod_{j=-R}^{R}\psi_{j}\psi_{j}^{\dagger} \right\rangle,
\ee where $\psi_{j},\psi_{j}^{\dagger}$ are annihilation and
creation operators of spinless fermions on the lattice site $j$.

Therefore, the PFFS in a spin chain corresponds to the EFP of the
corresponding Jordan-Wigner fermion theory. In general, we are
going to use the language of particles in this work, so we will
generically refer to the EFP even for spin systems, since all EFP
results are valid for the corresponding one-dimensional spin
systems as well.

\paragraph{Random Matrices:}

As a remark, we point out that the EFP (\ref{efpdef}) introduced
for a general one-dimensional quantum system is a well-known
quantity in the context of the spectra of random matrices
\cite{dysmehta}. Essentially, it is the probability of having no
eigenvalues in some range of the spectrum.

Consider, e.g., the joint eigenvalue distribution for the Circular
Unitary Ensemble (CUE). The CUE is defined as an ensemble of
$N\times N$ unitary matrices with the ensemble measure given by
the de Haar measure. Diagonalizing the matrices and integrating
out the unitary rotations, one obtains \cite{mehta}
\be
    \int DU \to \int \prod_{j=1}^{N} \de \theta_j \, \prod_{1\le j<k\le N}
    \left| \eu^{\ii \theta_j } - \eu^{\ii \theta_k} \right|^\beta,
\ee where $\beta = 2$ for the CUE and $\eu^{\ii \theta_j}$, with
$j=1,\ldots,N$, are the eigenvalues of a unitary matrix.

One can then read the joint eigenvalue distribution as
\be
    P_{N}(\theta_{1},\ldots,\theta_{N}) = {\rm const.} \prod_{1\le j<k \le N}
    \left|\eu^{\ii \theta_j} - \eu^{\ii \theta_k} \right|^\beta .
    \label{jed}
\ee

We can now introduce the probability of having no eigenvalues on
the arc $-\alpha<\theta<\alpha$ as
\be
    P(\alpha) = {1 \over {\cal N}} \int_{\theta_{j} \notin [-\alpha,\alpha]}
    \prod_{j=1}^{N} \de \theta_j \, \prod_{1\leq j<k\leq N}\left|
    \eu^{\ii \theta_j} - \eu^{\ii \theta_k} \right|^\beta.
\ee In the theory of Random Matrices, this quantity is known as
$E_{\beta}(0,\alpha)$, using the notations of \cite{mehta}, and is
clearly equivalent to the EFP once one identifies energy
eigenvalues and particles. For orthogonal, unitary, and symplectic
circular ensembles the joint eigenvalue distribution is given by
(\ref{jed}) with $\beta=1,2,4$ respectively, where $\beta = 2$
corresponds to free fermions, while $\beta = 1,4$ are particular
cases of the EFP in the Calogero-Sutherland model \cite{mehta}.

\section{Some implementations of One-Dimensional systems}

While one-dimensional physics has been an active sector of
research in theoretical studies, only recently laboratories around
the world have been able to build and study devices where the
motion of electrons is effectively limited to just one dimension
and where sufficient control is available to explore different
regimes and phases. This allows to finally test in experiments the
theoretical prediction on 1-D models. We present an overview of
some systems and some experiments where Luttinger liquid behavior
has been confirmed and discuss whether curvature effects, like the
correction to exact spin-charge separation, can be observed.

\paragraph{\bf GaAs/AlGaAs quantum wires:}

One dimensional quantum wires are realized by cleaved edge
overgrowth on a GaAs/AlGaAs heterostructures. These wires are in
the ballistic regimes and Luttinger Liquid behavior has been
confirmed by several experiments, for instance by measuring
deviation from exact quantization of the conductance against the
theoretical predictions \cite{GaAs1}. Recently, efforts have been
devoted in observing signatures of spin-charge separation in these
quantum wires \cite{GaAs}, but the results are so far
non-conclusive. The experiment was conducted on two parallel
ballistic wires, by measuring the tunneling current: the
measurements were in qualitative agreement with the theory, but
one of the two predicted charge modes was not observed. It is
conceivable that in the future a similar setup could be
implemented to test Coulomb drag effects \cite{CoulombDrag}, that
would carry a clear signs of spectrum curvature, but attempts in
this direction require a level of sophistication beyond the
technology available today.

\paragraph{\bf Carbon nanotubes:}

Single-wall and multi-wall carbon nanotubes are excellent examples
of effectively one-dimensional systems
\cite{nanotubes,nanotubes1}. Different realizations of the tubes
generated metallic, insulating, semi-metallic and semi-conducting
wires. Luttinger liquid behavior of metallic carbon nanotubes have
been demonstrated by measuring tunneling amplitudes into the wires
through STM experiments \cite{nanotubes}. The conductance and
differential conductance followed power law behaviors as functions
of temperatures and bias voltages, and the exponents were found to
be in good agreement with the theory. For semi-metallic wires the
description of electrons in terms of Majorana Fermions is almost
exact in that the spectrum is linear near the conical point with
very good approximation. In the metallic regimes, the Fermi points
move away from this point, but we do not expect it to be possible
to observe the effect of curvature in these systems with the
present technology.

\paragraph{\bf Organic conducting molecules:}

Some organic molecules, like polyacetylene, can be used as
one-dimensional wires. The main limitation in experiments with
organic molecules is the lack of control in the preparation of the
system, due to the fact that the molecules have a predefined
structure. Therefore, one has to look for the right molecule that
could fit a model and there is no a generic set-up that can be
used to test different regimes. Polyacetylene, for instance, is a
one-dimensional gapped insulator and therefore it cannot be used
to test Luttinger liquid behaviors \cite{solitonsAc}. Conductance
studies on Bechgaard salts fibers show the power law behaviors one
expects from a Luttinger liquid, but the data are hard to
quantitatively match with the theory, due to the complexity of the
systems \cite{organic}. Evidence for spin-charge separation
effects in these salts have been gathered by comparing charge and
thermal conductivity, but for a clear interpretation of the
results we would need to know these systems better than we
currently do.

\paragraph{\bf One-dimensional metallic chains:}

For a long time now, one dimensional metallic chains like $Au$
atoms on $Si(111)$, $Sr_2 CuO_3$ and $SrCuO_2s$ have been under
investigation, trying to measure evidences of spin-charge
separation, but results were not conclusive \cite{metallic}.
Recently, a clear signature of spin-charge separation has been
reported using Angle-Resolved PhotoEmission (ARPES) data
\cite{metallic1}. They reconstructed the spectral function, which
showed two distinct peaks corresponding to the spin and charge
contributions in quantitative agreement with the theoretical
prediction. There was, however, a significant portion of the
spectral function unaccounted by theoretical models, resulting in
an unexpected broadening of the peaks. It is possible that this
effect is due to the coupling between spin and charge degrees of
freedom arising from the spectrum curvature.

\paragraph{\bf Cold atomic gases:}

In recent years, impressing technological developments have
allowed to cool atomic systems confined in optical traps to
unprecedented low temperatures. Since the first observation of
Bose-Einstein condensation \cite{boseeinstcond}, temperatures have
kept dropping, allowing the observation of new physical phenomena.
An active field of research focuses on atomic gasses effectively
trapped in one-dimensional geometries. It is believed that many
solid state physics can be mimicked by these systems and that it
would be possible to study them at much lower temperatures than
their condensed matter counterparts \cite{coldfermigas-review}.
Most successful have been experiments with bosons, with the recent
realization of a Tonks-Girardeau gas, i.e. a gas of bosons with
such strong repulsion that at low enough temperatures and
densities exhibit fermionic behavior \cite{coldfermigas}. It is
also expected that a one-dimensional Tonks gas would exhibit
Luttinger Liquid behavior and some evidence is already available
supporting this prediction \cite{coldfermigas1}. Direct
experiments with fermionic gasses need to decrease the
temperatures by about two orders of magnitudes compared to the
current status in order to realize a Luttinger liquid system, but
the improvements in experimental techniques indicate that this
goal is probably not far in the future. Indeed, one-dimensional
cold atomic gasses are really promising for the investigation of
curvature effects of the spectrum in the near future, because
these systems allow for a more direct manipulation of the
excitations, compared to equivalent solid state systems.

\section{Outlook of the thesis work}

In Chapter \ref{XYModel} we introduce the spin-$1/2$ anisotropic
XY model, an important integrable spin chain model in one
dimension. In this chapter we analyze the model, study its rich
phase diagram and derive the fundamental correlators of the model.
The XY model is arguably the simplest integrable model with a
non-trivial phase diagram and a perfect candidate for the study of
the EFP.

In Chapter \ref{EFPinXY} we undertake this study and show that for
this system the EFP $P(n)$ can be expressed exactly as the
determinant of a $n \times n$ matrix. This matrix belong to a
class of very special matrices known as Toeplitz Matrices and the
asymptotic behavior of their determinant has been studied
extensively by mathematicians. We use the results of the theory of
Toeplitz determinants to calculate the asymptotic behavior of the
EFP in the different regions of the phase-diagram of the model. We
find that the EFP decays exponentially in most of the phase
diagram and only for the isotropic case studied in
\cite{shiroishi} is the decay Gaussian. On the critical lines we
observe an additional power-law correction to the exponential or
Gaussian decay. We employ a bosonization approach to interpret the
crossover from the Gaussian to the exponential decay with
universal exponents. These results are original and first appeared
in \cite{abanovfran}.

In Chapter \ref{HydroApp} we introduce the hydrodynamic approach.
We first analyze the hydrodynamic description of free fermions and
then proceed to show the technique in its generality and how to
obtain the internal energy of integrable systems from the Bethe
Ansatz. We also show that the linearization of the hydrodynamic
theory produces the traditional bosonization.

In Chapter \ref{HydroEFP} we show how to derive the leading
asymptotic behavior of the EFP for some Galilean invariant
systems. This calculation is performed at a semi-classical level,
where the EFP is calculated as the probability of a rare
fluctuation ({\it ``instanton''}). We demonstrate that
bosonization is insufficient to quantify the EFP. Using the
hydrodynamic approach we calculate the EFP to leading order for
Free Fermions and for Calogero-Sutherland particles.

In Chapter \ref{Spin-ChargeHydro} we analyze the prediction of
spin-charge separation in one dimension. First, we show the
limitations of the bosonization approach in dealing with the
correction to exact spin-charge separation. Then we develop a
hydrodynamic description for fermions with contact repulsion.
While our ultimate goal is to address the problem of evaluating
the corrections to exact spin-charge separation, this is still a
work in progress and we can only show how to implicitly construct
the hydrodynamic description of this model from its Bethe Ansatz
solution.

In Chapter \ref{ManyVortices} we address a problem very different
from the rest of the thesis. It is a single-particle problem in
two dimensions. We study the behavior of a scalar particle in a
medium filled with point-like magnetic fluxes (vortices) pinned on
the sites of a square lattice. By assuming the strength of the
fluxes to be all equal to half of the quantum flux unit, we are
able to construct a wavefunction of the system, showing that the
spectrum of such a system is discrete. Moreover, we are able to
show that a zero-energy particle entering such a medium would
decay exponentially. This means that an array of vortices could be
used to trap a particle and this bound state would have a
topological nature. As far as we know, this is the first time that
the existence of such a bound state is suggested.

Finally, in Chapter \ref{TConclusions} we will discuss the results
of this thesis, the problems that remain open and the directions
for further research.

We include appendices to supplement the analysis of the main text.

Appendix \ref{KnownRes} is a recapitulation of known results on
the EFP calculated for various integrable models.

Appendix \ref{ToeplitzApp} contains a brief review of the results
of the theory of Toeplitz determinants, which is extensively used
in the evaluation of the asymptotic behavior of the correlation
functions of the XY model, including the EFP we study in Chapter
\ref{EFPinXY}.

Appendix \ref{BetheInt} is a very brief introduction to the Bethe
Ansatz technique where we gather some results we need as an input
for the hydrodynamic description of integrable models.

In Appendix \ref{Integrability} we construct the conserved
quantities of gradient-less hydrodynamic theories. The
integrability of Galilean invariant hydrodynamic models without
gradient correction is a little-known fact discovered in the
1980's. Here we explicitly construct the integrals of motion for
the first time and we attempt to clarify the origin of this
integrability.


\chapter{The Spin-1/2 Anisotropic XY Model}
\label{XYModel}
The One-Dimensional Spin-1/2 Anisotropic XY spin chain is arguably
one of the simplest non-trivial quantum integrable models. The
reason for this simplicity lies in the fact that it can be reduced
to a free fermions model.

The XY model describes a one dimensional lattice system, where
each lattice site is occupied by a spin-1/2 quantum degree of
freedom interacting with its neighbors. The allowed interaction
involves only the {\it X} and {\it Y} components of the spins. In
addition, we consider the presence of an external transverse
magnetic field interacting with the {\it Z} component of the
spins.

Using what is known as a {\it ``Jordan-Wigner Transformation''}
the spin degrees of freedom can be mapped into spin-less fermions,
so that the model describes lattice fermions with a quadratic
Hamiltonian. The Hamiltonian being quadratic, a {\it ``Bogoliubov
Rotation''} defines {\it ``Bogoliubov Quasi-Particles''} in terms
of which the model reduces to lattice free fermions.

The apparent simplicity of the model does not mean that it is
trivial, as these quasi-particles are non-local in terms of the
original degrees of freedom. Therefore, every quantity that one
wants to calculate for the original model has a non-trivial
expression in terms of the free fermions.

A great simplification comes from the fact that for the XY model
these non-trivial expressions can often be expressed as
determinants of matrices with remarkable symmetries. These
matrices are known at {\it ``Toeplitz Matrices''} and a rich
mathematical literature has been devoted to the study of the
asymptotic behaviors of their determinant. We will make extensive
use of these results on Toeplitz determinants.

The anisotropic XY spin-1/2 chain in a transverse magnetic field
is defined by the Hamiltonian \be
   H = \sum_{i=1}^N \left[
   \left( {1 + \gamma \over 2} \right) \sigma_i^x \sigma_{i+1}^x +
   \left( {1 - \gamma \over 2} \right) \sigma_i^y \sigma_{i+1}^y \right]
   - h \sum_{i=1}^N \sigma_i^z,
   \label{spinham}
\ee where $\sigma_i^{\alpha}$, with $\alpha=x,y,z$, are the Pauli
matrices which describe spin operators on the $i$-th lattice site
of the spin chain and, for definiteness, we require periodic
boundary conditions: $\sigma_i^\alpha = \sigma_{i+N}^\alpha$ ($N
>> 1$).

We are interested in this model because of its rich phase-diagram.
The model depends on two parameters, the external magnetic field
$h$ and the anisotropy parameter $\gamma$ controlling the relative
strength of the interaction between the {\it X} and {\it Y}
components of the spins. By varying these parameters the system
crosses several quantum phase transitions (QPT). The rich
structure of its phase diagram is in contrast with the fair
simplicity of the model and makes it an excellent candidate for
examination in connection with QPT.

This is one of the reasons for which the XY model is one of the
most studied in the growing field of quantum computing and quantum
information.

In section \ref{xymodel} we introduce the model, and show its
different formulations in terms of spins and spinless fermions,
and we analyze its phase-diagram. In section \ref{xycorr} we
derive the fundamental correlators for the model. In this section
we will also give an overview of the results of McCoy and
co-authors \cite{mccoy} to better understand the structure of this
model in preparation for the next chapter where we will calculate
the EFP for the XY model \cite{shiroishi,abanovfran}.

The XY model was first solved by Lieb, Schultz and Mattis in
\cite{LSM-1961} without the external magnetic field. McCoy and
co-authors \cite{mccoy} were the first to develop the theory of
Toeplitz determinants in connection with the study of the XY spin
chain in the external field and to exploit these structures to
derive the fundamental correlators of the model.

\section{The model and its phases}
\label{xymodel}

The XY model described in (\ref{spinham}) has been solved in
\cite{LSM-1961} in the case of zero magnetic field and in
\cite{mccoy} in the presence of a magnetic field. We follow the
standard prescription \cite{LSM-1961} and reformulate the
Hamiltonian (\ref{spinham}) in terms of spinless fermions $\psi_i$
by means of a Jordan-Wigner transformation: \bea
   \sigma_j^+ & = &
   \psi_j^\dagger \: \eu^{\ii \pi \sum_{k<j} \psi_k^\dagger \psi_k} =
   \psi_j^\dagger \: \prod_{k<j}
   \left( 2 \psi_k^\dagger \psi_k - 1 \right),
   \label{JW1} \\
   \sigma_j^- & = &
   \eu^{- \ii \pi \sum_{k<j} \psi_k^\dagger \psi_k} \: \psi_j  =
   \prod_{k<j} \left( 2 \psi_k \psi_k^\dagger - 1 \right) \: \psi_j,
   \label{JW2} \\
   \sigma_j^z & = & 2 \psi_j^\dagger \psi_j - 1,
   \label{JW3}
\eea where, as usual, \be
   \sigma^\pm = { \sigma^x \pm i \sigma^y \over 2}.
   \label{spinpm}
\ee

The Hamiltonian in terms of these spinless fermions becomes:
\be
   H = \sum_{i=1}^N \left( \psi_i^\dagger \psi_{i+1} +
   \psi_{i+1}^\dagger \psi_i +
   \gamma \; \psi_i^\dagger \psi_{i+1}^\dagger +
   \gamma \; \psi_{i+1} \psi_i -
   2 h \; \psi_i^\dagger \psi_i \right)
   \label{realfermionH}
\ee
and in Fourier space it reads ($\psi_j = \sum_q \psi_q
\eu^{\ii q j}$):
\be
   H = \sum_q \left[
   2 \left( \cos \! q - h \right) \psi_q^\dagger \psi_q
   + \ii \gamma \sin \! q \: \psi_q^\dagger \psi_{-q}^\dagger
   - \ii \gamma \sin \! q \: \psi_{-q} \psi_q \right].
   \label{spinlessham}
\ee

We can now diagonalize this Hamiltonian by means of a Bogoliubov
transformation \be
   \chi_{q} = \cos \! {\vartheta_q \over 2} \: \psi_{q}
   + \ii \sin \! {\vartheta_q \over 2} \: \psi_{-q}^\dagger ,
   \label{bogtrans}
\ee which mixes the Fourier components with ``rotation angle''
$\vartheta_q$ defined by
\be
   e^{\ii \vartheta_q}  = \frac{1}{\varepsilon_q}(\cos q - h + \ii \gamma \sin q) .
   \label{rotangle}
\ee

In terms of these new Bogoliubov quasi-particles $\chi_q$ the
Hamiltonian (\ref{spinlessham}) has diagonal form
\be
   H= \sum_q \varepsilon_q \chi_q^\dagger\chi_q
   \label{quasipartham}
\ee with the quasiparticle spectrum
\be
   \varepsilon_q = \sqrt{ \left( \cos q - h \right)^2
   + \gamma^2  \sin^2 q}.
   \label{spectrum}
\ee

\begin{figure}
 \includegraphics[width=\columnwidth]{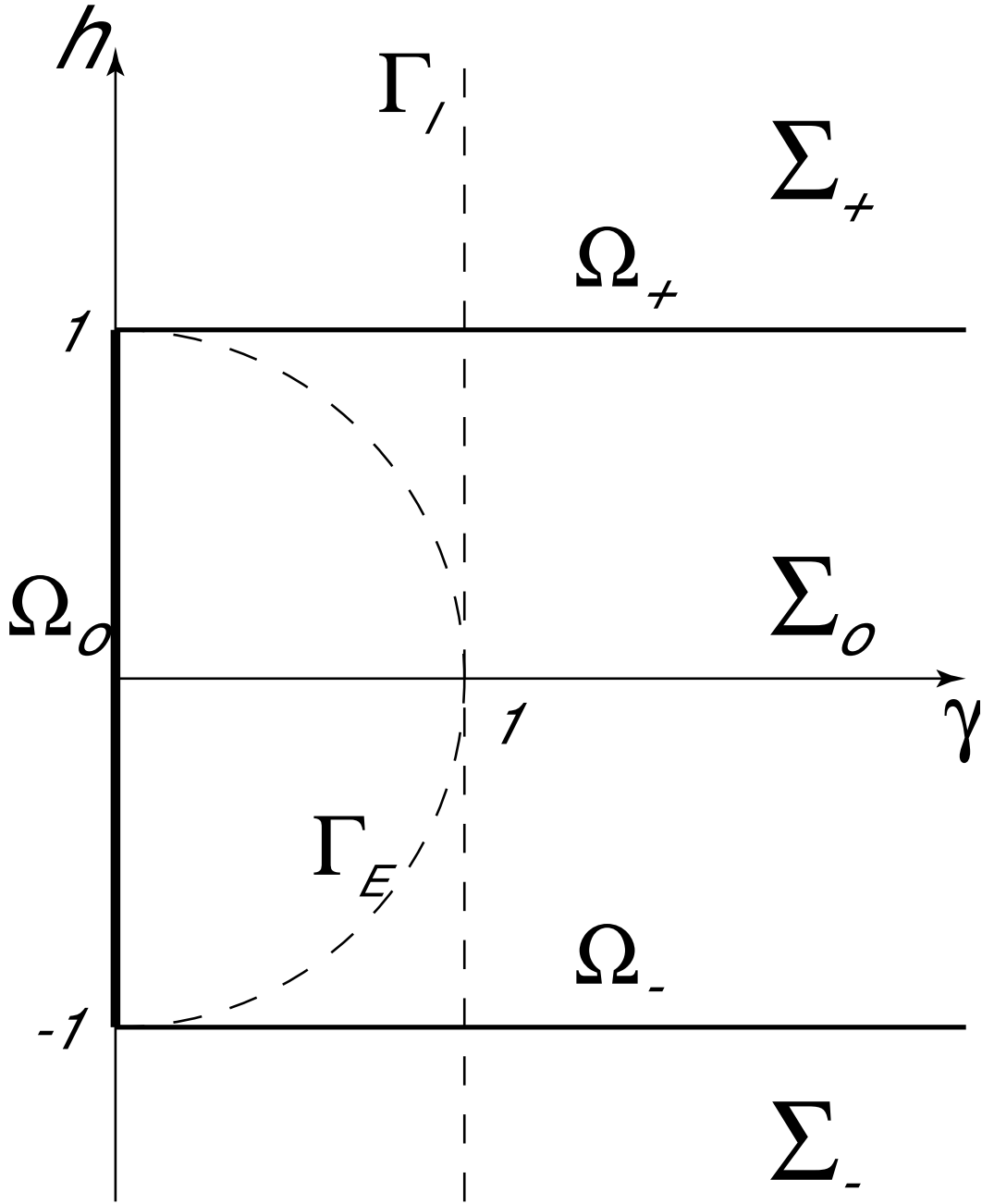}
\caption[Phase diagram of the XY Model] {Phase diagram of the XY
Model (only the part $\gamma \ge 0$ is shown). The theory is
critical for $h = \pm 1$ ($\Omega_\pm$) and for $\gamma = 0$ and
$|h| < 1$ ($\Omega_0$). The line $\Gamma_I$ represents the Ising
Model in transverse field. On the line $\Gamma_E$ the ground state
of the theory is a product of single spin states.}
   \label{phasediagram}
\end{figure}

A system is said to be {\it ``critical''} when its spectrum is
gapless, i.e. when one can excite particles from the Fermi Sea
without spending energy. When a system becomes critical, it
undergoes a {\it ``Quantum Phase Transition} (QPT)''. QPTs are
zero-temperature analogs of traditional phase transitions. QPTs
are characterized by singularities in thermodynamic quantities and
by correlators having a characteristic algebraic behavior. The
effective theory is scale invariant and in one dimension can be
described through Bosonization. QPTs are a very active field of
research, especially since experiments can reach low enough
temperatures where their signatures are observable.

We recognize from (\ref{spectrum}) that the theory is critical,
i.e. gapless, for $h=\pm 1$ or for $\gamma = 0$ and $|h| < 1$.

In Fig.~\ref{phasediagram} we show the phase diagram of the XY
model, which has obvious symmetries $\gamma \to -\gamma$ and $h
\to -h$\footnote{In the next chapter we will analyze the EFP for
the XY and we will see that this correlator breaks the latter
symmetry.}. The phase diagram has both critical and non-critical
regimes. Three critical lines $\Omega_0$ (Isotropic XY model:
$\gamma=0$, $|h|<1$) and $\Omega_\pm$ (critical magnetic field: $h
= \pm 1$) divide the phase diagram into three non-critical
domains, $\Sigma_-$, $\Sigma_0$, and $\Sigma_+$ ($h < -1$, $-1 < h
< 1$, and $h > 1$ respectively). Fig.~\ref{phasediagram} also
shows the line $\gamma=1$ ($\Gamma_I$) corresponding to the Ising
model in transverse magnetic field and the line $\gamma^2 + h^2 =
1$ ($\Gamma_E$) on which the wavefunction of the ground state is
factorized into a product of single spin states \cite{shrock}.

\section{The correlators of the model}
\label{xycorr}

In this section we review the derivation of the fundamental
correlators for the XY model at zero temperature following McCoy
and co-authors \cite{mccoy}.

The ground state $| 0 \rangle$ of the model in terms of the
Bogoliubov quasi-particles in (\ref{quasipartham}) is defined as
\be
   \chi_q | 0 \rangle =  0 \qquad \forall q
\ee i.e. it is the conventional ground state of free fermions. The
correlators for this theory are easily found to be
\bea
   \langle 0 | \chi^\dagger_q \chi_k | 0 \rangle & = &
   \delta_{k,q} , \\
   \langle 0 | \chi_q \chi^\dagger_k | 0 \rangle & = & 0 , \\
   \langle 0 | \chi_q \chi_k | 0 \rangle & = & 0 , \\
   \langle 0 | \chi^\dagger_q \chi^\dagger_k | 0 \rangle & = & 0 .
\eea

This vacuum is the ground state for the XY model, but it is not so
simple when expressed in terms of physical particles. The
Hamiltonian (\ref{spinlessham}) contains superconducting-like
terms, so its ground state is non-trivial. One can invert the
Bogoliubov transformation (\ref{bogtrans}) \be
   \psi_{q} = \cos \! {\vartheta_q \over 2} \: \chi_{q}
   - \ii \sin \! {\vartheta_q \over 2} \: \chi_{-q}^\dagger
\ee to calculate the fundamental correlators in terms of physical
fermions:
\bea
   \langle 0 | \psi^\dagger_q \psi_k | 0 \rangle =
   \sin^2 {\vartheta_q \over 2} \: \delta_{k,q}& = &
   {1 - \cos {\vartheta_q} \over 2} \: \delta_{k,q} , \\
   \langle 0 | \psi_q \psi^\dagger_k | 0 \rangle =
   \cos^2 {\vartheta_q \over 2} \: \delta_{k,q} & = &
   {1 + \cos {\vartheta_q} \over 2} \: \delta_{k,q} , \\
   \langle 0 | \psi_q \psi_k | 0 \rangle =
   - \ii \: \cos {\vartheta_q \over 2} \sin {\vartheta_q \over 2} \: \delta_{k,q} & = &
   - \ii \: {\sin {\vartheta_q} \over 2} \: \delta_{k,q} , \\
   \langle 0 | \psi^\dagger_q \psi^\dagger_k | 0 \rangle =
   \ii \: \cos {\vartheta_q \over 2} \sin {\vartheta_q \over 2} \: \delta_{k,q} & = &
      \ii \: {\sin {\vartheta_q} \over 2} \: \delta_{k,q} .
\eea

Now, the two-point fermionic correlators are easy to obtain by
Fourier transform. In the thermodynamic limit they read
\cite{LSM-1961,mccoy} \bea
   F_{jk} &\equiv& \ii\langle 0 | \psi_j \psi_k | 0 \rangle
   = - \ii\langle 0 | \psi_j^\dagger \psi_k^\dagger | 0 \rangle
   = \int_0^{2 \pi} {\de q \over 2\pi}\; \frac{\sin\vartheta_q}{2}
   \eu^{\ii q (j-k)},
  \label{F}
\\
   G_{jk} &\equiv& \langle 0 | \psi_j \psi_k^\dagger | 0\rangle
   =   \int_0^{2 \pi} {\de q \over 2 \pi}\; \frac{1+\cos\vartheta_q}{2}
   \eu^{\ii q (j-k)} .
 \label{G}
\eea These correlators will be fundamental in our calculation of
the EFP in the next chapter.

To calculate the correlation functions for the original spin chain
model (\ref{spinham}),
\be
   \rho^\nu_{lm} \equiv \langle 0 \left| \sigma^\nu_l \: \sigma^\nu_m \right| 0 \rangle
   \qquad \nu = x, y, z ,
   \label{rhocorr}
\ee we need more work. We follow \cite{LSM-1961} and express these
correlators in terms of spin lowering and raising operators
(\ref{spinpm}): \bea
   \rho^x_{lm} & = & \langle 0 \left| \left( \sigma^+_l + \sigma^-_l \right) \:
   \left( \sigma^+_m + \sigma^-_m \right) \right| 0 \rangle , \\
   \rho^y_{lm} & = & \langle 0 \left| \left( \sigma^+_l - \sigma^-_l \right) \:
   \left( \sigma^+_m - \sigma^-_m \right) \right| 0 \rangle , \\
   \rho^z_{lm} & = & \langle 0 \left| \left( 2 \sigma^+_l \sigma^-_l - 1 \right) \:
   \left( 2 \sigma^+_m \sigma^-_m - 1 \right) \right| 0 \rangle.
\eea and use (\ref{JW1}-\ref{JW3}) to write them using spinless
fermions operators.

For instance, let us consider $\rho^x_{lm}$:
\bea
   \rho^x_{lm} & = & \langle 0 \left| \left( \sigma^+_l + \sigma^-_l \right) \:
   \left( \sigma^+_m + \sigma^-_m \right) \right| 0 \rangle \nonumber \\
   & = &  \langle 0 | \left( \psi^\dagger_l + \psi_l \right) \:
    \prod_{i=l}^{m-1} \left( 2 \psi^\dagger_i \psi_i - 1 \right) \:
   \left( \psi^\dagger_m + \psi_m \right) | 0 \rangle \nonumber \\
   & = &  \langle 0 |\left( \psi^\dagger_l - \psi_l \right) \:
   \prod_{i=l+1}^{m-1} \left( 2 \psi^\dagger_i \psi_i - 1 \right) \:
   \left( \psi^\dagger_m + \psi_m \right) | 0 \rangle \nonumber \\
   & = &  \langle 0 | \left( \psi^\dagger_l - \psi_l \right) \:
   \prod_{i=l+1}^{m-1} \left( \psi^\dagger_i + \psi_i \right)
   \left( \psi^\dagger_i - \psi_i \right) \:
   \left( \psi^\dagger_m + \psi_m \right) | 0 \rangle ,
\eea
where we have used two identities
\be
   \sigma_j^+ =
   \psi_j^\dagger \: \eu^{\ii \pi \sum_{k<j} \psi_k^\dagger \psi_k} =
   \eu^{- \ii \pi \sum_{k<j} \psi_k^\dagger \psi_k} \: \psi_j^\dagger
\ee
and
\be
   \eu^{\ii \pi \psi_i^\dagger \psi_i} =
   2 \psi_i^\dagger \psi_i - 1 =
   \left( \psi_i^\dagger + \psi_i \right)
   \left( \psi_i^\dagger - \psi_i \right) =
   - \left( \psi_i^\dagger - \psi_i \right)
   \left( \psi_i^\dagger + \psi_i \right) .
\ee

Now we define the operators \bea
    A_i & \equiv \psi^\dagger_i + \psi_i \\
    B_i & \equiv \psi^\dagger_i - \psi_i
\eea
which allow us to write the correlators (\ref{rhocorr}) as
\bea
   \rho^x_{lm} & = & \langle 0 | B_l A_{l+1} B_{l+1} \ldots
   A_{m-1} B_{m-1} A_m | 0 \rangle \nonumber \\
   \rho^y_{lm} & = & (-1)^{m-1}\langle 0 | A_l B_{l+1} A_{l+1} \ldots
   B_{m-1} A_{m-1} B_m | 0 \rangle \nonumber \\
   \rho^z_{lm} & = & \langle 0 | A_l B_l A_m B_m | 0 \rangle .
   \label{rhocorr1}
\eea

We can use Wick's Theorem to expand these expectation values in
terms of two point correlation functions. By noticing that
\be
   \langle 0 | A_l A_m | 0 \rangle = \langle 0 | B_l B_m | 0 \rangle = 0
\ee
we write $\rho^z_{lm}$ as
\bea
    \rho^z_{lm} & = & \langle 0 | A_l B_l | 0 \rangle \langle 0 | A_m B_m | 0 \rangle
    -  \langle 0 | A_l B_m | 0 \rangle \langle 0 | A_m B_l | 0 \rangle \nonumber \\
    & = & H^2 (0) - H (m-l) H(l-m)
\eea
where
\be
   H (m-l) \equiv \langle 0 | B_l A_m | 0 \rangle = {1 \over 2}
   \int_0^{2 \pi} {\de q \over 2\pi}\; \eu^{\ii \vartheta_q} \eu^{\ii q (m-l)} .
   \label{H}
\ee

The other two correlators in (\ref{rhocorr1}) involve more terms.
It can be shown \cite{mccoy,LSM-1961} that the Wick's expansion
can be expressed as a Pfaffian of a matrix with elements given by
expectation values of each combination of two operators.

The Pfaffian of a matrix $M$ is defined as \cite{mehta2}
\be
   \Pf( {\bf M} ) \equiv
   \sum_P (-1)^P M_{p_1p_2} M_{p_3p_4} \ldots M_{p_{2n-1}p_{2n}},
   \label{pfaffiandef}
\ee
where $P=\{p_1,p_2,\ldots, p_{2n}\}$ is a permutation of
$\{1,2,\ldots,2n\}$, the sum is performed over all possible
permutations, and $(-1)^P$ is the parity of the permutation.

By using one of the fundamental properties of the Pfaffian:
\be
    \Pf( {\bf M} ) = \sqrt{\det( {\bf M} )}
\ee we can write the spin correlators (\ref{rhocorr1}) as $m-l
\times m-l$ matrix determinants:
\bea
    \rho^x_{lm} & = & \det \left| H (i-j) \right|_{i=l \ldots m-1}^{j= l+1 \ldots m} ,
    \label{rhoxmat} \\
    \rho^y_{lm} & = & \det \left| H (i-j) \right|_{i=l+1 \ldots m}^{j= l \ldots m-1}
    \label{rhoymat}
\eea with matrix elements given by (\ref{H}).

Matrices like (\ref{rhoxmat},\ref{rhoymat}) are very special.
Their entries depend only on the difference between the row and
column index, so that the same elements appear on each diagonal.
Therefore they look like:
\be
   \rho^x_{lm} = \rho^x (N) = \left| \matrix{  H(-1) & H(-2) & H(-3) & \ldots & H(-N) \cr
                             H(0) & H(-1) & H(-2) & \ldots & H(1-N) \cr
                             H(1) & H(0) & H(-1) & \ldots & H(2-N) \cr
                             \vdots & \vdots & \vdots & \ddots & \vdots \cr
                             H(N-2) & H(N-3) & H(N-4) &\ldots & H(-1) \cr
                     } \right|,
\ee \be
   \rho^y_{lm} = \rho^y (N) = \left| \matrix{  H(1) & H(0) & H(-1) & \ldots & H(2-N) \cr
                             H(2) & H(1) & H(0) & \ldots & H(3-N) \cr
                             H(3) & H(2) & H(1) & \ldots & H(4-N) \cr
                             \vdots & \vdots & \vdots & \ddots & \vdots \cr
                             H(N) & H(N-1) & H(N-2) &\ldots & H(1) \cr
                     } \right|,
\ee where $N=m-l$.

Matrices like (\ref{rhoxmat},\ref{rhoymat}) are known as {\it
``Toeplitz Matrices''} and a vast \hbox{mathematical} literature
has been devoted to the study of the asymptotic behavior of their
determinants ({\it ``Toeplitz Determinants''}). McCoy and
co-authors \cite{mccoy} were among the first to develop the theory
of Toeplitz determinants in connection to physical systems. They
showed that the fundamental correlators for the XY model can be
calculated in terms of Toeplitz Determinants. In the next chapter
we are going to calculate the Emptiness Formation Probability in
the XY model, a non trivial correlator, and we are going to show
that it also can be expressed as a Toeplitz determinant.

We will defer an appropriate discussion on the theory of Toeplitz
matrices to the next chapter, where we are going to use it to
calculate the EFP. Here, we are just going to recap the main
results of \cite{mccoy} on the asymptotic behavior of the
fundamental spin correlators (\ref{rhocorr}).

At zero temperature, the asymptotic evaluation of the Toeplitz
determinant in the different regions of the phase diagram for
$\rho^x (N)$ gives \cite{mccoy}
\be
   \rho^x (N) {\stackrel{N \to \infty}{\sim}} \cases{
   (-1)^N {2 \over 1 + \gamma} \left[ \gamma^2 (1 - h^2) \right]^{1/4}
   & $|h|<1$, $\gamma >0$ \cr
   (-1)^N {2 \gamma \over 1 + \gamma} \eu^{1/4}
   2^{1/12} A^{-3} (\gamma N)^{-1/4}
   & $|h|=1$, $\gamma >0$ \cr
   (-1)^N f_1(h,\gamma) {\lambda^N \over \sqrt{N}}
   & $|h|>1$, $\gamma >0$ \cr
   0 & $\gamma = 0$}
   \label{rhox}
\ee where $A = 1.282\ldots$ is the Glaisher's constant, $f_1
(h,\gamma)$ is some function (see \cite{mccoy}) and \be
   \lambda \equiv {h - \sqrt{ h^2 + \gamma^2 - 1} \over 1 - \gamma }.
\ee
For $\rho^y (N)$ it was found \be
   \rho^y (N) {\stackrel{N \to \infty}{\sim}} \cases{
   - (-1)^N f_2(h,\gamma) N^{-3} \lambda^{-2N}
   & $|h|<\sqrt{1-\gamma^2}$, $\gamma >0$ \cr
   0 & $|h|=\sqrt{1-\gamma^2}$, $\gamma >0$ \cr
   (-1)^N f_3(h,\gamma)  N^{-1}\left( {1 - \gamma \over 1 + \gamma} \right)^N
   & $\sqrt{1-\gamma^2}<|h|<1$, $\gamma >0$ \cr
   - (-1)^N {\gamma (1 + \gamma) \over 8} \eu^{1/4}
   2^{1/12} A^{-3} (\gamma N)^{-9/4}
   & $|h|=1$, $\gamma >0$ \cr
   - (-1)^N f_4(h,\gamma) N^{-3/2} \lambda^N
   & $|h|>1$, $\gamma >0$ \cr
   0 & $\gamma = 0$ . }
   \label{rhoy}
\ee

Finally, for $\rho^z (N)$ we have
\be
   \rho^z (N) {\stackrel{N \to \infty}{\sim}} \cases{
   m_z^2 - f_5(h,\gamma) N^{-2}\left( {1 - \gamma \over 1 + \gamma} \right)^N
   & $|h|<\sqrt{1-\gamma^2}$, $\gamma >0$ \cr
   m_z^2 & $|h|=\sqrt{1-\gamma^2}$, $\gamma >0$ \cr
   m_z^2 - {1 \over 2 \pi} \lambda^{-2N-2}
   & $\sqrt{1-\gamma^2}<|h|<1$, $\gamma >0$ \cr
   m_z^2 - (\pi N)^{-2}
   & $|h|=1$, $\gamma >0$ \cr
   1 - {1 \over 2 \pi} N^{-2} \lambda^{-2N}
   & $|h|>1$, $\gamma >0$ \cr
   m_z^2 - \left( {\sin (N \cos^-1 h) \over \pi N } \right)^2
   & $|h|<1$, $\gamma=0$ \cr
   {1 \over 4} & $|h|>1$, $\gamma=0$}
\ee where $m_z^2$ is the magnetization:
\be
   m_z = {1 \over \pi} \int_0^\pi {h - \cos q \over
   \sqrt{ (h - \cos q)^2 + \gamma^2 \sin^2 q } } \de q .
\ee


\chapter{The EFP for the XY Model}
\label{EFPinXY}
In the previous chapter we introduced the One Dimensional Spin-1/2
Anisotropic XY spin chain in a transverse magnetic field. We
studied its phase diagram and we calculated the fundamental
correlators of this model. Now we turn our attention to a
non-trivial correlator know as Emptiness Formation Probability
({\bf EFP}) which we are going to study for the XY model.

We introduced the EFP in Chapter 1, where we discussed its
significance and importance in the theory of integrable models and
in the general problem of calculating correlators in
one-dimensional theories.

The XY model is a very interesting model for the study of the EFP
for several reasons. As we saw in the previous chapter, the XY
spin chain is characterized by a very interesting phase diagram:
as we vary its two parameters (the anisotropy and the magnetic
field), we move through critical and gapped regions. Previous
studies of the EFP focused only on critical phases of various
systems and the XY model offers an opportunity to follow the
behavior of the EFP across a phase transition.

A second, more technical reason, lies in the relative simplicity
of the model. In the previous chapter we showed that the
fundamental correlators of the theory can be exactly expressed as
determinants of a very special class of matrices known as {\it
``Toeplitz Matrices''}. We show that the same property holds for
the EFP. This is of great advantage, since, despite being
considered the simplest correlator for integrable models, the EFP
in general does not have a simple expression. Using the theory of
{\it ``Toeplitz Determinants''}, we are able to calculate the
asymptotic behavior of the EFP in the various regions of the phase
diagram of the XY model.

Most of these results appeared first in \cite{abanovfran}. The
Toeplitz determinant approach was also used in Ref.
\cite{shiroishi} for the EFP in the case of the Isotropic XY model
(Eq. (\ref{spinham}) with $\gamma=0$). In this latter work it was
shown that the EFP decays in a Gaussian way for the critical
theory ($\gamma=0$, $-1 \le h \le 1$). This case corresponds to
one of the two critical lines in the $\gamma-h$ phase diagram of
the model (\ref{spinham}) discussed in the previous chapter. The
other line is the critical magnetization line(s) ($h = \pm 1$). In
the rest of the two-dimensional $\gamma-h$ phase diagram, the
model is non-critical.

For the XXZ spin chain in zero magnetic field, the EFP was found
to have a Gaussian decay $P(n)\sim e^{-\alpha n^{2}}$ as $n \to
\infty$ in the critical regime at zero temperature and exponential
$e^{-\beta n}$ at finite temperature
(\cite{KMST_gen-2002},\cite{KLNS-2002}).

A qualitative argument in favor of Gaussian decay was given in
Ref. \cite{abanovkor} within a field theory approach. It was
argued there that the asymptotics of the EFP are defined by the
action of an optimal fluctuation (instanton) corresponding to the
EFP. In the critical model, this fluctuation will have a form of
an ``$n\times n$'' droplet in space-time with the area $A \sim
n^{2}$ and the corresponding action $S \approx \alpha n^{2}$ which
gives the decay $P(n)\sim e^{-\alpha n^{2}}$. Similarly, at finite
temperature the droplet becomes rectangular (one dimension $n$ is
replaced by an inverse temperature $T^{-1}$) and the action cost
is proportional to $n$, giving $P(n) \sim e^{-\beta n}$. This
argument is based on the criticality of the theory\footnote{More
precisely, on the assumption that temporal and spatial dimensions
of an instanton scale similarly.} and it is interesting to
consider whether it could be extended to a non-critical theory. A
na\"ive extension of the argument would give the optimal
fluctuation with space-time dimensions $n \times \xi$ where $\xi$
is a typical correlation length (in time) of the theory. This
would result in $P(n) \sim e^{-\beta n}$ for non-critical
theories, similarly to the case of finite temperature in critical
regime. The rate of decay $\beta$ would be proportional to the
correlation length of the theory.

In this chapter we examine the relation between the asymptotic
behavior of the EFP and criticality using the example of the XY
model. Using Toeplitz determinant techniques, we obtain that the
EFP is asymptotically exponential in most of the phase diagram
according to the na\"ive expectations and that it is Gaussian only
at $\gamma=0$ in agreement with previous works on XXZ spin chains
and with Ref. \cite{shiroishi}. However, on the critical lines $h
=\pm 1$, in addition to the exponential decay, a pre-exponential
power-law factor arises, with a universal exponent. The power-law
prefactor is present in the isotropic case as well, but with a
different exponent. Using a bosonization approach, we will
interpret the transition from Gaussian to exponential decay.

The chapter is organized in the following way: in Section
\ref{EFPinXY} we explain how one can express the EFP as the
determinant of a Toeplitz matrix and review our results so that
readers who are not interested in derivations can skip the next
sections. In Section \ref{expsection} we analyze the exponential
decay of the EFP for the non-critical and critical phases of the
anisotropic XY Model. In Section \ref{pre-ex} we derive in detail
the asymptotic behaviors, including the pre-exponential factors,
of both non-critical and critical parts of the phase diagram. In
Section \ref{GammaESec} we study a special line of the phase
diagram on which the ground state is known exactly and compare the
explicit results one can obtain using the exact ground state with
the asymptotes of the EFP we derived in the previous sections. In
Section \ref{gammazero} we report on the already known results for
the EFP of the isotropic XY model \cite{shiroishi}. In Section
\ref{Crossover} we make contact with Ref. \cite{abanovkor} using a
bosonization approach to discuss the crossover as a function of
$n$ from the Gaussian to the exponential behavior of EFP for the
case of small anisotropy $\gamma$. The following section gives
some mathematical details on the calculation of the stationary
action in the bosonization approach of Section \ref{Crossover} and
can be skipped by the reader non interested in the mathematical
technique. Section \ref{FiniteEFP} presents the analysis of the
finite temperature behavior of the EFP, which gives an expected
exponential decay. Finally, Section \ref{Conclusions} will
summarize our results. For the reader's convenience we have
collected some results on asymptotic behavior of Toeplitz
determinants which are extensively used in the rest of the paper
in the Appendix \ref{ToeplitzApp}.

\section{EFP as a determinant of a Toeplitz Matrix}

We introduced the XY model in the previous chapter. In this
section we consider the correlator measuring the {\it
``Probability of Formation of Ferromagnetic Strings''} (PFFS)
\be
   P(n) \equiv
   \langle 0 | \prod_{i=1}^n {1 - \sigma_i^z \over 2} | 0 \rangle ,
   \label{EFPDef}
\ee at zero temperature (the non-zero temperature case is deferred
to section \ref{FiniteEFP}). This correlator measures the
probability that $n$ consecutive spins will be aligned downwards
in the ground state of the system.

In terms of the spinless fermions defined in the previous chapter
(\ref{spinlessham}), by direct substitution one can express the
PFFS (\ref{EFPDef}) as the expectation value over the spinless
fermions ground state \cite{shiroishi}
\be
   P(n) = \langle 0 | \prod_{j=1}^n \psi_j \psi_j^\dagger | 0 \rangle.
   \label{expect}
\ee This expression projects the ground state on a configuration
without particles on a string of length $n$ and hence gives the
meaning to the name {\it ``Emptiness Formation Probability''}.

Let us now introduce the $2n \times 2n$ skew-symmetric matrix
${\bf M}$ of correlation functions \be
   {\bf M} = \pmatrix{  {-\ii\bf F} &
                        {\bf G}  \cr
                        - {\bf G}  &
                        \ii{\bf F} \cr
                     },
\ee
where ${\bf F}$ and ${\bf G}$ are $n \times n$ matrices with
matrix elements given by $F_{jk}$ and $G_{jk}$ from
(\ref{F},\ref{G}) respectively.
Then, using Wick's theorem on the r.h.s of (\ref{expect}), we
obtain the EFP as the Pfaffian of the matrix ${\bf M}$
\be
    P(n) =  \Pf( {\bf M} ) ,
\ee where the Pfaffian was introduced in (\ref{pfaffiandef}).
Using one of the properties of the Pfaffian we have
\be
    P(n) =  \Pf( {\bf M} ) = \sqrt{\det( {\bf M} )}.
\ee

We can perform a unitary transformation
\be
   {\bf M'} = {\bf U M U}^{\dagger}
   = \pmatrix { 0 & {\bf S_n} \cr -{\bf S_n}^\dagger & 0 \cr}, \qquad
   {\bf U} = {1 \over \sqrt{2}} \pmatrix { {\bf I} & {\bf -I} \cr
                                           {\bf I} & {\bf I} \cr},
\ee
where ${\bf I}$ is a unit $n \times n$ matrix and ${\bf S_n}={\bf
G}+\ii{\bf F}$ and ${\bf S_n}^\dagger={\bf G}-\ii{\bf F}$.
This allows us to calculate the determinant of ${\bf M}$ as
\be
   \det ( {\bf M} ) = \det ( {\bf M'} ) =
   \det ( {\bf S_n} ) \cdot \det ( {\bf S_n}^\dagger )
   = \left|\det ( {\bf S_n} ) \right|^2.
   \label{detM}
\ee
The matrix ${\bf S_n}$ is a $n \times n$ Toeplitz matrix (i.e. its
matrix elements depend only on the difference of row and column
indices \cite{basor} like the matrices defined in the previous chapter
(\ref{rhoxmat},\ref{rhoymat})).
The generating function $\sigma(q)$ of a Toeplitz matrix is defined by
\be
    ({\bf S_n})_{jk}=\int_0^{2 \pi} {d q \over 2 \pi}\; \sigma (q)
    \eu^{\ii q(j-k)}
 \label{Tg}
\ee
and in our case can be found from (\ref{F},\ref{G}) as
\be
   \sigma(q) = {1 \over 2}\left( 1 + \eu^{\ii\vartheta_q} \right)
   = {1 \over 2} + {\cos q - h + \ii \gamma \sin q
   \over 2 \sqrt{(\cos q-h)^2+\gamma^2\sin^2q}} .
   \label{genfunc}
\ee

Thus, the problem of the calculation of the EFP
\be
   P(n) =  \left|\det ( {\bf S_n} ) \right|,
  \label{PnX}
\ee is reduced (exactly) to the calculation of the determinant of
the $n \times n$ Toeplitz matrix ${\bf S_n}$ defined by the
generating function (\ref{Tg},\ref{genfunc}). The representation
(\ref{PnX}) is exact and valid for any $n$. In our study we are
interested in finding the asymptotic behavior of (\ref{PnX}) at
large $n \to \infty$. \footnote{The reader might notice that our
generating function (\ref{genfunc}) is almost the same as the one
analyzed by Barouch et al. in \cite{mccoy} ($\sigma_{[13]}(q) =
{\cos q - h + \ii \gamma\sin q \over \sqrt{ (\cos q - h)^2 +
\gamma^2 \sin^2 q } } $). The only difference is the shift by the
unity in our expression. This difference changes dramatically the
analytical structure of the generating function, in particular,
its winding number around the origin, and requires a new analysis
of the generated Toeplitz determinants.}

Most of these results are derived using known theorems on the
asymptotic behavior of Toeplitz determinants. We have collected
these theorems in Appendix~\ref{ToeplitzApp}. In the following
sections we apply them to extract the corresponding asymptotes of
$P(n)$ at $n \to \infty$ in the different regions of the phase
diagram. Two major distinctions have to be made in this process.
For the critical isotropic ($\gamma = 0$) XY model, one applies
what is known as Widom's Theorem and finds a Gaussian behavior
with a power law prefactor \cite{shiroishi}. In the rest of the
phase diagram, we apply different formulations of what is known in
general as the Fisher-Hartwig conjecture, which always leads to an
exponential asymptotic behavior. As expected, we find a pure
exponential decay for the EFP in the non-critical regions.

For $h > 1$, the exponential decay is modulated by an additional
oscillatory behavior.

At the critical magnetizations $h = \pm 1$, we discover an
exponential decay with a {\it power law pre-factor}. Moreover, by
extending the existing theorems on Toeplitz determinants beyond
their range of applicability, for $h = \pm 1$ we obtain the first
order corrections to the asymptotics as a faster decaying power
law with the same exponential factor. For $h = 1$, the first order
correction is also oscillating, which indicates that the EFP has
an oscillatory behavior for $h \ge 1$.

\begin{figure}
 \includegraphics[width=\columnwidth]{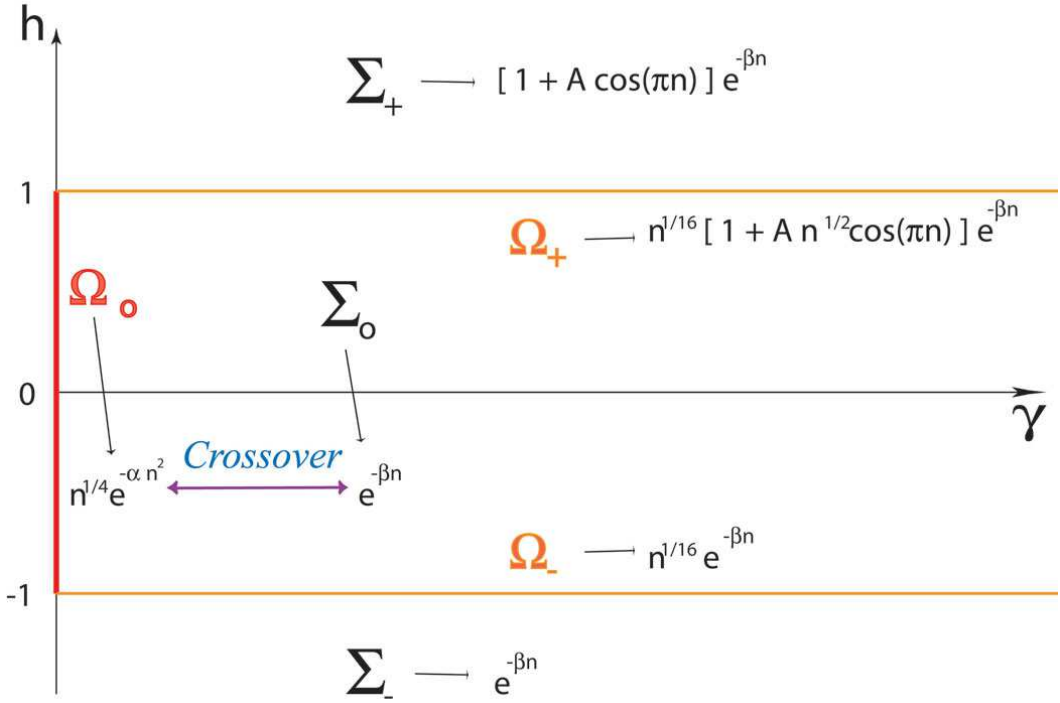}
\caption[EFP asymptotic behavior in the phase diagram of the XY
Model] {Asymptotic behavior of the EFP in the different regions of
the phase diagram of the XY Model (only the part $\gamma \ge 0$ is
shown). The theory is critical for $h = \pm 1$ ($\Omega_\pm$) and
for $\gamma = 0$ and $|h| < 1$ ($\Omega_0$). The line $\Gamma_I$
represents the Ising Model in transverse field. On the line
$\Gamma_E$ the ground state of the theory is a product of single
spin states.}
   \label{phasediagram-behavior}
\end{figure}

The reader who is not interested in the mathematical details of
our derivations can find the results in Figure
\ref{phasediagram-behavior} and in Table \ref{table1} and skip the
following sections to go directly to Sec.~\ref{Crossover}, where
we analyze the crossover between the Gaussian behavior at $\gamma
= 0$ and the asymptotic exponential decay at finite $\gamma$ using
a bosonization approach.

\section{Singularities of $\sigma(q)$ and exponential behavior of the
EFP}
\label{expsection}

To derive the asymptotic behavior of the EFP we rely on the
theorems on determinants of Toeplitz matrices.
These theorems depend greatly on the analytical structure of the
generating function (\ref{genfunc}), especially on its zeros and
singularities.

Setting $\gamma = 0$  in (\ref{genfunc}), we see that for the
Isotropic XY model the generating function has only a limited
support within its period $[0,2\pi]$.
This case is covered by what is known as Widom's Theorem and
will be considered in Section \ref{gammazero}.

\begin{figure}
   \dimen0=\textwidth
   \advance\dimen0 by -\columnsep
   \divide\dimen0 by 3
   \noindent\begin{minipage}[t]{\dimen0}
   \begin{flushleft}
   (a): ${\bf \Sigma_-}$
   \end{flushleft}
   \includegraphics[width=\columnwidth]{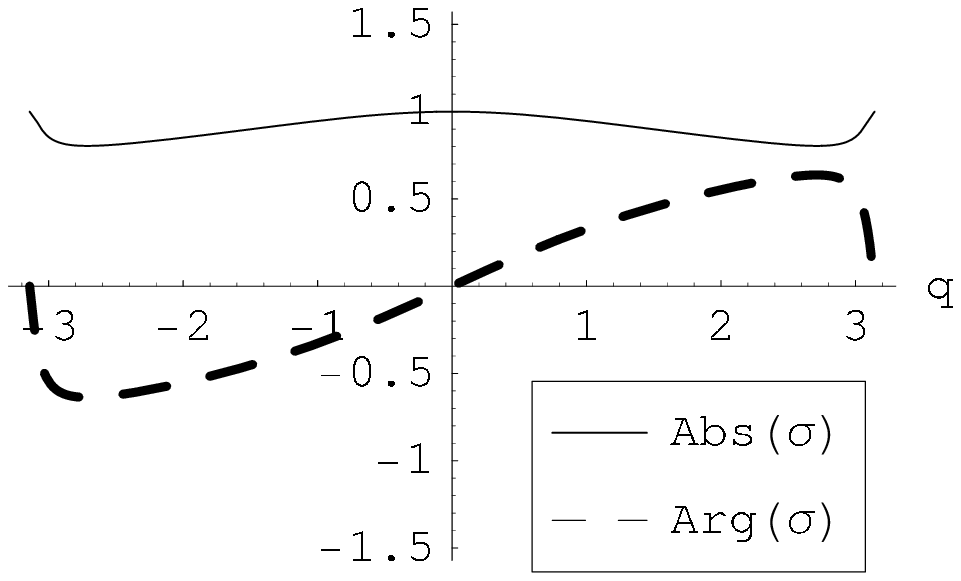}
   \end{minipage}
   \hfill
   \begin{minipage}[t]{\dimen0}
   \begin{flushleft}
   (b): ${\bf \Omega_-}$
   \end{flushleft}
   \includegraphics[width=\columnwidth]{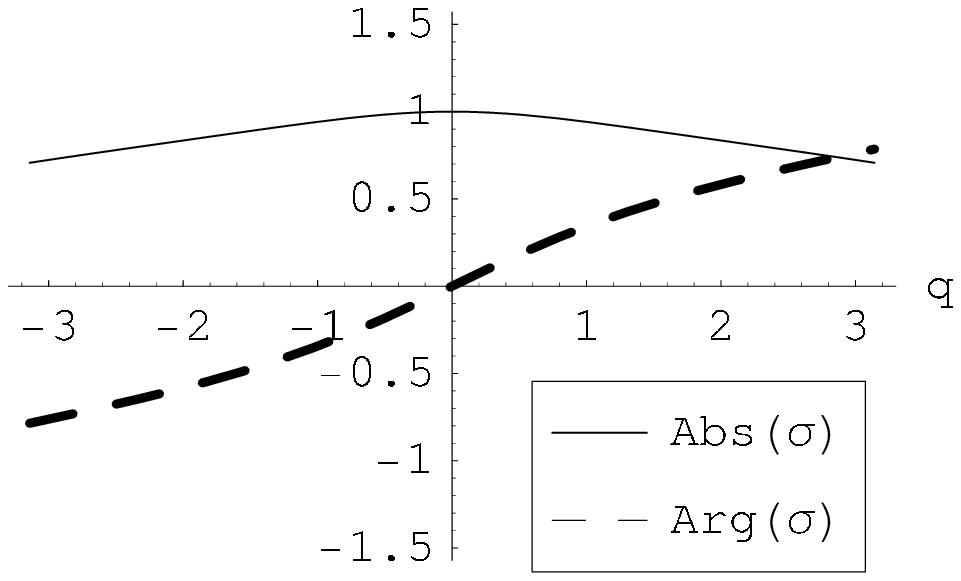}
   \end{minipage}
   \hfill
   \begin{minipage}[t]{\dimen0}
   \begin{flushleft}
   (c): ${\bf \Sigma_0}$
   \end{flushleft}
   \includegraphics[width=\columnwidth]{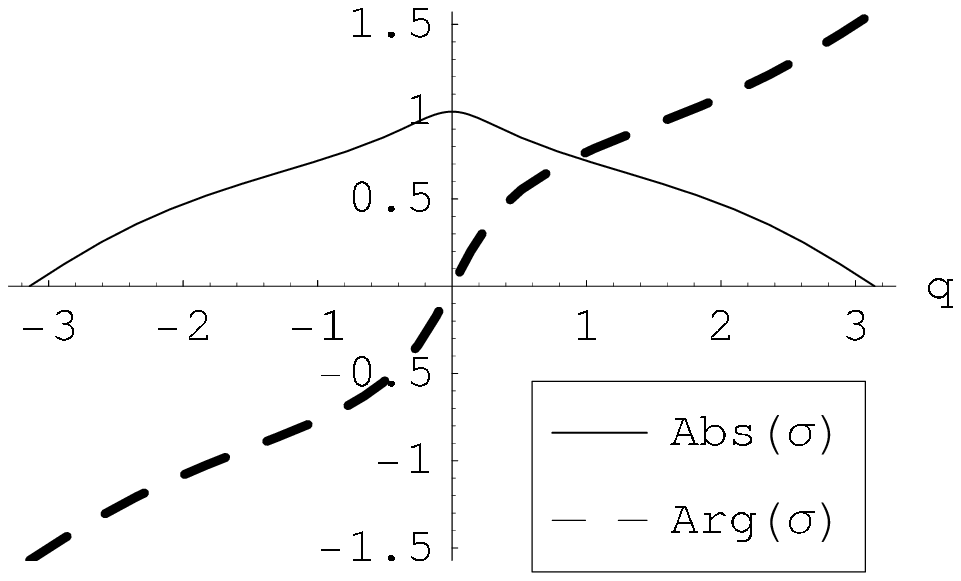}
   \end{minipage}
   \noindent\begin{minipage}[t]{\dimen0}
   \begin{flushleft}
   (d): ${\bf \Omega_+}$
   \end{flushleft}
   \includegraphics[width=\columnwidth]{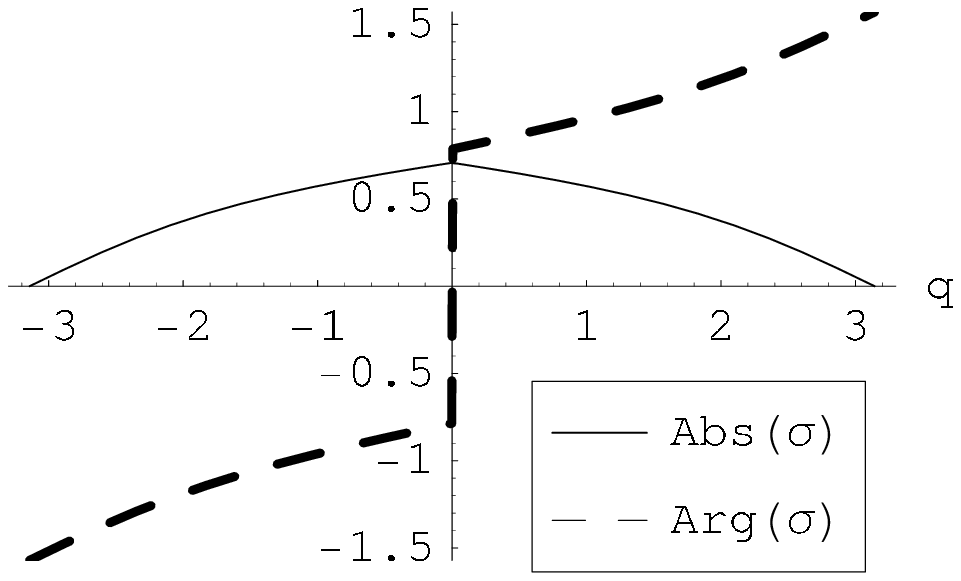}
   \end{minipage}
   \hfill
   \begin{minipage}[t]{\dimen0}
   \begin{flushleft}
   (e): ${\bf \Sigma_+}$
   \end{flushleft}
   \includegraphics[width=\columnwidth]{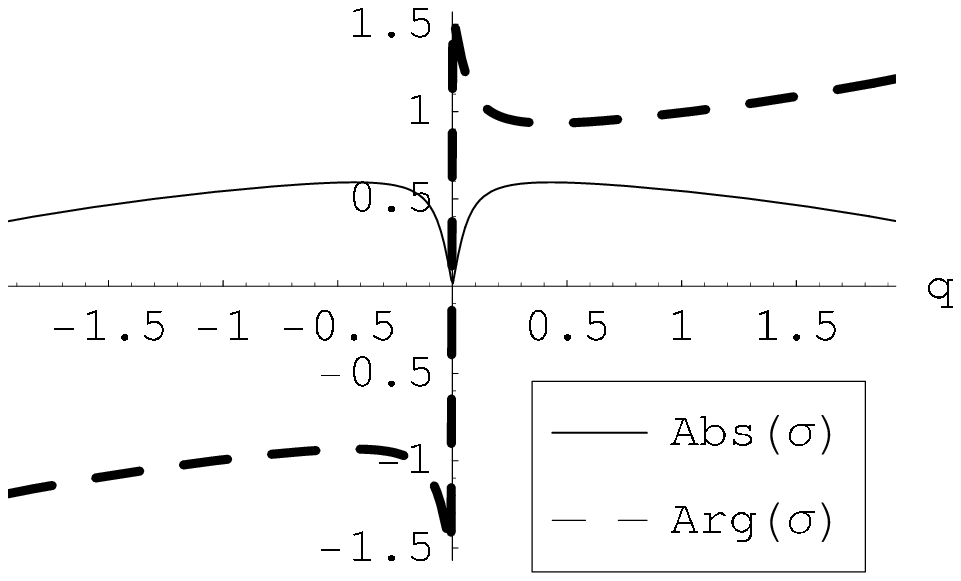}
   \end{minipage}
   \hfill
   \begin{minipage}[t]{\dimen0}
   \caption[Plot of the absolute value and argument of the generating
   function]
   {Plot of the absolute value and argument of the generating
   function (\ref{genfunc}) for $\gamma = 1.5$ at different values of
   $h$. From (a) to (e) $h = -1.1$, $-1$, $0.5$, $1$, $1.1$, respectively.}
   \label{GenFuncPlot}
   \end{minipage}
\end{figure}

In the remaining parts of the phase-diagram the generating function has
only pointwise singularities (zeros) as it is shown in
Fig.~\ref{GenFuncPlot}.
These cases are treated under a general (not yet completely proven)
conjecture known as the Fisher-Hartwig conjecture (FH), which
prescribes the leading asymptotic behavior of the Toeplitz determinant
to be exponential in $n$:
\be
   P(n) {\stackrel{n \rightarrow \infty}{\sim}} \eu^{-\beta n}.
\ee

While the pre-exponential factors depend upon the particulars of the
singularities of the generating function, the exponential decay rate is
given in the whole phase diagram ($\gamma \neq 0$) according to FH as
\bea
   \beta(h, \gamma) & = &
   - \int_0^{2 \pi} {\de q \over 2 \pi}\; \log  \left| \sigma(q) \right|
   \nonumber \\ & = &
   - \int_0^\pi {\de q \over 2 \pi}\;
   \log \left[ {1 \over 2} \left( 1 + {\cos q -h \over \sqrt{
   \left( \cos q - h \right)^2 + \gamma^2  \sin^2 q} } \right) \right] .
   \label{betagh}
\eea
The integral in (\ref{betagh}) is convergent for all $h$ and all
$\gamma \neq 0$ and $\beta(h,\gamma)$ is a continuous function of its
parameters.

\begin{figure}
  \includegraphics[width=\columnwidth]{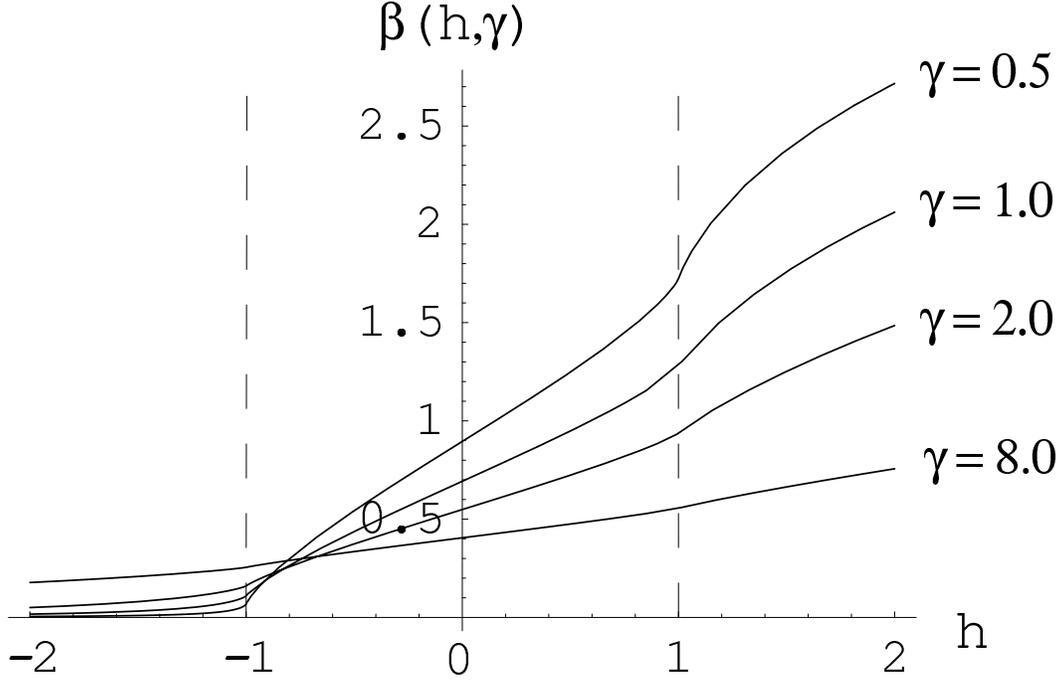}
\caption[Plot of the decay rate $\beta$ as a function of the
parameters $\gamma$ and $h$] {Plot of the decay rate $\beta$ as a
function of the parameters $\gamma$ and $h$. The function diverges
for $\gamma = 0$ and is continuous for $h = \pm 1$ (although it
has weak singularities at $h=\pm 1$).}
   \label{betagraph}
\end{figure}

In Fig.~\ref{betagraph}, $\beta(h,\gamma)$ is plotted as a function
of $h$ at several values of $\gamma$.
One can see that $\beta(h,\gamma)$ is continuous but has weak
(logarithmic) singularities at $h= \pm 1$.
This is one of the effects of the criticality of the model on
the asymptotic behavior of EFP.

These weak singularities are also a manifestation of the rich
analytical structure underlying $\beta(h,\gamma)$ and the
generating function (\ref{genfunc}). To understand these
structures, we interpret the periodic generating function
(\ref{genfunc}) as the restriction to the unit circle ($z=
\eu^{\ii \theta}$) of the complex function \be
   \sigma (z) \equiv {1 \over 2} \left( 1 +
   { p_1 (z) \over \sqrt{p_1 (z) \cdot p_2 (z)} } \right),
   \label{sigmaz}
\ee
where
\bea
   p_1 (z) = {1 + \gamma \over 2 z} (z - z_1) (z - z_2), \\
   p_2 (z) = {1 + \gamma \over 2 z} (z_1 z - 1) (z_2 z - 1)
\eea
with
\bea
   z_1 = {h - \sqrt{h^2 + \gamma^2 - 1} \over 1 + \gamma},
   \label{z1} \\
   z_2 = {h + \sqrt{h^2 + \gamma^2 - 1} \over 1 + \gamma}.
   \label{z2}
\eea

\begin{figure}
   \dimen0=\textwidth
   \advance\dimen0 by -\columnsep
   \divide\dimen0 by 3
   \noindent\begin{minipage}[t]{\dimen0}
   (a)
   \includegraphics[width=\columnwidth]{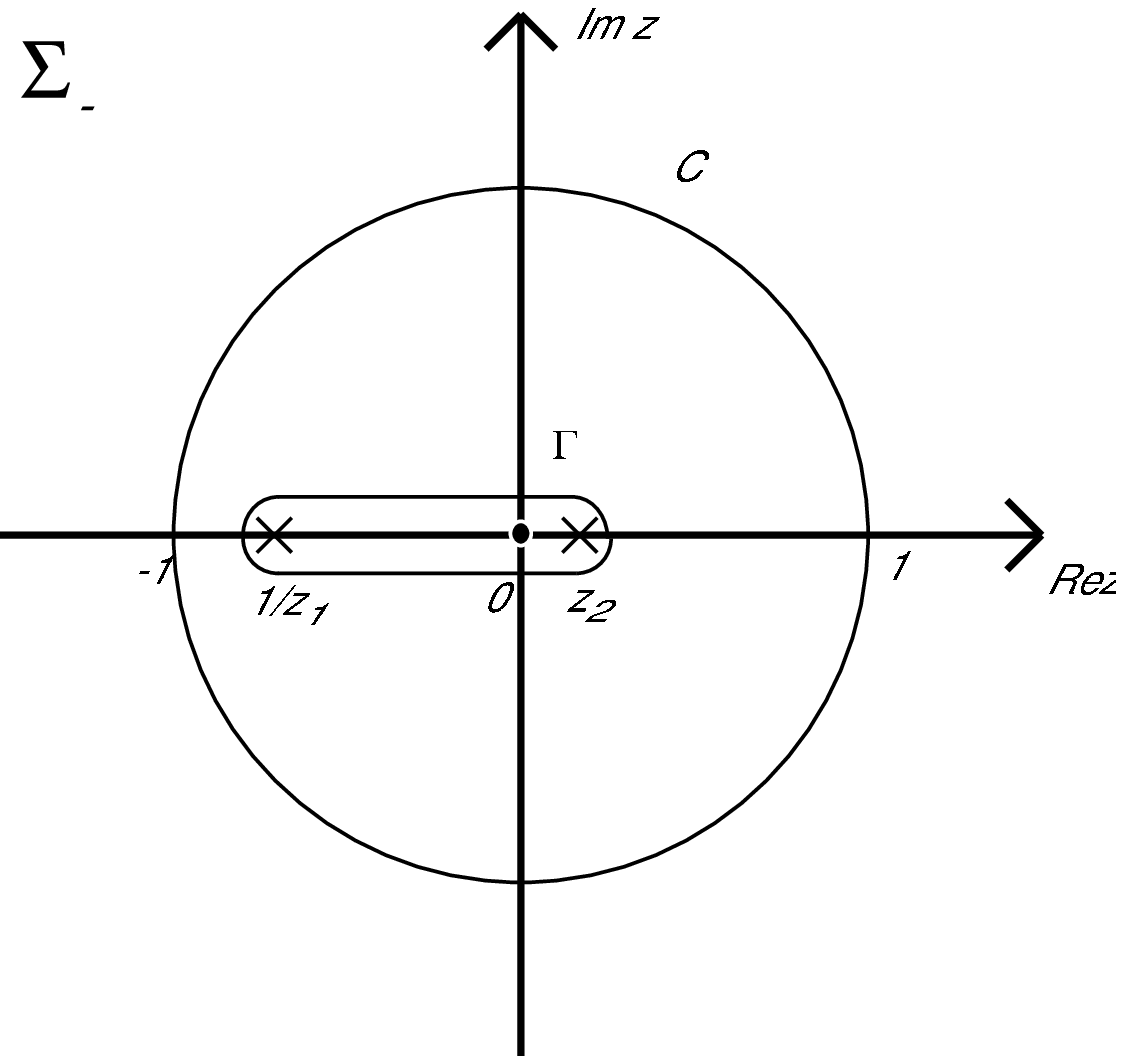}
   \end{minipage}
   \hfill
   \begin{minipage}[t]{\dimen0}
   (b)
   \includegraphics[width=\columnwidth]{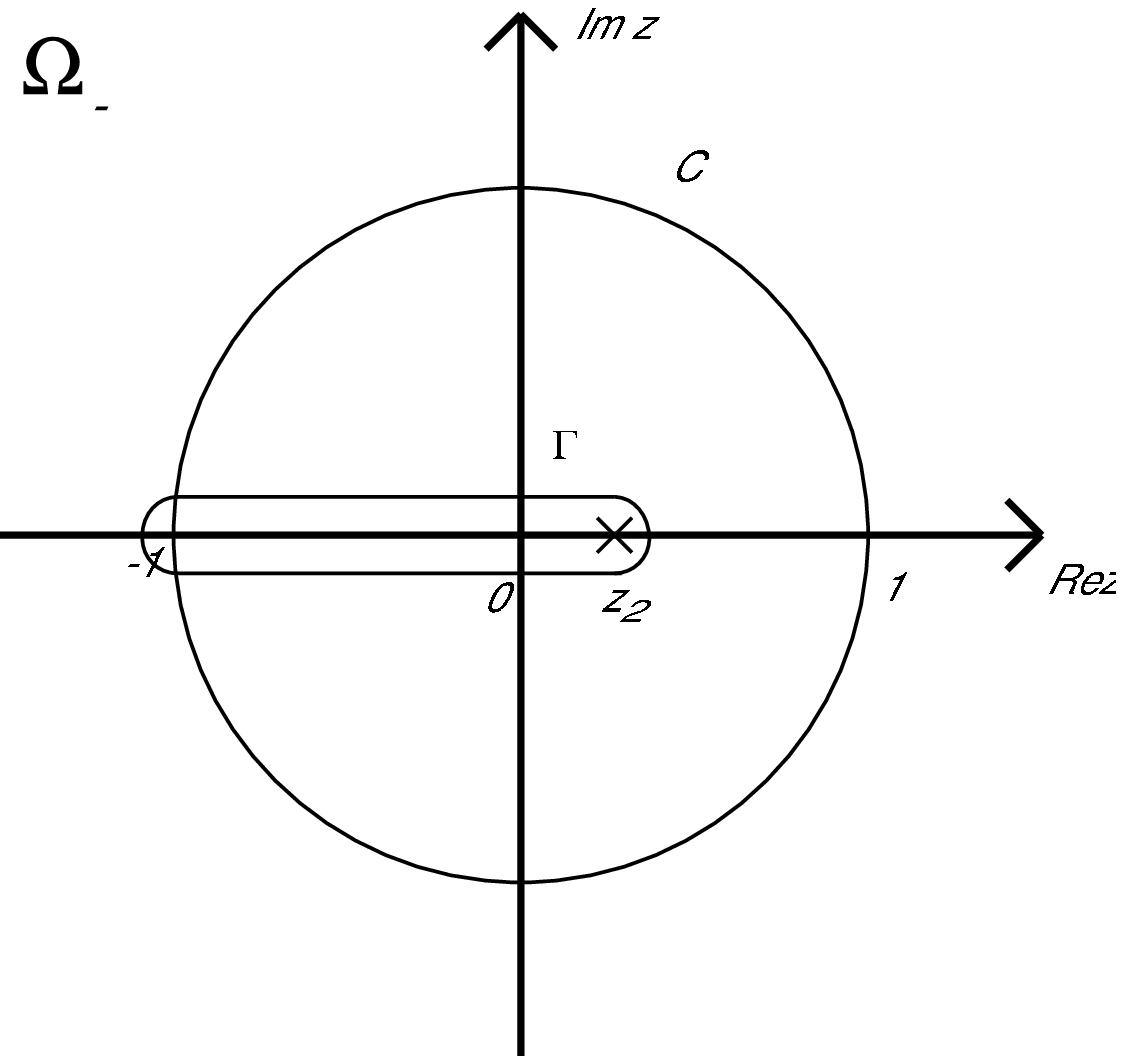}
   \end{minipage}
   \hfill
   \begin{minipage}[t]{\dimen0}
   (c)
   \includegraphics[width=\columnwidth]{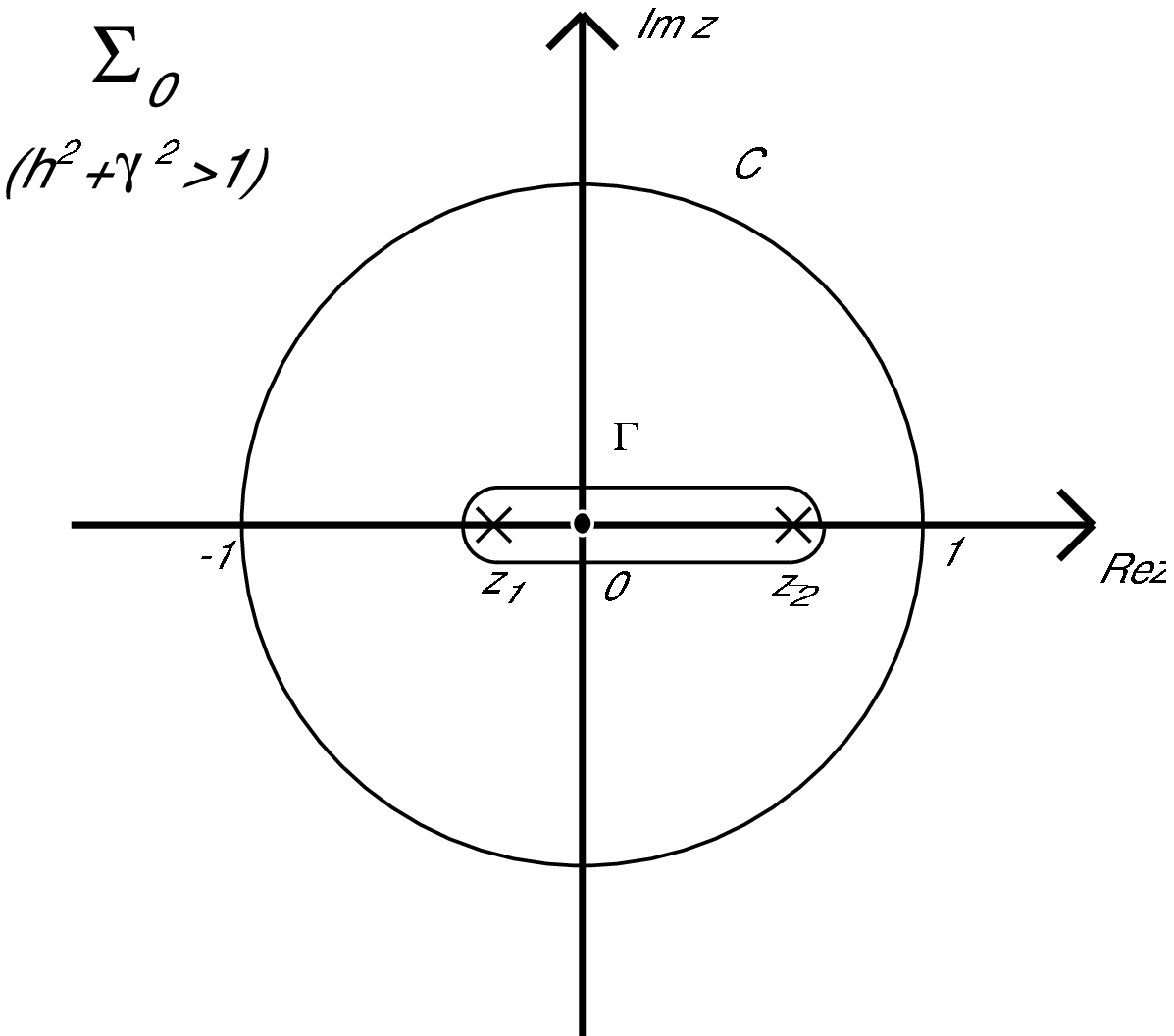}
   \end{minipage}
   \noindent\begin{minipage}[t]{\dimen0}
   (d)
   \includegraphics[width=\columnwidth]{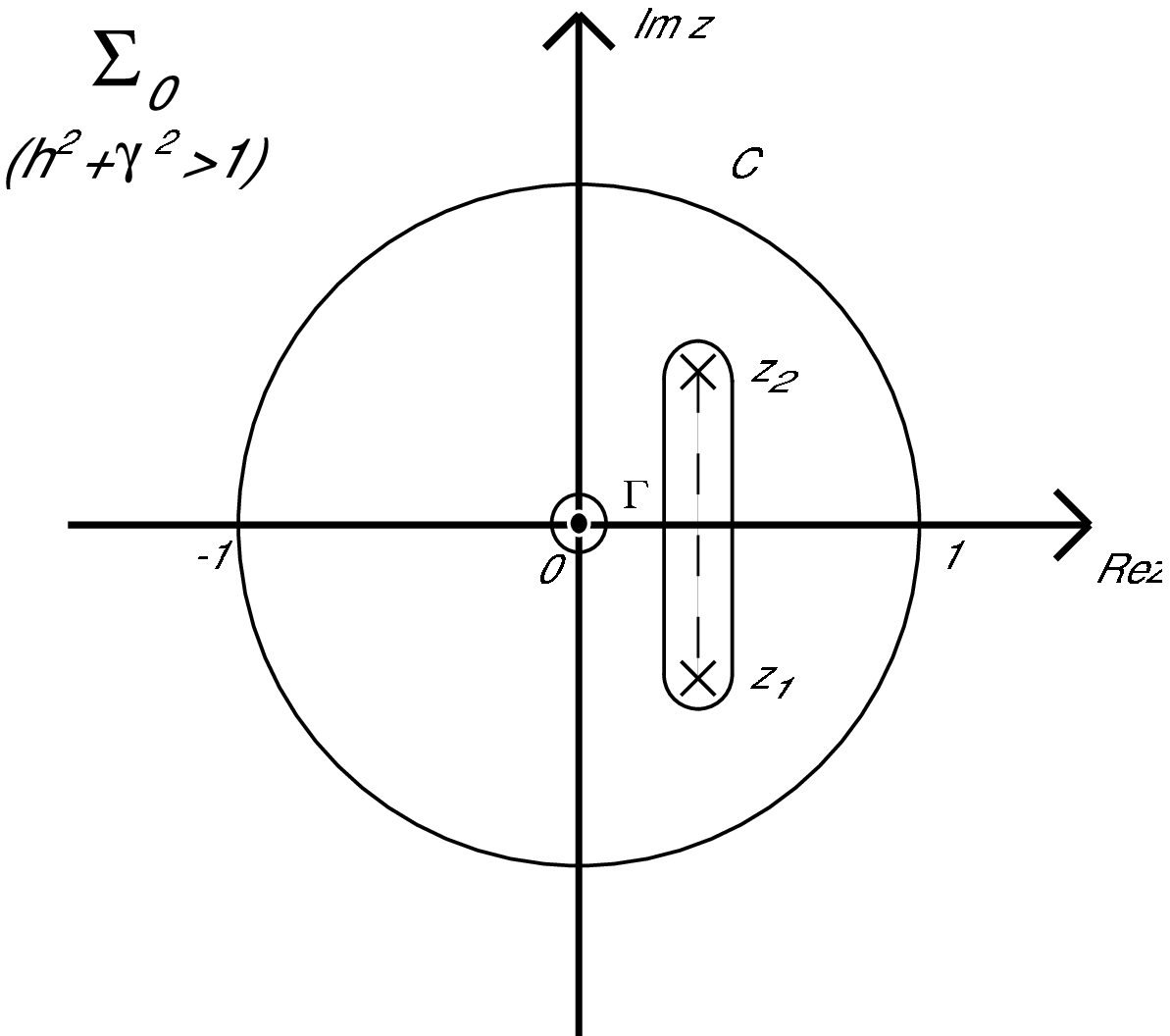}
   \end{minipage}
   \hfill
   \begin{minipage}[t]{\dimen0}
   (e)
   \includegraphics[width=\columnwidth]{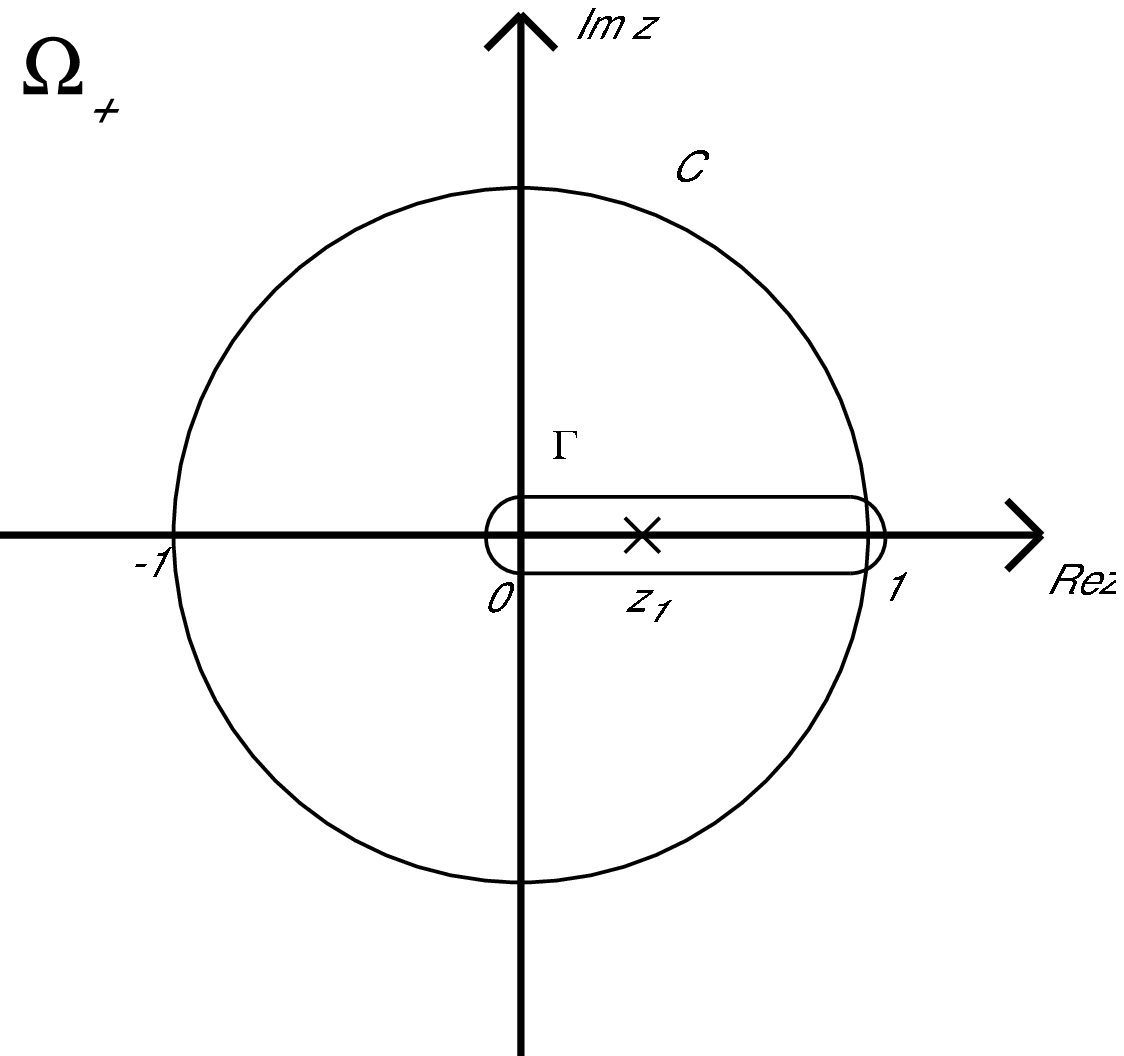}
   \end{minipage}
   \hfill
   \begin{minipage}[t]{\dimen0}
   (f)
   \includegraphics[width=\columnwidth]{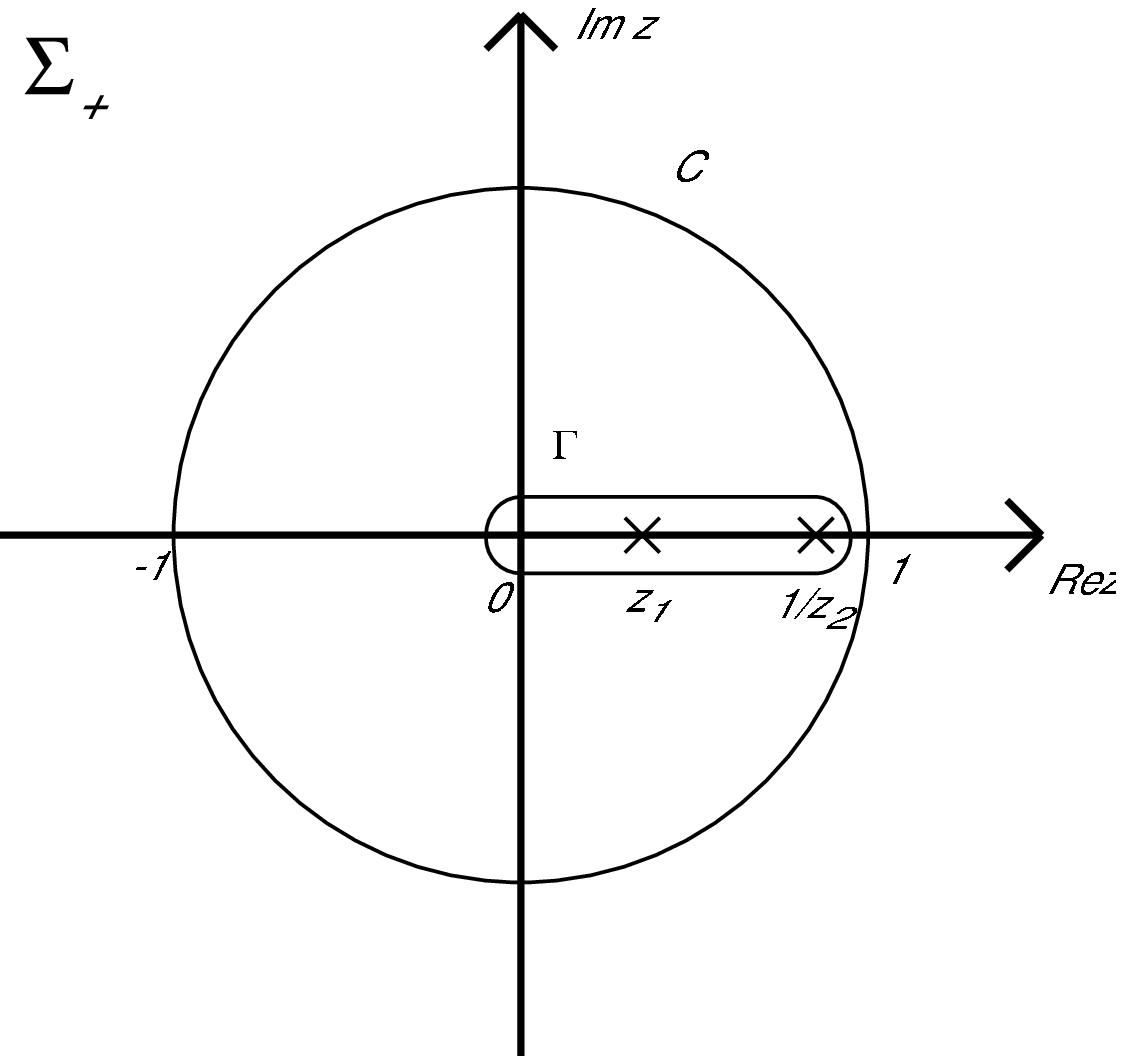}
   \end{minipage}
   \begin{minipage}[t]{\textwidth}
\caption[Branch cuts of the Logarithm in the definition of $\beta(h,\gamma)$]
{The integral in (\ref{betagh}) is performed over the unit circle
$C$. The analytical structure of the integrand allows for a deformation
of the contour of integration into $\Gamma$, which encloses a
logarithmic branching line, different in the various regions of the
phase-diagram (in (d), $\Gamma$ encloses also a simple pole at
the origin). The roots $z_1$ and $z_2$ were defined in (\ref{z1})
and (\ref{z2}).}
   \label{SigmaStruct}
   \end{minipage}
\end{figure}

The integral in (\ref{betagh}) can be regarded as a contour integral
over the unit circle of the function (\ref{sigmaz}). We can deform the
contour of integration taking into account the complex structure of the
integrand in the various regions (see Fig.~\ref{SigmaStruct}) and
express (\ref{betagh}) as a simpler integral on the real axis (after
partial integration and some algebra).

\subsection{The non-critical regions ($\Sigma_\pm$ and $\Sigma_0$)}

\subsubsection{$\Sigma_-$ ($h < -1$)}

For $h<-1$, the analytical structure of the integrand of (\ref{betagh})
is shown in Fig.~\ref{SigmaStruct}a.
We re-write the decay rate (\ref{betagh}) in this region as
\be
   \beta (h,\gamma)  = {1 \over 2}
   \ln \left[ {\sqrt{h^2 + \gamma^2 -1} - h \over \gamma +1 } \right]
   - \Lambda(h,\gamma) - \Delta(h,\gamma),
   \label{betaminus}
\ee
where
\bea
   \Lambda(h,\gamma) & \equiv & \ln \left| {1 \over 2} \left( 1 - {h \over |h|}
   \sqrt{ {1 - \gamma \over 1 + \gamma} } \right) \right|,
   \label{Lambda} \\
   \Delta(h,\gamma) & \equiv & \int^1_{|K|} {\de x \over 2 \pi}
   {1 \over \sqrt{(1-x^2) (x^2 - K^2)} }
   \left( x + {K \over x} \right)
   \ln \left| {x- a \over x + a } \right| ,
   \label{Delta}
\eea
with
\bea
   K & \equiv & { \sqrt{h^2 + \gamma^2 -1} - \gamma
   \over \sqrt{h^2 + \gamma^2 -1} + \gamma },
   \label{kappa}
   \\
   a & \equiv &
   {\sqrt{h^2 + \gamma^2 -1} - \gamma \over h - 1 } .
\eea
This decomposition of $\beta (h,\gamma)$ is especially useful
in analyzing the transitions between non-critical and critical
regimes.  In fact, we will see that the functions
$\Lambda(h,\gamma)$ and $\Delta(h,\gamma)$ defined above are
universal across the phase diagram (hence the need for the seemingly
redundant absolute values in our definitions).
The absolute value in the logarithm of the integrand is relevant for $\gamma > 1$,
since its argument changes sign and vanishes within the interval of integration
($a>K$ for $\gamma >1$).

\subsubsection{$\Sigma_0$ ($|h| < 1$)}

As before, we can express the contour integral defining
$\beta(h,\gamma)$ as a standard integral on the real axis.
For $|h| < 1$ and $h^2 + \gamma^2 > 1$, the structure of the integrand
is depicted in Fig.~\ref{SigmaStruct}c and the decay rate is simply
\be
   \beta(h,\gamma) =  - \Lambda(h,\gamma) - \Delta(h,\gamma),
   \label{beta0}
\ee
where $\Lambda(h,\gamma)$ and $\Delta(h,\gamma)$ have already been
defined in (\ref{Lambda},\ref{Delta}).

For $h^2 + \gamma^2 < 1$, the structure is quite different (see
Fig.~\ref{SigmaStruct}d). In this region the expression for
$\beta(h,\gamma)$ in terms of a real axis integral is complicated.
We can write the result as:
\bea
  \beta(h,\gamma) & = & - \Lambda(h,\gamma)
  - \sqrt{ 2 (1 + \gamma) (1 - h^2 - \gamma^2) } \times
  \nonumber \\
  && \times
  \int_0^1 {\de x \over 2 \pi}
  \left\{ 4 x \arctan \left[ {x \over h} \sqrt{1 - h^2 - \gamma^2} \right]
  \left( { A(x) + B(x) + C(x) \over \sqrt{1 - x^2} \sqrt{ \sqrt{q(x)} + p(x)} } \right)
  \right. \nonumber \\
  && \left. \qquad + {h \over |h|}
  2 x \ln \left[ {h^2 + (1 - h^2 - \gamma^2) x^2 \over (1 + \gamma)^2} \right]
  \left( { A(x) - B(x) - C(x) \over \sqrt{1 - x^2} \sqrt{ \sqrt{q(x)} - p(x)} } \right)
  \right\} \nonumber \\
\eea
where $\Lambda(h,\gamma)$ was defined in (\ref{Lambda}) and
\bea
   A(x) & \equiv & (1 + \gamma) { r(x) \over t(x)} , \\
   B(x) & \equiv & 2 (1 + \gamma) { s(x) \over t(x) \sqrt{q(x)}} , \\
   C(x) & \equiv & (\gamma^2 - 1 + 2 \gamma h^2) {1 \over \sqrt{q(x)} } ,
\eea
with
\bea
   p(x) & \equiv &
   (\gamma + 1)^3 - (3 \gamma + 1) h^2 + (\gamma - 1) (1 - h^2 -\gamma^2) x^2 , \\
   q(x) & \equiv &
   (\gamma + 1)^6 - 2 (3 \gamma^4 + 10 \gamma^3 + 12 \gamma^2 + 6 \gamma + 1) h^2
   + (9 \gamma^2 + 6 \gamma + 1) h^4 \nonumber \\
   && + 2 [\gamma^4 + 2 \gamma^3 - 2 \gamma -1 + (5 \gamma^2 + 2 \gamma +1) h^2]
   (1 - h^2 -\gamma^2) x^2 \nonumber \\
   && + (\gamma - 1)^2 (1 - h^2 -\gamma^2)^2 x^4 , \\
   r(x) & \equiv &
   (\gamma + 1)^2 - (2 \gamma + 1) h^2 + (1 - h^2 -\gamma^2) x^2 , \\
   s(x) & \equiv &
   (\gamma + 1)^4 - (3 \gamma^4 + 8 \gamma^3 + 9 \gamma^2 + 6 \gamma + 2) h^2
   + (5 \gamma^2 + 2 \gamma + 1) h^4 \nonumber \\
   &&  + [ (\gamma + 1)^2 - (3 \gamma^2 + 6 \gamma - 1) h^2 ] (1 - h^2 -\gamma^2) x^2 , \\
   t(x) & \equiv & [ (\gamma + 1 + h)^2 + (1 - h^2 -\gamma^2) x^2 ]
   [ (\gamma + 1 - h)^2 + (1 - h^2 -\gamma^2) x^2 ] .
   \nonumber \\
\eea

\subsubsection{$\Sigma_+$ ($h > 1$)}

A calculation similar to the previous ones (see Fig.~\ref{SigmaStruct}f)
gives the expression for the decay factor for $h > 1$:
\be
   \beta (h,\gamma)  = {1 \over 2}
   \ln \left[ {\sqrt{h^2 + \gamma^2 -1} + h \over \gamma + 1 } \right]
   - \Lambda(h,\gamma) - \Delta(h,\gamma),
   \label{betaplus}
\ee
where $\Lambda(h,\gamma)$ and $\Delta(h,\gamma)$ were introduced in
(\ref{Lambda}) and (\ref{Delta}).

One important difference exists in this region: as will be
discussed in length later in Section \ref{SigmaPSec}, in $\Sigma_+$
there are two equivalent representations of the generating function.
This ambiguity reflects on the value of $\beta$, in that the choice of
the representation for the generating function determines the branch cuts
in Fig.~\ref{SigmaStruct}.
We will see that we have to use both values of $\beta$, which differ
only by an imaginary constant:
\be
   \beta' = \beta + \ii \pi
   \label{betapluspi}
\ee
and this will add an oscillatory behavior to the EFP.

\subsection{The critical lines ($\Omega_\pm$)}

We can calculate the decay factor $\beta$ at $h=1$ ($\Omega_+$) from a
limiting procedure on (\ref{beta0}) or (\ref{betaplus}).
At $h = 1$, only $\Delta(h,\gamma)$ is nonvanishing, thus guaranteeing
the continuity of $\beta$ across the critical line.
From an appropriate limit of (\ref{Delta}), we calculate the decay
rate for $h = 1$ as
\be
   \beta(1,\gamma) = - \int_0^1 {\de x \over 2\pi}
   {1 \over \sqrt{1 - x^2} }
   \ln \left| {1 -\gamma x \over 1 + \gamma x} \right|
   - \ln \left| {1 \over 2}
   \left( 1 - \sqrt{ 1 - \gamma \over 1 + \gamma} \right) \right| .
   \label{betaPlus}
\ee
For $\gamma <1$, we can expand the logarithm in series and perform the
integral:
\be
   \beta(1,\gamma <1) = {1 \over \sqrt{\pi} }
   \sum_{n=0}^\infty { n! \over \Gamma(n + 1/2) }
   { \gamma^{2 n + 1} \over (2 n + 1)^2 }
    - \ln \left[ {1 \over 2}
   \left( 1 - \sqrt{ 1 - \gamma \over 1 + \gamma} \right) \right].
\ee
As discussed before in reference to (\ref{betaplus}), the definition of
$\beta(1,\gamma)$ is not unique and, as in (\ref{betapluspi}),
will generate again an oscillatory behavior for the EFP (see later in
Sec.~\ref{ssOmegap}).

The value of $\beta$ at $h = -1$ can also be obtained  from a limiting
procedure on (\ref{Delta})
\be
   \beta(-1,\gamma) = \int_0^1 {\de x \over 2\pi}
   {1 \over \sqrt{1 - x^2} } \ln \left| {1 -\gamma x \over 1 + \gamma x} \right|
   - \ln \left| {1 \over 2} \left( 1 + \sqrt{ 1 - \gamma \over 1 + \gamma} \right)
   \right| .
   \label{betaMinus}
\ee
Again, for $\gamma <1$, we can expand the logarithm in series to calculate the
integral:
\be
   \beta(-1,\gamma <1) = - {1 \over \sqrt{\pi} }
   \sum_{n=0}^\infty { n! \over \Gamma(n + 1/2) }
   { \gamma^{2 n + 1} \over (2 n + 1)^2 }
    - \ln \left[ {1 \over 2}
   \left( 1 + \sqrt{ 1 - \gamma \over 1 + \gamma} \right) \right].
\ee

As can be seen in Fig.~\ref{betagraph}, the decay factor $\beta$
is continuous across the critical lines, but has a discontinuity
of its derivative. As $\beta$ approaches the critical lines, it
actually shows a non-analytical behavior leading to a logarithmic
singularity:
\be
   \beta(h=\pm 1 + \epsilon,\gamma) = \beta(\pm 1,\gamma)
   + {\gamma \over \pi} \; \epsilon \ln | \epsilon | .
\ee
The derivative $d\beta/dh$ diverges logarithmically as
$h \to \pm 1$.

Moreover, one can easily notice from the difference between expression
(\ref{beta0}) and (\ref{betaplus}) that even the finite part of
the derivative of $\beta(h,\gamma)$ by $h$ is different if one
approaches the critical line $h = 1$ from above or below, due to the
appearance of the additional term in (\ref{betaplus}).
The same holds across the critical line $h = -1$, due to the presence
of the first term in (\ref{betaminus}), which doesn't appear in
(\ref{beta0}).

\section{The pre-exponential factors}
\label{pre-ex}

For $\gamma \ne 0$, the leading behavior of the EFP is always
exponential. However, the singularities of  the generating function
are different in different regions of the phase diagram and we must
therefore use different forms of the Fisher-Hartwig
conjecture in order to derive the pre-exponential factors and determine
the asymptotic behavior of $P(n)$.
We will now show how to obtain the results for each of the regions.

\subsection{The non-critical regions ($\Sigma_\pm$ and $\Sigma_0$)}
\label{non-critical}

\subsubsection{$\Sigma_-$ ($h < -1$)}
\label{SigmaMSec}

In this region ($\gamma \neq 0$, $h<-1$) the generating function
(\ref{genfunc}) is nonzero for all $q$ (see Fig.~\ref{GenFuncPlot}a):
this is the simplest case and can be treated using the (rigorously
proven) {\it Strong Szeg\"o Limit Theorem}, see (\ref{szego}).
It gives
\be
  P(n) =  \left| \det ({\bf S_n}) \right|
  {\stackrel{n \rightarrow \infty}{\sim}}
  {\it E_-} (h,\gamma) \eu^{-\beta(h,\gamma) n}
  \label{pnc}
\ee
with $\beta(h,\gamma)$ given by (\ref{betagh}) or (\ref{betaminus}) and
\be
   {\it E_-} (h, \gamma) =
   \exp\left(\sum_{k=1}^\infty  k \hat{\sigma}_k \hat{\sigma}_{-k}\right),
   \label{Esgh}
\ee
where $\hat{\sigma}_k$ is defined in (\ref{sigmak}) as the $k$-th
Fourier component of the logarithm of $\sigma$:
\bea
  \hat{\sigma}_k & \equiv & \int_0^{2 \pi} {\de q \over 2 \pi}\;
  \left[ \log \sigma(q) \right] \eu^{- \ii k q}
  \nonumber \\
  & = &
  \int_0^{2 \pi} {\de q \over 2\pi}\; \eu^{-\ii k q}
   \log \left( 1 + {\cos q - h + \ii \gamma \sin q \over
   \sqrt{ \left( \cos q - h \right)^2 + \gamma^2  \sin^2 q}} \right).
  \label{sigmahat}
\eea
The sum in (\ref{Esgh}) is convergent only for $\gamma \ne 0$ and for
$h < - 1$.
For $h \ge -1$, the generating function (\ref{genfunc}) develops
singularities which produce $1/k$ contributions to
(\ref{sigmahat}) that make the sum in (\ref{Esgh}) divergent.
Therefore, in the rest of the phase diagram these singularities
have to be treated to absorb the harmonic series contributions.
Consequently, each region of the phase diagram will involve a different
definition for the pre-exponential factor and the "regularization"
procedure will sometimes generate an additional power-law contribution.
The result is given by  the Fisher-Hartwig conjecture that we must use
in the remainder of the phase diagram.

\subsubsection{$\Sigma_0$ ($|h| < 1$)}
\label{Sigma0Sec}

As can be noticed from Fig.~\ref{GenFuncPlot}c, in $\Sigma_0$
($\gamma\neq 0$, $-1<h<1$) the generating function $\sigma(q)$ vanishes
and  its phase has a discontinuity of $\pi$ at $q = \pi$.
The asymptotic behavior of Toeplitz determinants with this type of
singularities in the generating function is given by FH, which is
actually proven for cases in which only one singularity is present.

We decompose the generating function as in (\ref{fishdec})
\be
   \sigma(q) = \tau(q) \eu^{{\ii \over 2} [(q - \pi) \mod 2 \pi - \pi]}
   \left( 2 - 2 \cos (q - \pi) \right)^{1/2}
   \label{param0}
\ee
and using (\ref{singexp}) we obtain
\be
   P(n) = \left| \det ({\bf S_n}) \right|
   {\stackrel{n \rightarrow \infty}{\sim}}
   {\it E_0 }(h,\gamma) \: \eu^{- \beta(h,\gamma) n}.
   \label{expbehavior}
\ee
The behavior is exponential as before with the  decay rate
$\beta(h,\gamma)$ from (\ref{betagh},\ref{beta0}), but the
pre-exponential factor is different. According to (\ref{fisherhartwig})
it is given by
\be
   {\it E_0 }(h,\gamma) \equiv { E[\tau] \over \tau_- (\pi) },
\ee
where, as in (\ref{szegoexp}) and (\ref{sigmak})
\bea
   E [\tau] & = &
   \exp\left(\sum_{k=1}^\infty  k \hat{\tau}_k \hat{\tau}_{-k}\right)
   \label{Etgh}
\eea
and
\be
   \hat{\tau}_k = \hat{\sigma}_k -  {(-1)^k \over k} \; \theta (k).
\ee
Here $\theta (k)$ is the usual Heaviside step function.
As we mentioned in the previous section, $\hat{\sigma}_k$ (\ref{sigmahat})
has $1/k$ contributions from singularities of $\sigma(q)$ and
the effect of the parametrization (\ref{param0}) is to cure
(remove) these harmonic series divergences of the prefactor of the EFP
in this regime.

\subsubsection{$\Sigma_+$ ($h > 1$)}
\label{SigmaPSec}

In $\Sigma_+$ ($\gamma\neq 0$, $h>1$), $\sigma(q)$ vanishes at $q = 0$
and $q = \pi$ and its phase presents two $\pi$ jumps at those points
(Fig.~\ref{GenFuncPlot}e).

In this case the application of FH leads to some ambiguity,
because there exist two representations of the kind
(\ref{fishdec}) and one obtains two values for $\beta(h,\gamma)$
using the two representations of the generating function:
$\beta_1=\beta$ and $\beta_2 = \beta +\ii \pi$, with $\beta$ from
(\ref{betagh}) or (\ref{betaplus}). This ambiguity is resolved by
the (yet unproven) generalized Fisher-Hartwig conjecture (gFH),
which gives EFP as a sum of two terms so that both values of
$\beta$'s are used (see in the Appendix \ref{gfhsec} or
\cite{basor}).

The two leading inequivalent parametrizations (\ref{fishgendec}) are:
\bea
   \sigma(q) & = & \tau^1(q) \eu^{{\ii \over 2}
   [(q - \pi)  \mod 2 \pi - \pi]}
   \left( 2 - 2 \cos (q - \pi) \right)^{1/2} \nonumber \\
   && \qquad \times \eu^{- {\ii \over 2} [q \mod 2 \pi - \pi]}
   \left( 2 - 2 \cos q \right)^{1/2} \\
   & = & \tau^2(q) \eu^{- {\ii \over 2} [(q - \pi) \mod 2 \pi - \pi]}
   \left( 2 - 2 \cos (q - \pi) \right)^{1/2} \nonumber \\
   && \qquad \times \eu^{{\ii \over 2} [q \mod 2 \pi - \pi]}
   \left( 2 - 2 \cos q \right)^{1/2}.
\eea
Application of (\ref{singgenexp}) gives the asymptotic behavior of the
determinants as
\be
   \left| \det ({\bf S_n}) \right|
   {\stackrel{n \rightarrow \infty}{\sim}}
   \left[ {\it E^1_+ }(h,\gamma) + (-1)^n {\it E^2_+}(h,\gamma) \right]
   \: \eu^{- \beta(h,\gamma) n}
\ee
with
\bea
   {\it E^1_+ }(h,\gamma) \equiv
   { E [\tau] \over \tau_+ (0) \tau_- (\pi) }, \\
   {\it E^2_+ }(h,\gamma) \equiv
   { E [\tau] \over \tau_+(\pi) \tau_- (0) }
\eea
and $\beta(h, \gamma)$, $E [\tau]$ defined in
(\ref{betagh},\ref{Etgh}) with
\be
   \hat{\tau}_k = \hat{\sigma}_k
   - {(-1)^k \over k} \; \theta (k) - {1 \over k} \; \theta (-k).
\ee
Once again, as in the previous section, the effect of the parametrization
is to remove the $1/k$ contributions to $\hat{\sigma}_k$ (\ref{sigmahat})
due to the singularities of the generating function.

We  conclude that the non-critical theory presents an
exponential asymptotic behavior of the EFP. In the region
$\Sigma_+$, however, the EFP in addition has even-odd oscillations
\be
   P(n) {\stackrel{n \rightarrow \infty}{\sim}} {\it E^1_+}(h,\gamma)
   \left[ 1 + A_+ (h,\gamma) \cos (\pi n) \right] \: \eu^{- \beta(h, \gamma) n},
   \label{oscbehavior}
\ee
where the exponential decay factor is given by (\ref{betaplus}).

The amplitude of the oscillations is
\bea
   A_+ (h,\gamma) & \equiv &
   { \tau_+ (0) \tau_- (\pi) \over \tau_- (0) \tau_+ (\pi)} \nonumber \\
   & = & { \tau (0) \over \tau (\pi)} \left( {\tau_- (\pi)
   \over \tau_- (0)} \right)^2 \nonumber \\
   & = & { h + 1 \over h - 1 } \exp \left( 4 \lim_{\epsilon \rightarrow 0}
   \oint {\de z \over 2 \pi} {\log \tau (z) \over z^2 - (1 + \epsilon)^2} \right),
   \label{AplusIn}
\eea
where we used (\ref{wienerhopf}), the definition of $\tau$ and
(\ref{wienint}).
We can deform the contour of integration as in Fig.~\ref{SigmaStruct}f
and calculate the integral in (\ref{AplusIn}) to obtain
\be
   A_+ (h,\gamma) = \sqrt{K(h,\gamma)} =
   {\sqrt{h^2 - 1} \over \sqrt{h^2 + \gamma^2 - 1} + \gamma} \; ,
   \label{Aplus}
\ee
where $K(h,\gamma)$ was defined in (\ref{kappa}).

\begin{figure}
  \includegraphics[width=\columnwidth]{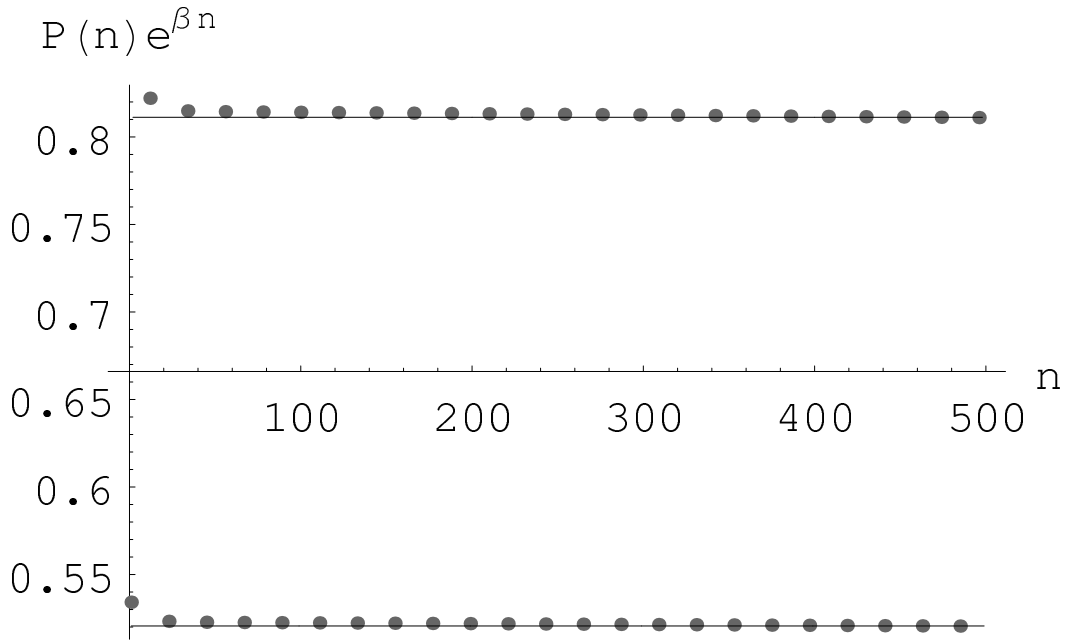}
\caption[Numerical vs. analytical behavior of the EFP at $\gamma
=1$, $h=1.1$]{Results of the numerical calculation of the Toeplitz
determinant are shown as points, $P(n) e^{\beta n}$ vs. $n$ at
$\gamma =1$, $h=1.1$. The value of $\beta$ is obtained numerically
from (\ref{betaPlus}). The solid line is the analytic result
$E(1+(-1)^n A)$ with $A=0.2182...$ from (\ref{Aplus}) and
$E=0.6659...$ obtained by fitting at large $n$. To make the plot
more readable we show only every 11th point (for
$n=1,12,23,\ldots$) of the numerical calculation of the
determinant. Note that the size of the points is not related to
the estimated error in the numerics, which is actually smaller.}
   \label{DetPlot1}
\end{figure}

Expression (\ref{oscbehavior}) for the EFP fits the numerical data
remarkably well (see Fig.~\ref{DetPlot1}) and this fact strongly
supports the generalized Fisher-Hartwig conjecture.

One can understand these oscillations as a result of ``superconducting''
correlations of real fermions described by the Hamiltonian (\ref{realfermionH}).
Fermions are created and destroyed in pairs of nearest neighbors.
At large magnetic fields, the oscillations are due to the fact that the
probability of having a depletion string of length $2k-1$ or $2k$ is very
similar.
Since the magnetic field in (\ref{realfermionH}) is essentially
a chemical potential for the fermions, the energy cost to destroy a pair
of particles is $4h$: at very big magnetic fields, the amplitude for a
pair destruction event is suppressed by a factor of $\gamma \over 4 h$,
i.e. a probability of $\gamma^2 \over 16 h^2$.
This means that the probability of depletion behaves like:
\bea
    P(2k-1) & \sim & 2 \left( {4 h \over \gamma} \right)^{-2k}
    \qquad {\rm and} \nonumber \\
    P(2k) & \sim & \left( {4 h \over \gamma} \right)^{-2k} \; ,
    \label{twoprob}
\eea
where the factor of two in the first expression is a simple
combinatorial factor.
The two probabilities in (\ref{twoprob}) can be combined in a
single expression:
\be
   P(n) = E \left[ 1 + A \cos(\pi n) \right]
   \left( {4 h \over \gamma} \right)^{- n} ,
   \label{Pnosc}
\ee
which is precisely (\ref{oscbehavior}), with
\be
   A = 1 - {\gamma \over h} + \Ord \left( { 1 \over h^2} \right) .
   \label{Aosc}
\ee

We can check the correctness of this interpretation by taking the
limit of (\ref{oscbehavior}) for $h >> 1, \gamma$.
From (\ref{betagh}) and (\ref{Aplus}) it is easy to find
\bea
   \beta(h \to \infty, \gamma) & = & \log {4 h \over \gamma}
   + \; \Ord \left( {1 \over h^2} \right) \\
   A_+ (h \to \infty, \gamma) & = & 1 - {\gamma \over h} + \Ord
   \left( {1 \over h^2} \right)
\eea
in agreement with (\ref{Pnosc},\ref{Aosc}).

\subsection{The critical lines ($\Omega_\pm$)}
\label{critical}

\subsubsection{$\Omega_+$ ($h = 1$)}
\label{ssOmegap}

For $h=1$ the generating function $\sigma(q)$ vanishes at $q = \pi$
and its phase has $\pi$ jumps at $q = 0, \pi$ (see
Fig.~\ref{GenFuncPlot}d).
As in the previous section, the existence of two singular points gives
rise to many terms of the form (\ref{fishgendec}).
However, in contrast to the $\Sigma_+$ region, the application of
gFH as in (\ref{singgenexp}) shows that all terms are
suppressed by power law factors of $n$ with respect to the leading one.

The leading term is generated by the parametrization:
\be
   \sigma(q) = \tau^1(q) \eu^{{\ii \over 2} [(q - \pi)  \mod 2 \pi - \pi]}
   \left( 2 - 2 \cos (q - \pi) \right)^{1/2}
   \eu^{- {\ii \over 4} [q \mod 2 \pi - \pi]}
\ee
and consists of an exponential decay with
$\beta(1, \gamma)$ from (\ref{betaPlus}) and a power
law contribution with critical exponent $\lambda = {1\over 16}$
\be
  \left| \det ({\bf S_n}) \right| \sim
  {\it E^1_1}(\gamma) n^{- {1 \over 16} } \: \eu^{- \beta(1, \gamma) n}
  \label{leading}
\ee
with
\be
   {\it E^1_1} (\gamma) \equiv E[\tau]
   G \left( {3 \over 4} \right) G \left( {5 \over 4} \right)
   {\tau_-^{1/4} (0) \over 2^{1/4} \tau_+^{1/4} (0) \tau_- (\pi) },
\ee
where $G$ is the Barnes G-function defined in (\ref{BGfun})
and $E [\tau]$ is defined as in (\ref{Etgh}) with
\be
   \hat{\tau}_k = \hat{\sigma}_k
   + \left( {1 \over 4} - (-1)^k \right) {1 \over k} \; \theta (k)
   - {1 \over 4k} \; \theta (-k) ,
\ee
with $\hat{\sigma}_k$ from (\ref{sigmahat}).

The next term (subleading at $n\to \infty$) is obtained from
the parametrization
\be
   \sigma(q) = \tau^2(q) \eu^{-{\ii \over 2} [(q - \pi)  \mod 2 \pi - \pi]}
   \left( 2 - 2 \cos (q - \pi) \right)^{1/2}
   \eu^{\ii {3 \over 4} [q \mod 2 \pi - \pi]}
\ee
and is given by
\be
  {\it E^2_1}(\gamma) (-1)^n n^{- {9 \over 16} } \: \eu^{- \beta(1, \gamma) n}
  \label{subleading}
\ee
with
\be
   {\it E^2_1}(\gamma) \equiv E[\tau]
   G \left( {1 \over 4} \right) G \left( {7 \over 4} \right)
   {\tau_+^{3/4} (0) \over 2^{3/4} \tau_-^{3/4} (0) \tau_+ (\pi) } .
\ee

Although the inclusion of the latter (subleading) term is somewhat beyond
even gFH, we write the sum of these two terms as a
conjecture for EFP at $h=1$
\be
   P(n) \sim {\it E^1_1}(\gamma) \:
   n^{- {1 \over 16} } \left[ 1 + (-1)^n A_1(\gamma)/n^{1 \over 2}
   \right] \: \eu^{- \beta(1, \gamma) n}.
 \label{pnomp}
\ee

\begin{figure}
  \includegraphics[width=\columnwidth]{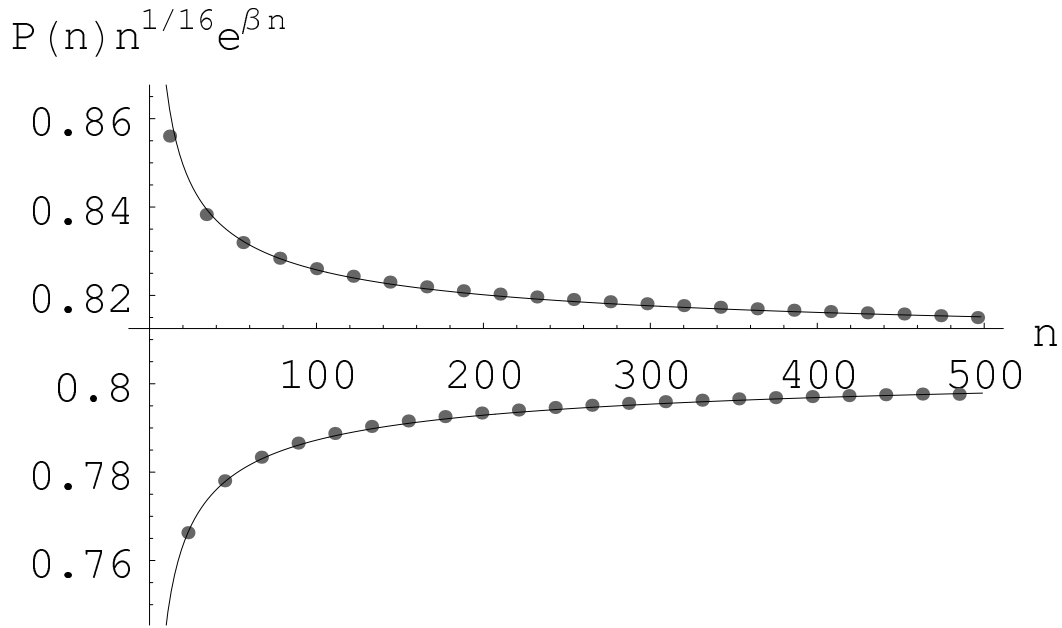}
\caption[Numerical vs. analytical behavior of the EFP at $\gamma
=1$, $h=1$] {Results of the numerical calculation of the Toeplitz
determinant are shown as points, $P(n) e^{\beta n} n^{1/16}$ vs.
$n$ at $\gamma =1$, $h=1$. The value $\beta=\log 2 +2G/\pi$ with
Catalan's constant $G$ is obtained from (\ref{betagh}). The solid
line is the analytic result $E(1+(-1)^n A/n^{1 \over 2})$ with
$A=0.2399...$ from (\ref{A1}) and $E=0.8065...$ as obtained by
fitting at large $n$. To make the plot more readable we show only
every 11th point (for $n=1,12,23,\ldots$) of the numerical results
on the determinant. Note that the size of the points is not
related to the estimated error in the numerics, which is actually
smaller.}
   \label{DetPlot2}
\end{figure}

As these results rely on our unproven conjecture, we present our
numerical data for this case in Fig.~\ref{DetPlot2}.
Indeed, we see that the form (\ref{pnomp}) is in good agreement with
numerics and this supports our hypothesis.

The amplitude of the oscillations is
\bea
   A_1 (\gamma) & \equiv & {1 \over \sqrt{2} } \;
   { G \left( {1 \over 4} \right) G \left( {7 \over 4} \right) \over
     G \left( {3 \over 4} \right) G \left( {5 \over 4} \right) } \;
   { \tau_+ (0) \tau_- (\pi) \over \tau_- (0) \tau_+ (\pi) } \nonumber \\
   & = & {1 \over \sqrt{2} } \; { \Gamma \left( {3 \over 4} \right)
   \over \Gamma \left( {1 \over 4} \right) } \;
   { \tau (0) \over \tau (\pi) }
   \left( { \tau_- (\pi) \over \tau_- (0) } \right)^2 \nonumber \\
   & = & { \Gamma \left( {3 \over 4} \right) \over
   \Gamma \left( {1 \over 4} \right) } \; {1 \over \gamma} \;
   \exp \left( 4 \lim_{\epsilon \rightarrow 0} \oint
   {\de z \over 2 \pi} {\log \tau (z) \over z^2 - (1 + \epsilon)^2} \right),
   \label{A1In}
\eea
where we used (\ref{wienerhopf}) and the identity
\be
   G(z+1) = \Gamma (z) G(z).
\ee
To calculate the integral we deform the contour of integration as in
Fig.~\ref{SigmaStruct}e and find
\be
   A_1 (\gamma) = { \Gamma \left( {3 \over 4} \right)
   \over \Gamma \left( {1 \over 4} \right) } \; {1 \over \sqrt{2 \gamma} }.
   \label{A1}
\ee

We conclude that at $h=1$ the EFP decays exponentially at $n\to\infty$
but with an additional power law pre-factor and a damped oscillatory
component.

{\it Remark.} It is curious to notice that the exponents $1/16$
and $9/16$ in (\ref{leading}) and (\ref{subleading}) remind us of
the scaling dimensions of spins $\sigma^x$ and $\sigma^y$.
\footnote{See Ref. \cite{mccoy} or (\ref{rhox},\ref{rhoy}), where
it was shown that the power laws for the $\sigma^x$ and $\sigma^y$
correlators are $1/4$ and $9/4$ respectively.} It looks as if the
EFP operator (\ref{EFPDef}), among other things, has inserted
square roots of spins transverse to the magnetic field at the ends
of the string.

\subsubsection{$\Omega_-$ ($h = -1$)}
\label{ssOmegam}

For $h=-1$ the generating function $\sigma(q)$ does not vanish but has
a phase discontinuity of $\pi$ at $q= \pi$.
We parametrize $\sigma(q)$ as
\be
   \sigma(q) = \tau^1(q) \eu^{- {\ii \over 4} [(q - \pi)  \mod 2 \pi - \pi]}
\ee
and apply FH to obtain
\be
 \label{pnft}
   P(n) \sim {\it E^1_{-1} (\gamma)} \:
   n^{- {1 \over 16} }  \: \eu^{- \beta(-1, \gamma) n}
\ee
with
\be
   {\it E^1_{-1}} (\gamma) \equiv E[\tau]
   G \left( {3 \over 4} \right) G \left( {5 \over 4} \right)
   {\tau_-^{1/4} (\pi) \over \tau_+^{1/4} (\pi)) },
\ee
where $\beta(-1, \gamma)$ and $E [\tau]$ are defined in
(\ref{betaMinus}) and (\ref{Etgh}) with
\be
   \hat{\tau}_k = \hat{\sigma}_k
   + {(-1)^k \over 4 k} \; \theta (k)
   - {(-1)^k \over 4 k} \; \theta (-k)
\ee
and $\hat{\sigma}_k$ from (\ref{sigmahat}).

We can stretch the gFH the same way as in the previous section
for $h=+1$ by considering the second parametrization
\be
   \sigma(q) = \tau^2(q) \eu^{\ii {3 \over 4} [(q - \pi)  \mod 2 \pi - \pi]}
\ee
which gives
\be
     P'(n) \sim {\it E^2_{-1} (\gamma)} \:
     n^{- {9 \over 16} }  \: \eu^{- \beta(-1, \gamma) n}
\ee
with
\be
   {\it E^2_{-1}} (\gamma) \equiv E[\tau]
   G \left( {1 \over 4} \right) G \left( {7 \over 4} \right)
   {\tau_+^{3/4} (\pi) \over \tau_-^{3/4} (\pi)) }.
\ee
Adding this subleading term to (\ref{pnft}) we obtain
\be
   P(n) \sim {\it E^1_{-1}}(\gamma) \:
   n^{- {1 \over 16} } \left[ 1 +  A_{-1}(\gamma)/n^{1 \over 2}
   \right] \: \eu^{- n \beta(-1, \gamma)}
   \label{pnomm}
\ee
with
\bea
   A_{-1} (\gamma) & \equiv &
   { G \left( {1 \over 4} \right) G \left( {7 \over 4} \right) \over
     G \left( {3 \over 4} \right) G \left( {5 \over 4} \right) } \;
   { \tau_+ (\pi) \over \tau_- (\pi) } \nonumber \\
   & = & { \Gamma \left( {3 \over 4} \right)
   \over \Gamma \left( {1 \over 4} \right) } \;
   { \tau_+ (\pi) \over \tau_- (\pi) }.
\eea
We propose (\ref{pnomm}) as an asymptotic form for EFP at $h=-1$.

\section{The line $\Gamma_E$: an exact calculation}
\label{GammaESec}

Before we conclude our analysis of the EFP with the study of the
isotropic XY model, let us check our results (\ref{pnc},\ref{betagh})
on the special line\footnote{We are grateful to Fabian Essler who
suggested us to check our results on this special line and pointed out
the reference \cite{shrock} to us.} in the phase diagram defined by
\be
   h^2 + \gamma^2 = 1 .
   \label{trajectory}
\ee
It was shown in Ref. \cite{shrock} that on this line the ground state
is a product of single spin states and is given by
\be
   \vert G \rangle = \prod_j \vert \theta, j \rangle = \prod_j
   \left[ \cos \left( {\theta \over 2} \right) | \uparrow, j \rangle +
   (-1)^j \sin \left( {\theta \over 2} \right) | \downarrow, j \rangle
   \right],
   \label{groundstate}
\ee
where $| \uparrow, j \rangle$ is an up-spin state at the lattice
site $j$, etc.
One can directly check that the state (\ref{groundstate}) is an
eigenstate of (\ref{spinham}) if the value of parameter $\theta$ is
\be
    \cos^2 \theta  =  \frac{1 - \gamma}{1 + \gamma}
\ee
and (\ref{trajectory}) is satisfied.
It is also easy to show \cite{shrock} that this state is, in fact, the
ground state of (\ref{spinham}).

The probability of formation of a ferromagnetic string in the
state (\ref{groundstate}) is obviously \be
   P(n) = \sin^{2n} \left( {\theta \over 2} \right)
    = \left( {1 \over 2} - {1 \over 2} { h \over |h|}
    \sqrt{ 1 - \gamma \over {1 +\gamma} } \right)^n,
 \label{pnexact}
\ee
which is an exact result on the line (\ref{trajectory}).
The value of  $\beta(h,\gamma)$ which immediately follows from this
exact result is
\be
   \beta(h = \pm \sqrt{1-\gamma^2},\gamma) =
   - \log\left( {1 \over 2} \mp {1 \over 2}
   \sqrt{ 1 - \gamma \over {1 +\gamma} } \right)
   = - \Lambda(h,\gamma) ,
 \label{betaexact}
\ee
where $\Lambda(h,\gamma)$ was defined in (\ref{Lambda}).

This is, indeed, consistent with (\ref{beta0}) since
under the condition (\ref{trajectory}) the function
$\Delta(h,\gamma)$ vanishes.
The integral (\ref{Delta}) defining $\Delta(h,\gamma)$ vanishes for (\ref{trajectory})
because the branching points (\ref{z1}) and (\ref{z2}) collapse to the
same point and therefore the region of integration shrinks to just one
point (\ref{Delta}).
In fact, the Toeplitz matrix (\ref{Tg}) generated by (\ref{genfunc})
becomes triangular on the line (\ref{trajectory}) with diagonal matrix
element $(S_n)_{jj} = \sin^2(\theta/2)$ and the determinant of
${\bf S_n}$ is exactly (\ref{pnexact}).

From the definitions of $\beta(h,\gamma)$, we see that the decay
factor consists of two terms, which now have a clear physical
meaning. The $\Lambda(h,\gamma)$ term is the factor we found above
in (\ref{betaexact}) and represents the contribution given by
un-entangled spins. The remaining part accounts for the
correlations between spins. Both $\Delta(h,\gamma)$ and the
correlation functions given by (\ref{F}) and (\ref{G}) vanish on
the line (\ref{trajectory}).

Finally, it should be noted that there are actually two ground
states for the theory on this special line:
\bea
   \vert G_+ \rangle & = & \prod_j \vert \theta, j \rangle = \prod_j
   \left[ \cos \left( {\theta \over 2} \right) | \uparrow, j \rangle +
   (-1)^j \sin \left( {\theta \over 2} \right) | \downarrow, j \rangle
   \right] \\
   \vert G_- \rangle & = & \prod_j \vert \theta, j \rangle = \prod_j
   \left[ \cos \left( {\theta \over 2} \right) | \uparrow, j \rangle -
   (-1)^j \sin \left( {\theta \over 2} \right) | \downarrow, j \rangle
   \right]
\eea These two states break the translational symmetry and are
orthogonal in the thermodynamic limit. The reason for this
degeneracy of the ground state is yet not well understood.

If we were to consider a linear combination of these states, any
local operator (involving only a finite number of lattice sites)
would have vanishing cross-terms and very likely the same
expectation value on either state (though one can design an
operator which would violate the latter property). Therefore, for
our present interest, evaluation of the EFP on any combination of
these two ground states would yield the same result
(\ref{pnexact}) and we don't need to concern ourselves with this
degeneracy.

The existence of these two ground states is instead of great
interest in the study of quantum information applied to the XY
model. We recently noticed this degeneracy and this helped us
understand a previously known formula on the entropy of a block of
neighboring spins \cite{frankor}. This entropy is known to
quantify the degree of entanglement of two subsystems of a system
(in this case the block of neighboring spins and the rest of the
chain). We remark in passing that many of the approaches to the
calculation of the entropy for the XY model use Toeplitz
determinant representations and the theory on Toeplitz determinant
for deriving their asymptotic behavior, in the spirit of the
present work.

\section{The critical line $\Omega_0$ ($\gamma = 0$) and the Gaussian
behavior}
\label{gammazero}

The case $\gamma = 0$, corresponding to the Isotropic XY Model, has
been studied in Ref. \cite{shiroishi}.
For $\gamma=0$ the generating function (\ref{genfunc}) is reduced
to the one found in \cite{shiroishi}.

For $|h| < 1$, the generating function $\sigma(q)$ has a limited
support between $[-\cos^{-1} h, \cos^{-1} h]$. To find the asymptotic
behavior of the determinant of the Toeplitz matrix one can apply
Widom's Theorem \cite{widomsupp} and obtain \cite{shiroishi}
\be
   P(n) \sim
   2^{5 \over 24} \eu^{3 \zeta'(-1)} (1-h)^{- {1 \over 8}}
   n^{- {1 \over 4}} \left( {1 + h \over 2} \right)^{n^2 \over 2}.
 \label{pnsh}
\ee
We see that in this case, the EFP decays as a Gaussian with an
additional power-law pre-factor.

In a different context, the formula (\ref{pnsh}) appeared also in
\cite{dysmehta} as a probability of forming a gap in the spectrum of
unitary random matrices.
This is not unexpected, since the joint eigenvalue distribution of
unitary random matrices is known to coincide with the distribution of free
fermions in the ground state.

For $|h| > 1$, the theory is no longer critical and the ground state
is completely polarized in the $z$ direction, giving a trivial EFP
$P(n)=0$ for $h>1$ and $P(n)=1$ for $h<-1$.

\section{Crossover between Gaussian and exponential behavior: a Bosonization approach}
\label{Crossover}

In order to understand qualitatively the crossover between the
Gaussian asymptotic behavior at $\gamma = 0$ and the exponential
decay for $\gamma \ne 0$, we employ a bosonization approach
similar to the one used in \cite{abanovkor}, which will be
sketched in Chapter \ref{HydroEFP}. In the limit $\gamma \ll 1$ we
consider the continuum limit of (\ref{spinlessham}), bosonize the
fermionic fields, and write the Euclidean action of the theory as
${\cal S}=\int \de x \, \de \tau {\cal L}$, where $\tau \equiv \ii
t$ is the imaginary time and the Lagrangian is
\be
   {\cal L} = {1 \over 2} \left[ \left( \partial_\mu \vartheta \right)^2
   - {2 \gamma \over \pi} \cos \left( \sqrt{4 \pi} \vartheta \right)
   \right] ,
   \label{bosoniclag}
\ee where we also rescaled the time $\tau \to v_F \tau$, with $v_F
\equiv 2 \sqrt{1 - h^2}$, the Fermi velocity at $\gamma = 0$.

This is a Sine-Gordon theory for the ``conjugate field''
$\vartheta(x,\tau)$, which describes the imaginary time dynamics
of our 1-D system. In terms of $\vartheta$ the density of fermions
is given by $\rho = {1 \over \sqrt{\pi}} \partial_{\tau} \vartheta
+ \rho_0$, where $\rho_{0}=k_{F}/\pi$ is the density of fermions
in the ground state.

In the field theory approach, the EFP (see Ref. \cite{abanovkor}) in the
limit $n \to \infty$ would be given with exponential accuracy by the
probability of an instanton $P(n) \sim e^{-{\cal S}_{0}}$, where
${\cal S}_{0}$ is the action of the instanton.
Here the instanton is the solution of the classical equations of motion
of (\ref{bosoniclag}) which corresponds to the formation of an emptiness
of length $n$ at the time $\tau=0$.
Unfortunately, the EFP instanton involves large deviations of the
density of fermions from the equilibrium density $\rho_{0}$ and is beyond
the bosonization approach as the derivation of (\ref{bosoniclag}) relies
on the linearization of the fermionic spectrum near the Fermi points.

Following \cite{abanovkor}, we are going to slightly generalize
our problem by considering the depletion formation probability
instead of the EFP requiring
\be
   \left. \rho \right|_{\tau=0, \; 0 < x <n} = \rho_{0} + {1 \over \sqrt{\pi}}
   \left. \partial_\tau \vartheta (x,\tau) \right|_{\tau=0, \; 0 < x <n}
   = \rho_{0}-\bar\rho ,
   \label{boundarycond}
\ee
where $\bar{\rho}$ is some constant.
The original EFP problem corresponds to $\bar{\rho} = \rho_{0}$.
Here, instead, we consider the probability of weak depletion, i.e.
\be
   - {1 \over \sqrt{\pi}} \left. \partial_\tau \vartheta (x,\tau) \right|_{\tau=0, \; 0 < x <n}
   =\bar{\rho} << \rho_{0} .
   \label{PFWFS}
\ee
We study the latter using an instanton approach to (\ref{bosoniclag})
and infer the (qualitative) behavior of the original EFP from this weak
limit.

To simplify the problem further, we assume that the instanton
configuration is completely confined to one of the wells of the Cosine
potential in (\ref{bosoniclag}) and that the field $\vartheta$ is small enough to
allow for an expansion of the Cosine:
\be
   {\cal S} \approx {1 \over 2 } \int \de x \; \de \tau \left[ \left(
   \partial_\mu \vartheta \right)^2 + 4 \gamma \; \vartheta^2 \right] .
   \label{expandedact}
\ee
In this formulation, the anisotropy parameter $\gamma^{1/2}$ plays the
role of the mass of the bosonized theory.
The probability we are looking for is given by the action ${\cal S}_{0}$ of the
classical field configuration which satisfies the Euler-Lagrange
equation (in this case a Klein-Gordon equation in two dimensions)
with the boundary condition (\ref{boundarycond})
\be
   P_{\bar{\rho}} (n) = \eu^{-{\cal S}_0} .
   \label{actionprob}
\ee

In the limit $\gamma = 0$, the theory is massless and scale invariant.
In \cite{abanovkor} it was shown that, due to the scale invariance, the
action of the instanton is quadratic in $n$.
The instanton configuration in this case is essentially a droplet of
depletion in space-time with dimensions proportional to $n$ both in the
space and time direction, in order to satisfy the boundary condition
(\ref{boundarycond}).
This result is consistent with the Gaussian asymptotic behavior
prescribed by Widom's theorem (see Sec.~\ref{gammazero}).

In the massive case, a finite correlation length $\xi \sim
\gamma^{-1/2}$ is generated and we observe a crossover. For string
lengths $n$ smaller than the correlation length $\gamma^{-1/2}$,
the instanton action is not sensitive to the presence of the
finite correlation length and is still quadratic in $n$ (giving a
Gaussian decay for EFP). In the asymptotic limit of string lengths
greater than $\gamma^{-1/2}$, the time dimension of a depletion
droplet is of the order of $\xi$ (instead of $n$ as in the
massless limit): the action is linear in $n$ and the probability
has an exponential behavior.\footnote{This picture is very similar
to the one for massless theory at finite temperature. In the
latter the inverse temperature plays the role of the correlation
length \cite{abanovkor} (see \ref{singintapp}).}

\begin{figure}
   \dimen0=\textwidth
   \advance\dimen0 by -\columnsep
   \divide\dimen0 by 2
   \noindent\begin{minipage}[t]{\dimen0}
   \includegraphics[width=\columnwidth]{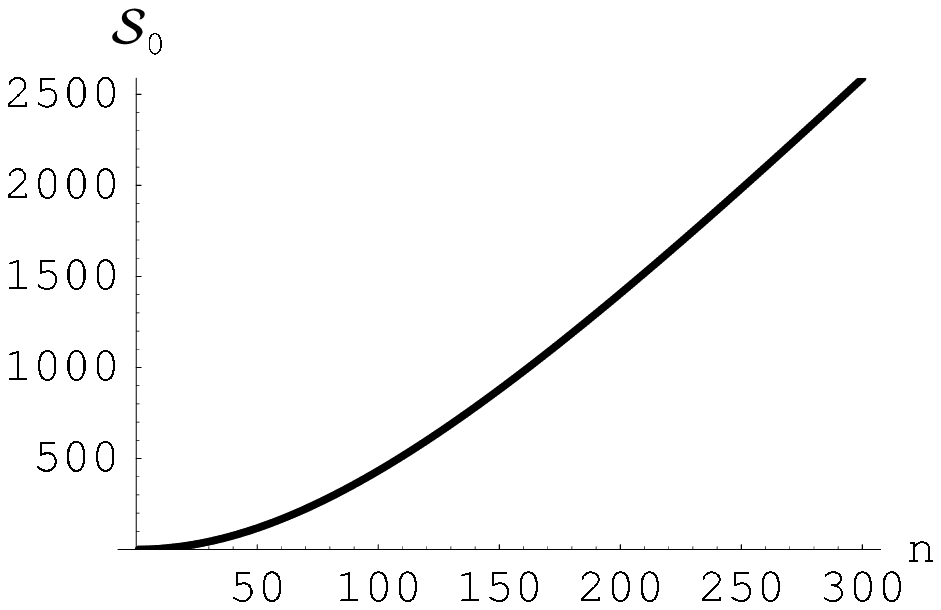}
   \caption[Plot of the value of the stationary action ${\cal S}_{0}$ vs.
   the string length $n$]
   {Plot of the value of the stationary action ${\cal S}_{0}$ vs.
   the string length $n$. The action   ${\cal S}_{0}$
   is obtained from (\ref{statact}) with $f(y)$ given by  the
   numerical solution of the singular integral equation (\ref{singinteq}).
   The graph depicts ${\cal S}_{0}(n)$ for $m=2\sqrt{\gamma}=0.01$,
   $\bar{\rho}=0.2$.  The crossover takes place around $ n \sim 2/m = \sqrt{1/\gamma} =200$.}
   \label{actfig1}
   \end{minipage}
   \hfill
   \begin{minipage}[t]{\dimen0}
   \includegraphics[width=\columnwidth]{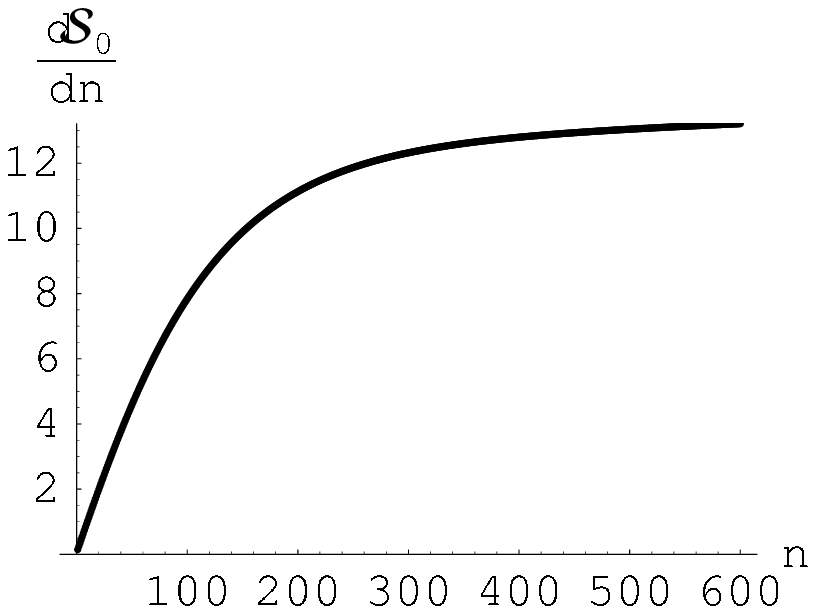}
   \caption[Plot of the derivative of the stationary action]
   {Plot of the derivative $d{\cal S}_{0}/dn$  with ${\cal S}_{0}$ from
   (\ref{statact}). The plot corresponds to $m=0.01$, $\bar{\rho}=0.2$
   and clearly shows a crossover from the quadratic to the linear behavior
   at $n \sim 2/m = \sqrt{1/\gamma} = 200$.}
   \label{actfig2}
   \end{minipage}
\end{figure}

In the next section we show how to solve the integral equation
corresponding to the boundary problem
(\ref{boundarycond},\ref{expandedact}) and present its numerical
solution and some analytical results. Figures \ref{actfig1} and
\ref{actfig2} clearly show the crossover between a quadratic
behavior of the stationary action for small $n$ to a linear
asymptotic one for $n \to \infty$.

\section{Calculation of the stationary action in the bosonization approach}
\label{singintapp}

In the previous section  we have formulated the XY model near
$\gamma = 0$ in terms of the bosonic field with Lagrangian
(\ref{expandedact}). It was also pointed out that, instead of the
EFP, we are interested in the Probability of Formation of Weakly
Ferromagnetic Strings (PFWFS) and that we are going to calculate
this probability in the saddle point approximation. In this
section we are going to show how to solve the integral equation
that solves this problem.

This is a fairly technical section and can be easily skipped if
the reader is not interested in the mathematical techniques
required to tackle this problem.

We consider a configuration of the field (instanton) which
satisfies the boundary condition imposed by the PFWFS
(\ref{boundarycond},\ref{PFWFS})
\be
   \left. \partial_\tau \vartheta (x,\tau) \right|_{\tau=0,0<x<n} = \sqrt{\pi} \bar{\rho}
   \label{BoundaryCond}
\ee
and that minimizes the action, i.e. that satisfies the Euler-Lagrange
equations:
\be
   \left( \partial_\mu \partial^\mu - m^2 \right) \vartheta = 0.
   \label{kleingordon}
\ee
The latter equation is the Klein-Gordon equation with the mass
given by $m^2 \equiv 4 \gamma$ (see (\ref{expandedact})). The
PFWFS will be found from the value of the action ${\cal S}_{0}$
corresponding to this instanton configuration (\ref{actionprob}).
In this appendix we calculate the stationary action needed in
Sec.~\ref{Crossover}.

We now solve the differential equation (\ref{kleingordon}) with non-trivial
boundary condition (\ref{BoundaryCond}) by recasting it as the integral
equation:
\be
   \vartheta (x,\tau) = {1 \over 2 \pi} \int_0^n \partial_t K_0
   \left( m \sqrt{(x-y)^2 + \tau^2} \right) f(y) \; \de y,
\ee
where $K_0 (x,x';\tau,\tau')$ is the modified Bessel function
of 0-th order -- the kernel of the differential operator
(\ref{kleingordon}) in two dimensions. We impose the boundary
condition (\ref{BoundaryCond}) by requiring that the ``source''
$f(y)$ satisfies
\bea
   \left. \partial_\tau \vartheta(x,0) \right|_{0<x<n} & = &
   \lim_{\tau \to 0} {1 \over 2 \pi} \int_0^n \left\{
   K_2 \left( m \sqrt{(x-y)^2 +\tau^2} \right) { m^2 \tau^2 \over (x-y)^2
   + \tau^2 } \right. \nonumber \\
   && \quad \left. - \,
   K_1 \left( m \sqrt{(x-y)^2 + \tau^2} \right) { m \over \sqrt{(x-y)^2
   + \tau^2 } } \right\} f(y) \; \de y
   \nonumber \\
   & = & \sqrt{\pi} \bar{\rho} .
   \label{inteq}
\eea
This is the integral equation on $f(y)$ we have to solve.

Once the limit $\tau \to 0$ is taken, the kernel in Eq.
(\ref{inteq}) is singular. We isolate the singularity by rewriting
equation (\ref{inteq}) as: \be
   {\de \over \de x} \; {1 \over \pi} \; \dashint_0^n {f(y) \over x - y} \; \de y
   + \lim_{\tau \to 0} \; {1 \over \pi} \int_0^n G_0 (x,\tau;y) f(y)\; \de y = 2 \sqrt{\pi} \bar{\rho}
\ee
with
\bea
   G_0 (x,\tau;y) & \equiv & {(x-y)^2 - \tau^2 \over (x-y)^2 + \tau^2} +
   K_2 \left( m \sqrt{(x-y)^2 +\tau^2} \right) { m^2 \tau^2 \over (x-y)^2 +\tau^2 }
   \nonumber \\
   && -
   K_1 \left( m \sqrt{(x-y)^2 + \tau^2} \right) { m \over \sqrt{(x-y)^2 + \tau^2 } }
   \label{kernel0}
\eea or, after integration over $x$, as
\be
   {1 \over \pi} \; \dashint_0^n {f(y) \over x - y} \; \de y
   + {1 \over \pi} \int_0^n G (x;y) f(y)\; \de y = 2 \sqrt{\pi} \bar{\rho} \; x
   \label{singinteq}
\ee
with
\be
   G(x;y) \equiv \lim_{\tau \to 0} \int_0^x G_0 (x_1,\tau;y) \; \de x_1 .
\ee
We have recasted Eq.~(\ref{inteq}) in the standard form for a
singular integral equation (\ref{singinteq}).
Once we have the solution of this equation, we can calculate the
action corresponding to this instanton as
\be
   {\cal S}_0 = { \sqrt{\pi} \bar{\rho} \over 2} \int_0^n f(y) \; \de y.
   \label{statact}
\ee

We solved the singular integral equation (\ref{singinteq})
numerically and we computed the corresponding action
(\ref{statact}). The results of this calculation are presented as
a plot of the action ${\cal S}_{0}$ vs. $n$ in Fig.~\ref{actfig1},
where we notice the crossover from a quadratic to a linear
behavior (corresponding to a crossover from Gaussian to
exponential behavior for the probability, (\ref{actionprob})) as
we expected. To confirm the nature of this crossover, in
Fig.~\ref{actfig2} we plot $d{\cal S}_{0}/dn$ and we see that it
starts linearly and then saturates asymptotically as it should.

In the limit $n \ll 1/m$, we can expand the Bessel functions in the kernel
(\ref{kernel0})
\be
   G_0 (x,\tau;y) =  -{m^2 \over 2} \left( {\tau^2 \over (x-y)^2 + \tau^2} +
   {1 \over 2} \ln \left[ (x-y)^2 + \tau^2 \right]
   + \ln {m \over 2} + G - {1 \over 2} \right) + \ldots,
   \label{G0exp}
\ee
where $G$ is Catalan's constant.
Then we solve the singular integral equation (\ref{singinteq}) to
first order by first transforming it into a regular integral equation.

In \cite{musk}, Chap. 14, Sec. 114 it is explained that a singular
integral equation like (\ref{singinteq}) is equivalent to
\be
   f(x) + {1 \over \pi \ii} \int_0^n N(x;y) \; f(y) \; \de y
   = 2 \bar{\rho} \sqrt{ \pi x(n-x)},
   \label{reginteq}
\ee
where the new kernel is
\be
   N(x;y) \equiv {\sqrt{x(n-x)} \over \pi \ii} \; \dashint_0^n
   {G(y';y) \over \sqrt{y'(n-y')} (x-y')} \; \de y' .
\ee

\begin{figure}
   \includegraphics[width=\columnwidth]{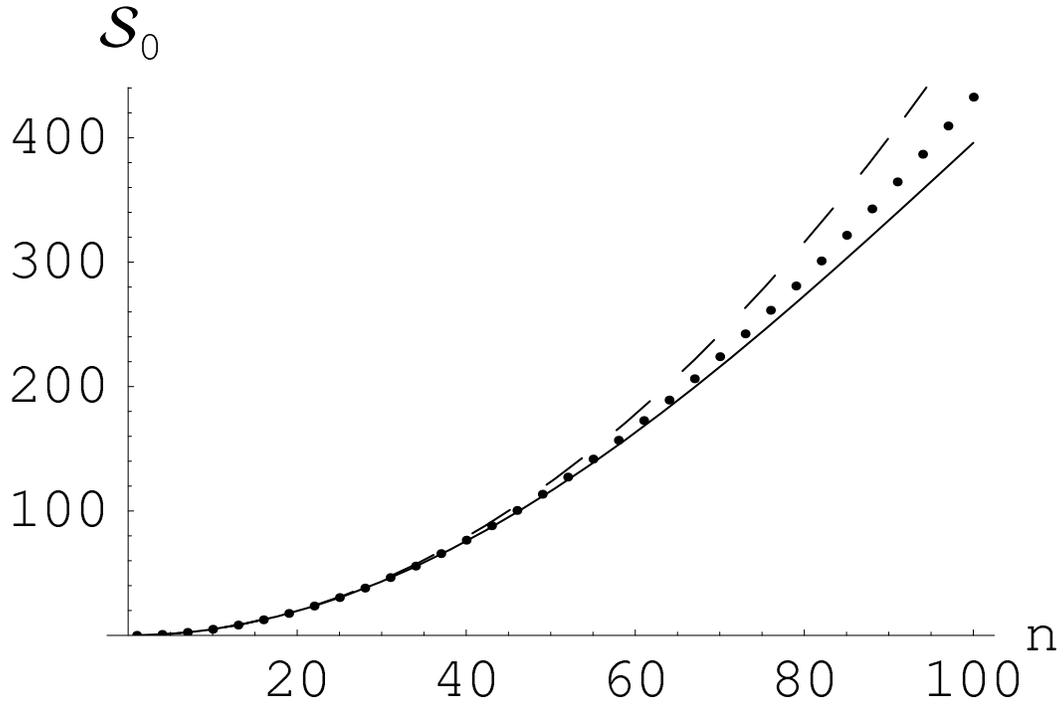}
   \caption[Numerical vs. analytical plot of the stationary action]
   {The solid line is the plot of the stationary action
(\ref{ansol}) against $n$. This analytical solution is valid for
$n << 1/m$ and corresponds to $m = 0.01$ and $\bar{\rho} = 0.2$.
The dotted line represents the value of the action (\ref{statact})
with the source given by numerical solution of the singular integral
equation (\ref{singinteq}).
The dashed line corresponds to the zeroth-order, pure Gaussian, solution,
i.e. (\ref{ansol}) with $m \equiv 0$, which we include for comparison.
We see that the inclusion of first order correction almost doubles the
range in which the analytical solution is accurate.}
   \label{actcomp}
\end{figure}

Using (\ref{G0exp}), we can explicitly calculate the integral
defining $N(x;y)$ in terms of elementary functions and after some
algebra the integral equation (\ref{reginteq}) results in a long,
but essentially simple, regular integral equation. Its solution is
\bea
   f(x) & = & \sqrt{\pi} \bar{\rho} \left[ 2 + {m^2 n^2 \over 8} \left(
   \ln {m \, n \over 8} + G - {3 \over 2} \right) \right] \sqrt{x(n-x)}
   \nonumber \\
   &&  \quad - \sqrt{\pi} \bar{\rho} \; {m^2 n^2 \over 4} \left(x - {n \over 2} \right)
   \tan^{-1} \sqrt{ x \over n - x}.
\eea
The corresponding stationary action (\ref{statact}) is
\be
   {\cal S}_0 = n^2 \: {\pi^2 \bar{\rho}^2 \over 8} \left[ 1 +
   {m^2 n^2 \over 16} \left( \ln {m \; n \over 8} + G - 2 \right) \right] .
   \label{ansol}
\ee The first term in (\ref{ansol}) corresponds to the Gaussian
decay of PFWFS we expect in the limit of $m=0$. In
Fig.~\ref{actcomp}, we compare this analytical result for the
action with the numerical result of Fig.\ref{actfig1}. In the
plot, we include the pure Gaussian decay (the first term in
(\ref{ansol})), which already gives a remarkable agreement for
small $n$. The full solution (\ref{ansol}) extends this agreement
further for larger $n$. For $m=0$ ($\gamma=0$), (\ref{ansol})
reproduces the result calculated in \cite{abanovkor} for $h=0$.

\section{Emptiness Formation Probability at finite temperature}
\label{FiniteEFP}

Finally, we consider what happens at finite temperature ($T>0$).
The correlators (\ref{F}) and (\ref{G}) become \bea
   F_{jk}^T & \equiv & \ii \langle \psi_j \psi_k \rangle_T
   = - \ii \langle \psi_j^\dagger \psi_k^\dagger \rangle_T =
   \int_0^{2 \pi} {\de q \over 2 \pi} {1 \over 2} \sin \vartheta_q
   \tanh {\varepsilon_q \over 2 T}  \eu^{\ii q (j-k)},
   \nonumber \\
   G_{jk}^T & \equiv & \langle \psi_j \psi_k^\dagger \rangle_T
   = \int_0^{2 \pi} {\de q \over 2 \pi} {1 \over 2}
   \left( 1+ \cos \vartheta_q  \tanh {\varepsilon_q \over 2 T} \right)
   \eu^{\ii q (j-k)},
\eea where the additional factor takes care of the thermal
average.

The EFP is expressed by \be
   P(n) \equiv {1 \over Z} \Tr \left\{ \eu^{- \frac{H}{T}}
   \prod_{j=1}^n {1 - \sigma_i^z \over 2} \right\},
   \label{EFPNonZeroTDef}
\ee and in the spinless fermion formalism it becomes
\be
   P(n) = \langle \prod_{i=1}^n \psi_i \psi_i^\dagger \rangle_T.
\ee

We again use Wick's Theorem (or its thermal version, called
Bloch-de Dominicis theorem \cite{todakubo}) to express it as a
Pfaffian. The calculation proceeds the same way as for zero
temperature and the EFP can be represented as \be
   P(n) = |\det({\bf T_n})|,
\ee where ${\bf T_n}$ is the $n \times n$ Toeplitz matrix
generated by the function \be
   t(q) = {1 \over 2}\left( 1 + \eu^{\ii \vartheta_q}
   \tanh {\varepsilon_q \over 2 T} \right) ,
\ee where the ``rotation angle'' $\vartheta_q$ and the spectrum
$\varepsilon_q$ were defined in (\ref{rotangle}) and
(\ref{spectrum}) respectively.

The generating function $t(q)$ is never-vanishing and has zero
winding \hbox{number.} Therefore, for $T > 0$ we can apply the
standard Szeg\"o Theorem to obtain \be
   P(n) {\stackrel{n \rightarrow \infty}{\sim}} E(h, \gamma, T)
   \eu^{- n \beta(h, \gamma, T)},
  \label{expbehT}
\ee where \bea
   \beta(h, \gamma, T) & = & -\int_0^{2 \pi} {\de q \over 2 \pi}\;
   \log \left| t(q) \right| \nonumber \\
   & = & - {1 \over 2} \int_0^{2 \pi} {\de q \over 2 \pi}\;
   \log \left[ {1 \over 2} \left( 1 + {\cos q -h \over \varepsilon_q}
   \tanh {\varepsilon_q \over 2 T} \right) \right],
 \\
   E(h, \gamma, T) & = &  \exp \left( \sum_{k=1}^\infty k
   \hat{t}_k \hat{t}_{-k} \right)
\eea with \be
   \hat{t}_k = \int_0^{2 \pi} {\de q \over 2 \pi}
   \eu^{-\ii k q} \log \left[ {1 \over 2} \left(
   1 + {\cos q - h + \ii \gamma \sin q \over \varepsilon_q}
   \tanh { \varepsilon_q \over 2 T} \right) \right],
\ee and $\varepsilon_q$ is given as in (\ref{spectrum}) by \be
   \varepsilon_q = \sqrt{ (\cos q - h)^2 + \gamma^2 \sin^2 q}.
\ee

As can be expected from simple thermodynamic considerations, at
finite temperature the behavior is always purely exponential
asymptotically. As it was shown in \cite{abanovkor}, at finite but
very low temperatures one can observe a crossover from the zero
temperature behavior at short string lengths $n$ to the
exponential behavior (\ref{expbehT}) in the limit of very large
$n$. This crossover occurs at a length scale of the order of the
inverse temperature.

\begin{landscape}
\begin{table}
 \begin{center}
 {\large
 \centering
   \noindent\begin{tabular}{|c|c|l|l|l|l|}
     \hline
     \multicolumn{6}{|c|}{ } \\
     \multicolumn{6}{|c|} {\bfseries EFP for the Anisotropic XY model} \\
     \multicolumn{6}{|c|}{ } \\
     \hline
      Region  &  $\gamma$, $h$  &  $P(n)$  &  Eq.  & Section & Theorem \\
     \hline
     \hline
      $\Sigma_-$  &  $h<-1$  &  $ E\, \eu^{-n \beta}$  &  \ref{pnc}  &  \ref{SigmaMSec}  &  Szeg\"o \\
     \hline
      $\Sigma_0$  &  $-1<h<1$  & $ E\, \eu^{-n \beta }$  &  \ref{pnc}  &  \ref{Sigma0Sec}  &  FH  \\
     \hline
      $\Sigma_+$  &  $h>1$  &
       $ {\it E} \left[ 1 + (-1)^n A \right] \: \eu^{-n \beta}$  &  \ref{oscbehavior}  &  \ref{SigmaPSec}  &  gFH \\
     \hline
      $\Gamma_E$  &  $\gamma^2+h^2=1$  & $ E\, \eu^{-n \beta}$  &  \ref{pnexact}  &  \ref{GammaESec}  &  Exact \\
     \hline
     \hline
      ${\bf \Omega_+}$  &  $h=1$  &
       $ {\it E} \: n^{- {1 / 16} } \left[ 1 + (-1)^n A/\sqrt{n} \: \right] \: \eu^{- n \beta}$  &  \ref{pnomp}  &  \ref{ssOmegap}  &  gFH  \\
     \hline
      ${\bf \Omega_-}$  &  $h=-1$  & $ {\it E} \: n^{- {1 / 16} } \left[ 1 +  A/\sqrt{n} \: \right]  \: \eu^{- n \beta}$  &  \ref{pnomm}  &  \ref{ssOmegam}  &  gFH \\
     \hline
      ${\bf \Omega_0}$  &  $\gamma=0$,\, $|h|<1$  & $ E\, n^{-1/4}e^{-n^2 \alpha}$  &  \ref{pnsh}  &  \ref{gammazero}  & Widom  \\
     \hline
   \end{tabular} }
   \caption[Asymptotic behavior of the EFP in different regimes]
   {Asymptotic behavior of the EFP in different regimes.
    The exponential decay rate $\beta$ is given by Eq. (\ref{betagh})
    for all regimes. The regions in boldface are the critical ones.
    The coefficients $E,A$ are functions of $h$ and $\gamma$
    whose explicit expressions are provided in the text.
    Relevant theorems on Toeplitz determinants are collected in the
    \ref{ToeplitzApp}.}
    \end{center}
   \label{table1}
\end{table}
\end{landscape}

\section{Discussion and conclusions}
\label{Conclusions}

The asymptotic behavior of the Emptiness Formation Probability
$P(n)$ as $n \to \infty$ for the Anisotropic XY model in a
transverse magnetic field as a function of the anisotropy $\gamma$
and the magnetic field $h$ has been studied. We have summarized
our results in Table \ref{table1}. These asymptotic behaviors were
first presented in \cite{abanovfran}.

Our main motivation has been to study the relation between the
criticality of the theory and the asymptotics of the EFP. Let us
now consider the results on the critical lines ($\Omega_0$ and
$\Omega_\pm$ in Fig. \ref{phasediagram}). The Gaussian behavior on
$\Omega_0$ ($\gamma = 0$, $|h| < 1$) is in accord with the
qualitative argument of Ref. \cite{abanovkor} using a field theory
approach. In $\Sigma_0$ ($\gamma \ne 0$, $|h| < 1$) the asymptotic
decay is exponential. We proposed a physical interpretation of the
crossover between the two asymptotes using a bosonization analysis
of the region of small $\gamma$: we suggest that there is an
intermediate regime of Gaussian decay for the string lengths
smaller than $1 / \sqrt{\gamma}$ which crosses over to the
exponential behavior for longer strings.

On the critical lines $\Omega_\pm$, the decay of the EFP is
exponential instead of Gaussian, and apparently contradicts the
qualitative picture of Ref. \cite{abanovkor}. The reason for this
disagreement is that although at $h=\pm 1$ the model can be
rewritten in terms of massless {\it quasiparticles} $\chi$ defined
in (\ref{bogtrans}), we are still interested in the EFP for the
``original'' Jordan-Wigner fermions $\psi$. In terms of $\chi$
this correlator has a complicated (nonlocal) expression very much
different from the simple one (\ref{expect}). From the technical
point of view, the difference is that in the qualitative argument
in favor of a Gaussian decay of EFP for critical systems there is
an implicit assumption that the density of fermions (or
magnetization) is related in a local way to the field responsible
for the critical degrees of freedom (free boson field $\phi$).
This assumption is not valid on the lines $h=\pm 1$. The theory is
critical on those lines and can be described by some free field
$\phi$. However, the relation between the magnetization and this
field is highly nonlocal and one can not apply the simple argument
of \cite{abanovkor} to the XY model at $h=\pm 1$.

Although EFP at the critical magnetic field does not show a
Gaussian behavior, there is an important difference between the
asymptotic behavior of EFP on and off critical lines. Namely, a
power-law pre-factor $n^{-\lambda}$ appears on all critical lines.
For the XY model it is universal (i.e. $\lambda$ is constant on a
given critical line) and takes values $\lambda=1/4$ for $\gamma=0$
\cite{shiroishi} and $\lambda =1/16$ on the lines $h=\pm 1$. It
would be interesting to understand which operators determine these
particular ``scaling dimensions'' of the EFP (see the remark at
the end of Section \ref{ssOmegap}).

At $h\ge 1$ the use of gFH predicts even-odd oscillations of $P(n)$.
We compared the predicted oscillations to numerical calculations of
Toeplitz determinants and found a very good agreement (see
Figs. \ref{DetPlot1},\ref{DetPlot2}).
We proposed a physical interpretation of the oscillations as coming from
pair correlations of spins which can be clearly seen as superconducting
correlations in the fermionic representation (\ref{realfermionH}).

In some parts of the phase diagram ($\Sigma_{+}$, $\Omega_{\pm}$)
we used the so-called {\it generalized Fisher-Hartwig conjecture}
\cite{basor} which is not yet proven.
However, our numeric calculations support the analytical results (see
Figures \ref{DetPlot1} and \ref{DetPlot2}).
We note that to the best of our knowledge this is the first physically
motivated example where the original Fisher-Hartwig conjecture fails
and its extended version is necessary.
\footnote{We note that recently the theory on Toeplitz
determinants has been used and extended with new results in order
to calculate yet one more important physical quantity. We refer
the interested reader to \cite{korepin03}, \cite{korepin04} and
\cite{keating}, where the entanglement for the XY Spin chain and
for Random matrix models have been calculated. } We also suggest
that the gFH could be used to find the subleading corrections to
the asymptotic behavior, as we did for $h = \pm 1$ in
(\ref{pnomp},\ref{pnomm}). This novel hypothesis is supported by
our numerics and it would be interesting to confirm it
analytically.

In conclusion, we notice that it was straightforward to generalize
our results for nonzero temperature. The only modification is that
at $T \ne 0$ the thermal correlation functions must be used
instead of (\ref{F},\ref{G}). Then, the generating function
(\ref{genfunc}) is non-singular everywhere and we have an
exponential decay of $P(n)$ in the whole phase diagram according
to the standard Szeg\"o Theorem and standard statistical mechanics
arguments. We presented results for $T \ne 0$ in
Section~\ref{FiniteEFP}.


\chapter{The Hydrodynamic Approach}
\label{HydroApp}
Very often it is possible to treat a quantum many body system as a
continuum. A description of this kind views the system as a fluid,
where the motion of the individual particles is not important, but
one is interested in their {\it collective} behavior. This idea
has a long history (see
\cite{landau1941,forster,halperin,abanovHydro} for instance).

We consider a one-dimensional system at zero temperature, in the
thermodynamic limit. We will take a ``semi-classical'' limit,
bringing Planck's constant $\hbar$ to zero and the number of
particles $N$ to infinity in such a way that their product stays
constant. This limit allows us to describe the system only in
terms of its local density and velocity.

These two fields are clearly not sufficient to describe all states
of the whole system, but they will provide an accurate
approximation of it for some sectors of the theory. For instance
we cannot consider configurations where the particles occupy
disconnected regions of the phase-space (see Figure
\ref{FF-AllowedPhaseSpace}).

The hydrodynamic approach can be applied to exactly integrable
systems, where the description in terms of just two fields is
essentially exact, once we limit our attention to the `allowed'
sectors of the theory.

The traditional hydrodynamic approach developed in the 1960's
\cite{forster} addresses finite temperature systems and studies
the time evolution of conserved quantities in many particle
models. We concentrate instead on zero-temperature dynamics.

The conservation laws guarantee that a perturbation or a
fluctuation in the density of a conserved quantity of the system
will not disappear on a length scale of the order of the
interparticle distance, but will diffuse on a macroscopic length
scale. The diffusion equation is the main dynamical equation
describing the evolution of these quantities. This description
relies heavily on non-equilibrium processes taking place at a
microscopical level to average into a diffusive motion. Once the
dynamical equation is established, it is relatively simple to
calculate some asymptotics of correlation functions for the
conserved quantities using statistical mechanics methods.

It is clear from this brief explanation that the hydrodynamic
description is valid in the long wavelength approximation, with
wavelengths longer than any other length scale of the system, so
that diffusion averages have meaning and one can use the continuum
description of the system.

We are taking a zero temperature limit, where quantum fluctuations
take the role of thermal fluctuations.

Hydrodynamic equations are non-linear differential equations and
their solution and treatment is not easy. If one linearizes them,
the resulting theory is what is known in studying one-dimensional
models as {\it ``Bosonization''}. Bosonization is a renowned tool,
with a long history, to calculate correlators for the strongly
interacting one dimensional systems. However, there are
\hbox{phenomena} which cannot be captured by a linearized theory.
For instance the calculation of the Emptiness Formation
Probability is beyond the linear approximation. In the next
chapter we show how one can calculate the EFP to leading order for
a number of systems using the hydrodynamic approach.

Before we can analyze any system in particular, in section
\ref{HydroPreliminaries} we introduce some definitions and show
the basic structure of the quantities of interest. Then, in
Section \ref{HydroFF} we start explaining our approach on the
example of the simplest possible system: Free Fermions (FF). This
will allow us to warm up slowly to Section \ref{HydroLag} where we
generalize the approach to other systems with a derivation based
on a Lagrangian Formalism. In section \ref{HydroBos} we show how a
linearized hydrodynamics reduces to the standard bosonization.

Finally, in section \ref{HydroBethe} we show how the Bethe Ansatz
technique can be used to derive the hydrodynamic Hamiltonian of an
integrable system. The Bethe Ansatz allows one to calculate
exactly, although implicitly, the wavefunctions of an integrable
system. Even more importantly, it provides fundamental information
on the thermodynamics of the theory and an easier access to the
conserved quantities of the model. For a short summary of the
Bethe Ansatz, we refer the reader to Appendix \ref{BetheInt}.

In Appendix \ref{Integrability} we show that if we neglect
gradient corrections in a Galilean invariant system, the resulting
hydrodynamic theory possesses a double infinite series of
conserved quantities. It is still unclear whether these conserved
quantities are enough to make the theory integrable in the sense
of Liouville (i.e. the phase-space can be exactly factorized into
a series of tori described by the action-angle variables). The
answer is probably negative. But what is even more puzzling is the
underlying symmetry guaranteeing this abundance of conserved
quantities and its physical meaning. We suggest that origin of
this quantities should be connected to that of free fermions
system, but we are unable to prove this conjecture at this moment.

Zero temperature hydrodynamics was first developed by Landau in
\cite{landau1941}. The collective description approach was
successfully used in the context of lower-dimensional string
theories and two-dimensional quantum gravity \cite{jevicki}. Our
treatment is based on the work of A.G. Abanov and collaborators
and it was described in \cite{abanovHydro}.

\section{Some preliminaries}
\label{HydroPreliminaries}

Let us start by introducing some definitions.

We want to develop a hydrodynamic description of a system in terms
of its local density
\be
  \rho (x) \equiv \sum_{j=1}^N m \delta (x-x_j)
  \label{microrho}
\ee
and current
\bea
  j (x) & \equiv & {1 \over 2} \sum_{j=1}^N \left\{ p_j , \delta (x-x_j) \right\}
  \nonumber \\
  & = &  - \ii {\hbar \over 2} \sum_{j=1}^N \left\{ {\partial \over \partial x_j},\delta (x-x_j) \right\} ,
  \label{microj}
\eea
where the sums are performed over the positions $x_j$ of
every particle in the system and
\be
   \left\{ A, B \right\} \equiv A \cdot B + B \cdot A
\ee
is the {\it anti-commutator}. The velocity is then defined as
\be
   v \equiv {1 \over 2} \left( {1 \over \rho} j + j {1 \over \rho} \right).
   \label{jrhov}
\ee From this point onward we will always intend the velocity and
current to be properly symmetrized, but we will not write it
explicitly for the sake of brevity (e.g. (\ref{jrhov}) will be
written as $v = j / \rho$).

In a Hamiltonian formalism the dynamics of the system is encoded
in the Hamiltonian (expressed as a function of the hydrodynamic
parameters $\rho$ and $v$) and in the commutation relation between
$\rho$ and $v$.

A ``natural'' assumption for the form of a hydrodynamic
Hamiltonian is
\be
   H = \int \rho(x) \de x \left[ m {v^2 (x) \over 2} + \epsilon \left( \rho(x) \right) \right],
   \label{HydroH1}
\ee where $\epsilon(\rho)$ is the internal energy per particle as
a function of the local density of particles. The integration
measure is equivalent to a sum over all particles, while the first
term is the traditional macroscopic kinetic term. The second term
in the Hamiltonian is some sort of potential term that includes
all the interaction, but also the effect of the quantum
statistics. We are going to see that this form is indeed the
correct one in the next sections.

A very important point is how we take the thermodynamic limit. We
are going to see later that the internal energy $\epsilon(\rho)$
is often an essentially quantum quantity, which vanishes in the
classical limit ($\hbar \to 0$). In such case, we are interested
in taking the number of particles to infinity in a way that the
internal energy stays finite as $\rho \to \infty$ and $\hbar \to
0$. We are going to show how to derive the Hamiltonian for
different systems in the next sections.

Assuming standard commutation relation between position and
momentum, \be
   [x_i,p_j] \equiv x_i p_j - p_j x_i = \ii \hbar \delta_{ij} ,
\ee
from their microscopical definitions
(\ref{microrho},\ref{microj}) we can calculate
\be
   \left[ \rho(x), j(y) \right] = \ii \hbar \rho(y) \partial_x \delta (x-y)
\ee
or, using (\ref{jrhov}),
\be
   \left[ \rho(x), v(y) \right] = - \ii \hbar \delta' (x-y) ,
   \label{rhovcomm}
\ee where $\delta' (x)$ denotes the derivative of the Dirac's
delta-function. Eq. (\ref{rhovcomm}) is the canonical commutation
relation for hydrodynamic parameters (see, for instance,
\cite{landau}) and, in connection with the Hamiltonian of the
system (\ref{HydroH1}), will specify the time evolution of $\rho$
and $v$:
\bea
   \rho_t & = & {\ii \over \hbar} [H , \rho]
   \nonumber \\
   & = & - \partial_x ( \rho \, v ) ,
   \label{rhotHam} \\
   v_t & = & {\ii \over \hbar} [H, v]
   \nonumber \\
   & = & - \partial_x \left( {v^2 \over 2} + (\rho \epsilon)_\rho \right) .
   \label{vtHam}
\eea

\section{The simplest example: Free Fermions}
\label{HydroFF}

Let us consider a one dimensional system of Free Fermions (FF),
without internal degrees of freedom. The local density of
particles is given by \be
   \rho(x) = m \int {\de k \over 2 \pi \hbar} .
\ee

Let us assume that the system moves with some velocity $v(x)$,
which itself is a slow function of $x$. This means that the left
and right Fermi points are also functions of $x$ so that:
\be
   \rho(x) = m \int_{k_L (x)}^{k_R (x)} {\de k \over 2 \pi \hbar} =
   m {k_R (x) - k_L (x) \over 2 \pi \hbar} .
   \label{rhoFF}
\ee

The momentum density of the system is then given by
\be
   P (x) \equiv j(x) = \int_{k_L (x)}^{k_R (x)} k \; {\de k \over 2 \pi \hbar}
   = {k^2_R (x) - k^2_L (x) \over 4 \pi \hbar},
   \label{JFF}
\ee where we recognized that the momentum of the system is in fact
its current
\be
   j(x) = \rho(x) \; v (x).
\ee From this equation and using (\ref{rhoFF},\ref{JFF}), we
express the local velocity of the system as
\be
   v(x) = {k_R (x) + k_L (x) \over 2 m}.
   \label{vFF}
\ee Equations (\ref{rhoFF},\ref{vFF}) give us the hydrodynamic
parameters of the free fermions fluid $\rho$ and $v$ in terms of
the left and right Fermi points $k_{L,R}$.

\begin{figure}
 \includegraphics[width=\columnwidth]{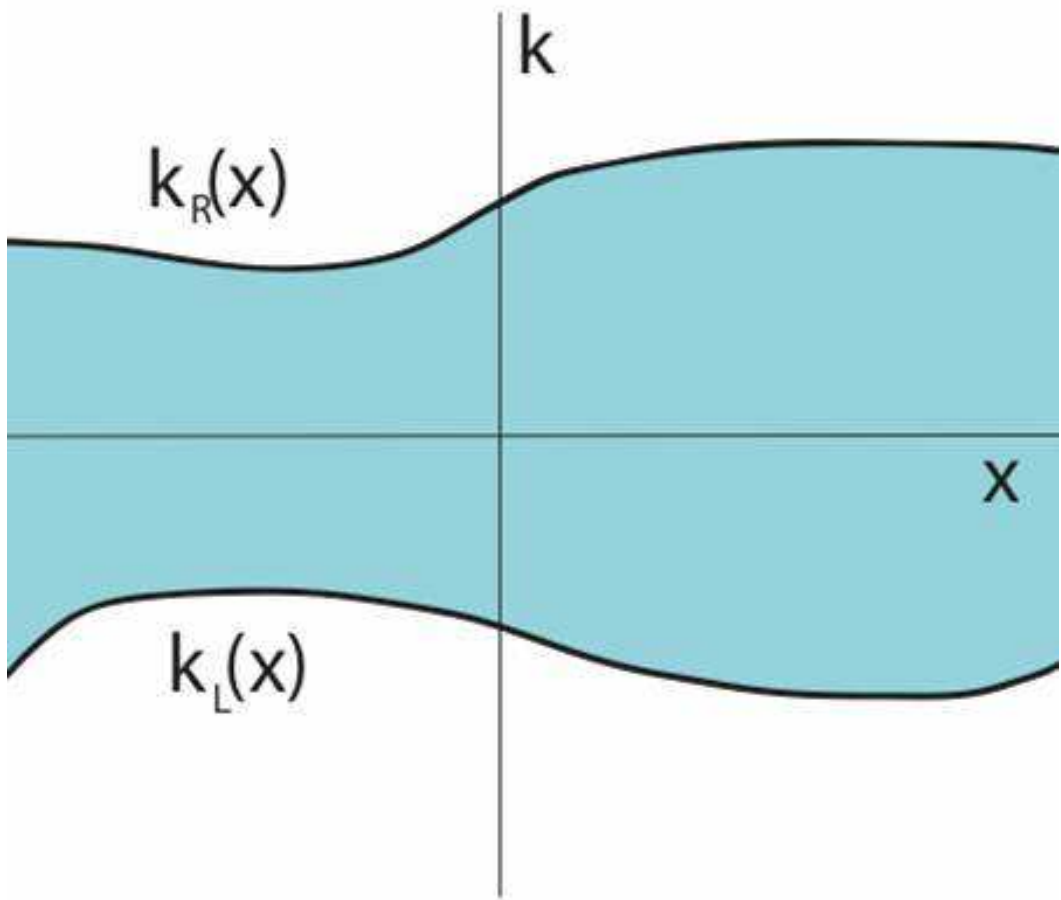}
 \caption[Phase Space depiction for the hydrodynamics of free fermions]
 {Depiction of the typical free fermions phase space configuration we can describe in
our hydrodynamic description. At zero temperature, the particles
are confined within the right and left Fermi Points (dark area)
and these move in space (and time, not depicted here).}
 \label{FF-PhaseSpace}
\end{figure}

\begin{figure}
 \includegraphics[width=\columnwidth]{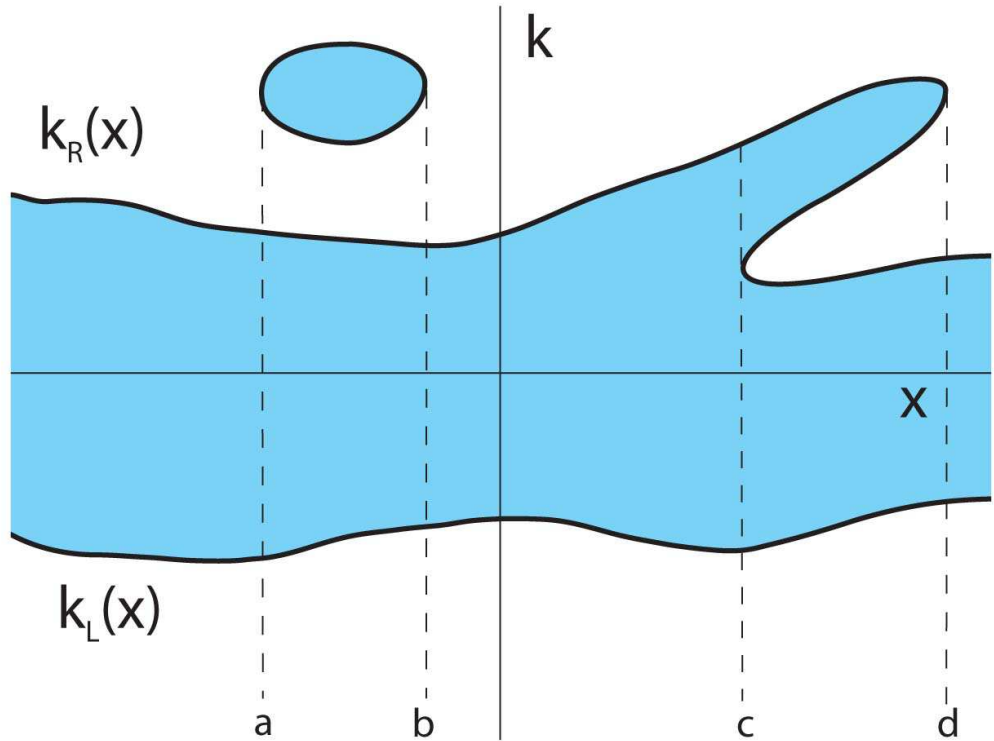}
 \caption[The hydrodynamic description is applicable only in some sector of the free fermions Phase Space]
 {Phase space depiction of a free fermions system. The hydrodynamic description
 can be applied only in star-like configurations like the one between points ${\bf b}$
 and ${\bf c}$. Between ${\bf a}$ and ${\bf b}$ and between ${\bf c}$ and ${\bf d}$
 the world line of the right Fermi Points $k_R (x)$ is not following a quadratic
 profile and the description of the system in terms of just two fields (density and
 velocity) fails.}
 \label{FF-AllowedPhaseSpace}
\end{figure}

We invert (\ref{rhoFF},\ref{vFF}) and get the right and left Fermi
momenta as functions of the density and velocity at each point in
space and time: \be
  k_{R,L} (x,t) = m \; v(x,t) \pm {\hbar \pi \over m} \; \rho (x,t).
  \label{kRLFF}
\ee This is one of the most important limitations of our approach.
For the Fermi points to be functions of the coordinates, they have
to be single valued. This means that we can only describe
star-like configurations (a.k.a. quadratic profiles) that in
phase-space look like in Figure \ref{FF-PhaseSpace}. We cannot
address states similar to the one in Figure
\ref{FF-AllowedPhaseSpace}, where disconnected regions of
phase-space are populated or where the world-line of $k_{R,L}$
goes back above or below itself.

The Hamiltonian of a free fermions system is \be
   H = \int_{k_L}^{k_R} {k^2 \over 2 m} \; {\de k \over 2 \pi \hbar} = {k^3_R - k^3_L \over 12 \pi \hbar m}
\ee and, by using (\ref{kRLFF}), can be written in terms of the
density and velocity of the system as
\be
   E(x) = H = {\rho(x) v^2 (x) \over 2} + {\hbar^2 \pi^2 \over 6 m^4} \rho^3 (x).
   \label{HFF}
\ee We are now in a position to discuss what kind of thermodynamic
limit we are interested in. As we increase the number of
particles, we let the density grow as well. As we approach the
semi-classical limit with $\hbar \to 0$, we let the density grow
so that the product of the two quantity remains constant:
\be
   \hbar \rho \to {\rm const}
   \label{thermlimit}
\ee This means that the energy per particle remains finite and
that the kinetic and potential terms (respectively the first and
the second term in (\ref{HFF})) have comparable magnitude. To make
this point more apparent, let us rescale the velocity as
\be
   v \to {m \over \hbar} \; v
\ee so that the Hamiltonian becomes
\be
   H = {\hbar^2 \over m^2} \left( {\rho v^2 \over 2} + {\pi^2 \over 6 m^2} \rho^3 \right).
\ee

From this point on, we will set constants $\hbar = m =1$, this
means that all quantities are going to be expressed in units of
$\hbar$ and $m$: \be
   H (x) = {\rho(x) v^2 (x) \over 2} + {\pi^2 \over 6} \rho^3 (x).
\ee Note that the second term on the right hand side gives the
internal energy of the system as $\epsilon(\rho)= {\pi^2 \over 6}
\rho^2$ and that this contribution comes only from the Pauli
principle, since we are describing free fermions. For this reason
it is not correct to identify the internal energy with a pure
potential term.

We showed in the previous section that the microscopic description
of the system (\ref{microrho},\ref{microj},\ref{jrhov}) imposes
the following commutation relation between density and velocity:
\be
   \left[ \rho(x), v(y) \right] = -\ii \delta' (x-y) .
\ee Using this relation, the Hamilton equations give us the
dynamics of the system:
\bea
    \rho_t (x) = \left[ H(x), \rho(y) \right] & = & -  \partial_x (\rho v) ,
    \nonumber \\
    v_t (x) = \left[ H(x), v(y) \right] & = & -
    \partial_x \left( {v^2 \over 2} + {\pi^2 \over 2} \rho^2 \right).
    \label{FFEq}
\eea We recognize the first equation as the continuity equation
that expresses particle conservation and relates density and
velocity of the system as dependent quantities
\be
   \partial_t \rho + \partial_x ( \rho v) =0 .
   \label{ContEqFF}
\ee The second dynamical equation is:
\be
   \partial_t v + v \partial_x v = - \pi^2 \rho \partial_x \rho
   \label{EulerEqFF}
\ee and is known as the {\it ``Euler equation''} in the classical
theory of fluids.

We have shown in the simple case of free fermions how to construct
a collective description of the system in terms of local density
and velocity, which represents its hydrodynamic description.

It is worth noticing here that we derived
(\ref{ContEqFF},\ref{EulerEqFF}) essentially semi-classically. For
the free fermion case (and only for this case), it can be shown
\cite{jevicki} that a quantum treatment of the theory gives
exactly the same results, i.e. one only needs to consider
commutation relations instead of Poisson bracket, all the
functions are promoted to be operators (we put ``hats'' on top of
the density, velocity and Hamiltonian) and all equations have to
be interpreted as operator equations.

In the next section we generalize this construction to some
interacting systems.

\section{Lagrangian formulation of Hydrodynamics}
\label{HydroLag}

As we showed in section \ref{Crossover}, the leading behavior of
the EFP can be easily calculated in the semiclassical
approximation. To this end it is useful now to turn to a
Lagrangian formalism to calculate the hydrodynamic action and the
partition function of the quantum theory \cite{abanovHydro}.

We consider the partition function
\be
   Z = \int {\cal D} u \; \eu^{\ii {\cal S} [u] } ,
\ee where ${\cal S} [u]$ is the Action
\be
   {\cal S} [u] = \int \de x \int \de t {\cal L} (u,\dot{u})
\ee of some field or collection of fields $u(x,t)$ and where
$\dot{u} \equiv \partial_t u$.

Eventually, in the next chapter, we are interested in calculating
the EFP as a rare fluctuation in the equilibrium configuration, an
{\it ``instanton''}, and this configuration will take place in
imaginary time. To do this, we will need to formulate the
Euclidean theory with the Euclidean action
\be
   {\cal S}_E [u] = \int \de x \int_{-1/2T}^{1/2T} \de \tau {\cal L} (u,\partial_\tau u),
   \label{EuclideanAction}
\ee with periodic boundary conditions in the imaginary time $\tau
= \ii t$ defined by the temperature of system $T$.

Let us first consider a system with Galilean
invariance\footnote{Lattice systems are more complicated and have
not been treated in this thesis for lack of time, but their
analysis will be addressed shortly.}. The requirement of Galilean
invariance restricts considerably the form of the lagrangian to
\be
   {\cal L} (\rho, v) = {\rho v^2 \over 2} + \rho \epsilon (\rho) + \ldots
   \label{HydroL}
\ee where the first term is the kinetic energy of the system and
is the only one that can depend on the velocity and the second
term is the internal energy of the fluid, to be determined by the
fluid's equation of state. $\epsilon (\rho)$ is the internal
energy per particle at given density $\rho$. Other terms can be
included in the lagrangian and were here denoted by the dots:
these terms include spatial derivatives of density and velocity
and can be neglected if density and velocity gradients are
sufficiently small.

Particle conservation requires the continuity equation
\be
   \partial_t \rho + \partial_x j =0
   \label{ContEq}
\ee to be satisfied, where $j = \rho v$. We can interpret this
equation as a constraint on the fields, relating $\rho$ and $v$ to
each other. We can solve this constraint by introducing the {\it
``particle displacement field''} $u(x,t)$ such that
\bea
   \rho & \equiv & \rho_0 + \partial_x u , \nonumber \\
   j & \equiv & - \partial_t u,
   \label{rhoju}
\eea  where $\rho_0$ is the equilibrium value of the density at
infinity, so that the displacement field can satisfy standard
boundary conditions and vanish at infinity. (One could define the
displacement field in terms of the microscopic theory like in
(\ref{microrho},\ref{microj}) as \be
   u(x) \equiv \sum_{j=1}^N \theta (x - x_j) - \rho_0 x
\ee where the sum is over the $x_j$ position of all particles in
the system and $\theta (x)$ is the Heaviside Step-function.)

One can then write the Lagrangian in terms of the displacement
field:
\be
   {\cal L} (u,\dot{u}) = {\dot{u}^2 \over 2 (\rho_0 + u_x) } +
   (\rho_0 + u_x) \epsilon (\rho_0 + u_x)
   \label{uLag}
\ee and look for the field configuration minimizing the action
${\cal S} [u]$. We can write the Euler-Lagrange equation for this
Lagrangian in terms of the physical field:
\be
   \partial_t v + v \partial_x v = -
   \partial_x \partial_\rho \left[ \rho \epsilon (\rho) \right] ,
   \label{EulerEq}
\ee where $\rho$ and $v$ were defined in terms of $u(x,t)$ in
\ref{rhoju}. This is the Euler equation for a one dimensional
fluid \cite{landau} and reduces to (\ref{EulerEqFF}) for the free
fermions case ($\epsilon (\rho) = {\pi^2 \over 6} \rho^2$).

We see that, once the internal energy $\epsilon (\rho)$ in
(\ref{HydroL}) is known from the equation of state, one has
everything to calculate the Lagrangian of a one-dimensional
Galilean invariant system. In section \ref{HydroBethe} we are
going to show how to compute the internal energy $\epsilon (\rho)$
as a function of the density for integrable models, using the
Bethe Ansatz solution. Alternatively, for non-integrable systems
one can compute $\epsilon (\rho)$ from numerics or from other
phenomenological methods.

The problem is not equally straightforward for systems without
Galilean invariance, like for lattice models. Nonetheless, a
hydrodynamic description can be developed in such cases as well.
In chapter \ref{Spin-ChargeHydro} we will consider systems with
more than one type of particles, namely fermions with spin, and
their hydrodynamics will not be as simple as the one for Galilean
invariant models.

\section{Bosonization as a linearized hydrodynamics}
\label{HydroBos}

If the deviations from the equilibrium state are small, we can
expand the Lagrangian (\ref{uLag}) around $\rho = \rho_0$ and $
j=0 $ as
\be
   {\cal L} (u,\dot{u}) \sim {1 \over 2 \rho_0} \left( \dot{u}^2 +
   v_{s0}^2 \, u_x^2 \right) ,
\ee where we defined the sound velocity at equilibrium as
\be
   v_{s0}^2 \equiv \rho \left. \partial_\rho^2
   \left( \rho \epsilon (\rho) \right) \right|_{\rho=\rho_0} .
   \label{vs0}
\ee

By scaling the time as $v_{s0} t \to t$ we can write the
linearized Action as
\be
   {\cal S} \sim {v_{s0} \over \rho_0} \int \de^2 x {1 \over 2}
   \left( \partial_\mu u \right)^2,
   \label{BosAction}
\ee where $\de^2 x \equiv \de x \de t$ and $\mu = x,t$.

The linearized hydrodynamics is described by the familiar
quadratic action of bosonization, giving the Laplace equation
\be
   \Delta u = 0
\ee as the equation of motion.

\section{Hydrodynamics from the Bethe Ansatz}
\label{HydroBethe}

The Bethe Ansatz is a powerful tool to analyze integrable systems.
For the reader unfamiliar with this technique, we review in
Appendix \ref{BetheInt} the basic formulas that we need to develop
our hydrodynamic approach. In fact, once one is familiar with the
machinery, its use in the hydrodynamic formalism is quite
straightforward.

The Bethe Ansatz solution is based on the distribution of the
quasi-momenta $\tau (q)$. This distribution is found as the
solution of an integral equation known as the Bethe Equation
(\ref{intbetheeq}): \be
   \tau (q) + \int_{-k}^k K (q-q') \tau (q') \de q' = {1 \over 2 \pi} ,
   \label{intfortau}
\ee where the Kernel $K (q-q')$ encodes the interactions of the
system.

For the ground state of the system, the limits of integration in
the integral equation are chosen to be symmetric, i.e. from $-k$
to $k$. But we are interested in a different sector of the theory,
a sector characterized by a total finite momentum of the
system\footnote{One can think as the ground state of the theory as
coming from a grand-canonical ensemble approach. One then
constructs the partition function by adding to the Hamiltonian a
vector potential coupled to the total momentum of the theory, in
addition to the standard chemical potential coupled to the number
of particles.}.

We are guided by our experience with the free fermions model in
section \ref{HydroFF}\footnote{One can derive the free fermions
case from the general formalism we are developing in this section
using the Bethe Ansatz. One has just to keep in mind that for free
fermions, the kernel in (\ref{intfortau}) is equal to zero,
therefore the momentum distribution is $\tau = 1/2 \pi$.} and we
assume asymmetric limits of integration for the Bethe equation:
\be
   \tau (q) + \int_{k_L}^{k_R} K (q-q') \tau (q') \de q' = {1 \over 2 \pi} .
   \label{BetheEq}
\ee

The number of particles and momentum per unit length of the system
are given by (\ref{BetheN},\ref{BetheP}). If we assume them to be
space dependent, we can identify them as the density and current
we need for our hydrodynamic description:
\bea
   \rho (x) & = & \int_{k_L (x)}^{k_R (x)} \tau (q) \de q ,
   \label{BetheRho} \\
   j (x)  & = & \int_{k_L (x)}^{k_R (x)}q \; \tau (q) \de q.
   \label{BetheJ}
\eea

The energy of the system is given by
\be
   H(x) = \int_{k_L (x)}^{k_R (x)} {q^2 \over 2} \tau (q) \de q .
\ee This is an implicit expression for the Hamiltonian, since it
depends on the momentum density $\tau (q)$, which in turn is a
function of $k_R$ and $k_L$.

We could invert (\ref{BetheRho},\ref{BetheJ}) and express $k_R$
and $k_L$ as a function of $\rho$ and $j$ (as we did for the free
fermions) and then use this relation to express the Hamiltonian as
a function of the hydrodynamic parameters density and current.

We can actually do better. For a system like this we can use
Galilean invariance to boost the reference frame. In practice, we
perform a change of variable in the integrations by shifting the
integration variable as \be
   q' = q - v
\ee where
\be
   v \equiv {k_R + k_L \over 2}
\ee so that equations (\ref{BetheRho},\ref{BetheJ}) become
\bea
   \rho (x) & = & \int_{-k (x)}^{k (x)} \tau' (q') \de q' \\
   j (x)  & = & \int_{-k (x)}^{+k (x)} (q'+v) \; \tau' (q') \de q' = \rho(x) v(x)
\eea where we use the fact the the function $\tau' (q') = \tau
(q+v)$ is even and where
\be
   k = {k_L - k_R \over 2}.
\ee

More importantly, this change of variable writes the Hamiltonian
as
\bea
   H & = & {\rho v^2 \over 2} + \int_{-k}^{k} {q'^2 \over 2} \tau' (q') \de q'
   \nonumber \\
   & = & {\rho v^2 \over 2} + \rho \epsilon(\rho)
   \label{HydroH}
\eea
where
\be
   \epsilon(\rho) \equiv {1 \over \rho} \int_{-k}^{k} {q'^2 \over 2} \tau' (q') \de q' .
\ee

This is precisely the expression we derived in generality in
(\ref{HydroH1}) or (\ref{HydroL}), but now we have a microscopic
way to calculate the internal energy function $\epsilon (\rho)$,
from the Bethe Ansatz. Moreover, the formalism developed in this
section will prove very powerful when applied to more complicated
systems as we are going to see in Chapter \ref{Spin-ChargeHydro}.

To conclude, we recall once more that from the commutation
relation (\ref{rhovcomm}) and using (\ref{rhotHam},\ref{vtHam}),
we get the equations of motion (\ref{ContEq},\ref{EulerEq}).


\chapter{The EFP from Hydrodynamics}
\label{HydroEFP}
We now turn back to the problem of calculating the correlator
known as {\it Emptiness Formation Probability}. In Chapter
\ref{EFPinXY} we calculated the asymptotics of this correlation
function in the XY model. This calculation was facilitated by the
specific structure of the model which allowed us to express the
EFP exactly as the determinant of a matrix.

We argued in Chapter \ref{Introduction} for the importance of the
EFP in the theory of integrable models. There, we also pointed out
that the EFP can provide us with important insights in the general
problem of calculating correlators that involve large deviations
from the equilibrium configuration. We propose the hydrodynamic
approach to address this problem.

We introduced our hydrodynamic approach in the previous chapter;
here we are going to show how this formalism helps us in
calculating the leading asymptotic behavior of the EFP for some
integrable models possessing a simple Galilean invariance.

We are going to calculate the EFP as the probability of a rare
fluctuation of the theory, an {\it ``instanton''} which depletes a
region of particles.

In section \ref{instantons} we are going to explain how we are
setting up the calculation and draw some general conclusion. In
section \ref{BosFail} we are going to attempt the calculation
using the linearized hydrodynamics (also know as {\it
Bosonization}) and show that this approximation is only enough to
produce the correct qualitative, but not quantitative result. In
section \ref{EFPAsympt} we are going to manipulate the action of
the theory to show that we can extract the leading behavior of the
EFP from the asymptotic behavior of the instanton solution. In
section \ref{FFEFP} we are going to use this result to calculate
the EFP for the free fermions hydrodynamics we constructed in
section \ref{HydroFF}. Then, in section \ref{CSEFP} we will
consider Calogero-Sutherland particles and calculate the EFP in
the hydrodynamic approach.

These results were presented for the first time in
\cite{abanovHydro}. In the next chapters we are going to develop a
hydrodynamic description for more complicated models and we will
show for the first time how to calculate the EFP for these models.

\section{EFP as an instanton configuration}
\label{instantons}

We consider the Lagrangian formulation of the hydrodynamic theory
introduced in section \ref{HydroLag}. We are going to work in
Euclidean space, i.e. in imaginary time, because we are not
studying the dynamical evolution of the system, but a property of
the ground state.

We write the action of the Euclidean hydrodynamic theory as
\be
    Z = \int {\cal D} u \; e^{-{\cal S}_E [u]},
 \label{Z}
\ee where ${\cal S}_E [u]$ was defined in (\ref{EuclideanAction}).

Our approach is essentially the same as the one used in section
\ref{Crossover}. The asymptotic behavior of the EFP is defined as
a rare fluctuation that depletes a region of length $2 R$ from
every particle. We interpret this configuration $u(\tau,x)$ as the
result of a collective motion where all particles move away so
that at some time $\tau=0$ we have no particles in the spatial
interval $\left[-R,R\right]$.

This trajectory $u(\tau,x)$ is a solution of the equations of
motion satisfying the EFP boundary conditions:
\be
    \rho(\tau=0;-R<x<R) =0,
    \label{bcefp}
\ee and standard boundary conditions at infinity
\bea
    \rho &\to& \rho_{0}, \;\;\;\; x,\tau\to\infty,
    \nonumber \\
    v &\to& 0, \;\;\;\; x,\tau\to\infty,
    \label{bcinf}
\eea

Then with exponential accuracy, the EFP is calculated as
\be
    P(R) \sim e^{-{\cal S}_{opt}} ,
  \label{prinst}
\ee where ${\cal S}_{opt}$ is the value of the action
(\ref{EuclideanAction}) on the configuration $u(x,t)$.

In the introduction of Chapter \ref{EFPinXY} we already argued
that we can estimate the qualitative dependence on $R$ of the
stationary action ${\cal S}_{opt}$ on very general grounds. The
argument went as follows: in order to satisfy the EFP boundary
conditions \ref{bcefp}, the instanton solution needs to open a gap
in its density that perturbs the configuration both in space and
time. The spatial extent of this disturbance is clearly of the
order of the gap to be opened, i.e. $R$.

If we assume that the effective fluid we are describing is
compressible and there are no other relevant length scales in the
system, then the typical time scale of the disturbance has to be
of the order of $R/v_{s}$, where $v_{s}$ is the sound velocity at
$\rho=\rho_{0}$. Therefore, the (space-time) ``area'' of the
disturbance is of the order of $R^{2}$, the action ${\cal
S}_{opt}\sim R^{2}$ and we conclude that the decay of the EFP
(\ref{prinst}) is going to be Gaussian:
\be
   P(R) \sim e^{-\alpha R^{2}},
   \label{gauss}
\ee where $\alpha$ is some (non-universal) constant depending on
the details of (\ref{EuclideanAction}).

In general, this behavior will persist until another length scale
emerges to compete with $R$. At finite but sufficiently low
temperatures, the temporal extent of the instanton $R/v_{s}$ is
smaller than the inverse temperature $R/v_{s}\ll 1/T$ and the
instanton does not feel the effect of the temperature, resulting
in am intermediate Gaussian decay of EFP as in (\ref{gauss}).
However, as the temperature increases, or as we consider bigger
$R$, the periodic boundary conditions in time become relevant and
the $1/T$ scale defines the temporal size of the instanton, so
that the space-time area of the disturbance scales as $R$ and one
obtains \be
  P(R) \sim e^{-\gamma R}.
  \label{exp}
\ee

Similarly, if some other term in the theory
(\ref{EuclideanAction}) drives the system away from criticality,
the hydrodynamic approach will describe an incompressible fluid.
The correlation length of the density fluctuations will have the
same effect of a relevant length scale as just discussed and the
EFP will decay exponentially as in (\ref{exp}) even at zero
temperature if $R$ is bigger than this correlation length. In both
cases, as one increases $R$, one would observe a crossover from
Gaussian to exponential behavior, with the crossover happening
when $R \sim v_s \over T$ or then $R$ is comparable with the
relevant correlation length.

Once again, this is the same argument of \cite{abanovkor} and we
have shown the latter crossover phenomenon in (\ref{Crossover}).

Before we proceed to any actual calculation, for the sake of
generality, we would like to consider a slightly different problem
from the Emptiness Formation Probability (\ref{bcefp}). We are
going to impose the Depletion Formation Probability (DFP)
conditions on the instanton configuration: \be
   \rho(\tau=0;-R<x<R) =\bar\rho,
   \label{bcdfp}
\ee where $\bar\rho$ is some constant density. Setting
$\bar\rho=0$ we obtain the EFP problem while for $\bar\rho$ close
to $\rho_{0}$ one can use the bosonization technique to calculate
$P(R;\bar\rho)$.

In conclusion, we would like to find a solution to the equations
of motion that satisfy the boundary conditions
(\ref{bcdfp},\ref{bcinf}) and then use this solution to calculate
the value of the stationary action ${\cal S}_{opt}$. In Euclidean
space, due to the different sign convention, the Euler equation is
\be
   \partial_{\tau} v + v \partial_{x} v =
   \partial_{x} \partial_{\rho} [ \rho \epsilon(\rho) ],
   \label{euler}
\ee and the continuity equation reads
\be
   \partial_{\tau}\rho+\partial_{x}j=0.
   \label{continuity}
\ee

\section{Linearized hydrodynamics or bosonization}
\label{BosFail}

A simple qualitative analysis would tell us why we cannot
calculate the EFP using a bosonization approach. In fact, one of
the most important approximations done in deriving the
bosonization description of a system is to assume a linear
spectrum. This is definitely a very reasonable assumption when one
wants to look at low energy excitations close to the Fermi points,
but the EFP clearly picks up contributions from every point in the
spectrum, since it imposes a strong deviation from the equilibrium
distribution in the configuration space. As we pointed out in
Chapter \ref{Introduction}, the inability of Bosonization to
describe the EFP is one of the main motivations for our study of
this correlator, since it is a poster child of the advantage of
using a hydrodynamic description over bosonization for certain
calculations.

However, we can try to calculate the DFP (\ref{bcdfp}) in the
bosonization approach if
\be
  {\rho_0 - \bar{\rho} \over \rho_0 } \ll 1,
  \label{smbos}
\ee i.e., if we consider the probability of formation of a only
small constant density depletion along the long string $-R<x<R$.

In this case the deviation of the density from $\rho_{0}$ is small
almost everywhere, as we will show below, and therefore we can
assume that only particles close to the Fermi surface will be
involved in the creation of the depletion and bosonization is
applicable.

As we showed in section \ref{HydroBos}, we can derive bosonization
by linearizing the hydrodynamic action (\ref{EuclideanAction})
into (\ref{BosAction}). The corresponding linearized equation of
motion for the displacement field $u(\tau,x)$ is the Laplace
equation \be
   \Delta u =0
   \label{laplace}
\ee which has to be solved with the DFP boundary condition
\be
  u(\tau=0,x) = -(\rho_{0}-\bar\rho) x, \quad \mbox{for}\; -R<x<R.
  \label{bcdfpbos}
\ee

In \cite{abanovHydro}, this problem was solved by inspection and
the correct instanton solution was found to be \be
   u(x,t) = -(\rho_0 - \bar{\rho} ) \,
   \Re \left( z_0 - \sqrt{ z_0^2 - R^2 } \right),
   \label{bossol}
\ee with the complex notation
\be
   z_0 \equiv  x + \ii v_{s0} \tau
\ee and where $v_{s0}$ was defined in (\ref{vs0}) as
\be
   v_{s0}^2 \equiv \rho \left. \partial_\rho^2
   \left( \rho \epsilon (\rho) \right) \right|_{\rho=\rho_0}
   \footnote{It is important to remark that bosonization is
universal and that it has the same form for every theory. The
specificity of each model is captured by the parameter $v_{s0}$.}
. \ee

One can check that indeed, at $\tau=0$, $-R<x<R$ the complex
coordinate $z_0$ is real and square root in (\ref{bossol}) is
purely imaginary so that (\ref{bcdfpbos}) is satisfied. At
space-time infinity $z_0 \to \infty$ we have
\be
  u(x,t) \approx -{\alpha \over z_0} - {\bar{\alpha} \over \bar{z_0} }
  \label{asymptu}
\ee
with
\be
  \alpha = \bar{\alpha} \equiv {1 \over 4} \left( \rho_0 - \bar{\rho} \right) \, R^{2}
  \label{alphabos}
\ee and where we used $\bar{z_0} \equiv x - \ii v_{so} \tau$ to
denote complex conjugation.

In terms of the original hydrodynamic parameters (\ref{rhoju}), we
obtain from (\ref{asymptu}) that as $z_0 \to \infty$
\bea
  \rho & \approx & \rho_0 + {\alpha \over z_0^2 } + {\bar{\alpha} \over \bar{z_0^2} },
  \nonumber \\
  v & \approx & -\ii {v_{s0} \over \rho_0 } \left( {\alpha \over z_0^2}
  - {\bar{\alpha} \over \bar{z_0^2} } \right),
  \label{rvbosinf}
\eea
which obviously satisfy the boundary conditions at infinity (\ref{bcinf}).

By plugging this instanton solution (\ref{bossol}) into
(\ref{BosAction}), we can calculate the stationary action for the
DFP problem as
\be
  {\cal S}_{DFP} = {1 \over 2} {v_{s0} \over \pi \rho_0}
  \left[ \pi ( \rho_0 - \bar{\rho} ) \, R \right]^2.
  \label{dfpbos}
\ee

Let us end this section by noticing that the instanton solution
(\ref{bossol}) is singular close to the ends of the string $t=0$,
$x=\pm R$ and therefore gradients of $u$, i.e. density and
velocity, diverge. This is not consistent with our
\hbox{approximation} justifying the bosonization approach, i.e.
that the solution does not deviate too much from its equilibrium
value. However, in \cite{abanovHydro} the corrections coming from
those areas were estimated to contribute only with terms of higher
order in the small parameter (\ref{smbos}) to the action
(\ref{dfpbos}).

\section{EFP through the asymptotics of the solution}
\label{EFPAsympt}

In \cite{abanovHydro}, Abanov showed that the calculation of the
value of the stationary action $S_{opt}$ (\ref{prinst}) can be
extracted from the asymptotics of the EFP solution of the
hydrodynamic equations, using a Maupertui principle.

The first step is to calculate the variation of the action
(\ref{EuclideanAction}) with respect to the displacement field $u$
\bea
    \delta {\cal S}_E  & = & \int \de^2 x \, \Big\{ -\partial_{\tau} (v \delta u)
    - \partial_{x} \left[ \left( {v^2 \over 2 } - \partial_{\rho} (\epsilon\rho)
    \right) \delta u \right]
    \nonumber \\
    && + \delta u \left[ \partial_{\tau} v + v \partial_{x} v
    - \partial_{x} \partial_{\rho} (\epsilon\rho) \right] \Big\}.
    \label{dS1}
\eea

Note that we kept surface terms (full derivatives) in addition to
the last term which produces the equation of motion (\ref{euler}).

Let us now calculate the derivative of the action ${\cal S}_E$
with respect to the equilibrium background density $\rho_{0}$ and
evaluate it on the EFP (or DFP) solution that saturates the
equations of motion:
\be
    \partial_{\rho_0} {\cal S}_{opt}
    = \int \de^2 x \, \left\{ - \partial_{\tau} (v \partial_{\rho_0} u)
    - \partial_{x} \left[ \left( {v^2 \over 2} - \partial_{\rho} (\epsilon\rho)
    \right) \partial_{\rho_0} u \right] \right\}.
    \label{dS2}
\ee

We can rewrite the full derivative terms in (\ref{dS2}) as a
boundary term
\bea
    \partial_{\rho_0} {\cal S}_{opt}
    = \oint \, \left[ v \partial_{\rho_{0}} u\, \de x
    +\left( \partial_{\rho} (\epsilon\rho) - {v^2 \over 2} \right)
    \partial_{\rho_0} u \, \de \tau \right],
    \label{dS3}
\eea where the integral is taken over an infinitely large contour
around the $x-\tau$ plane.

Equation (\ref{dS3}) gives us the value of the (derivative of the)
stationary action from boundary terms lying at infinitely large
$x$ and $\tau$. If we assume the asymptotic behavior for the
instanton solution to be
\bea
  u(x,t) & \approx & -{\alpha \over z_0} - {\bar{\alpha} \over \bar{z}_0 } ,
  \nonumber \\
  \rho & \approx & \rho_0 + {\alpha \over z_0^2 } + {\bar{\alpha} \over \bar{z}_0^2 },
  \nonumber \\
  v & \approx & -\ii {v_{s0} \over \rho_0 } \left( {\alpha \over z_0^2}
  - {\bar{\alpha} \over \bar{z}_0^2 } \right),
  \label{asybeh}
\eea like in (\ref{asymptu},\ref{rvbosinf}), after some algebra we
conclude
\be
   \partial_{\rho_0} {\cal S}_{opt} = 2 \pi {v_{s0} \over \rho_0}
   ( \alpha + \bar{\alpha} ).
   \label{main}
\ee

The problem of calculating the leading behavior of the EFP (DFP)
is then reduced to the evaluation of the asymptotic behavior of
the instanton solution. Once one knows $\alpha$ in (\ref{asybeh}),
by integrating equation (\ref{main}) one finds the desired
behavior of the EFP (DFP) problem.

We can check this result with what we found in the previous
section from bosonization. Substituting into (\ref{main}) the
value of $\alpha$ obtained in (\ref{alphabos}) we obtain
\be
   \partial_{\rho_0} {\cal S}_{opt} = \pi {v_{s0} \over \rho_0}
  (\rho_0 - \bar{\rho}) \, R^2
\ee which is equivalent to (\ref{dfpbos}) up to higher order terms
in the  small parameter (\ref{smbos}).

\section{EFP for Free Fermions}
\label{FFEFP}

In section \ref{HydroFF} we analyzed in details the hydrodynamic
description of free fermions. The hydrodynamic equations for this
system are
\bea
   \partial_{\tau} \rho + \partial_{x}(\rho v) & = & 0,
   \nonumber \\
   \partial_{\tau} v + v \partial_{x} v & = & \pi^{2} \rho \partial_{x} \rho.
   \label{hff}
\eea with the sound velocity
\be
    v_s = v_F = k_F = \sqrt{ \rho \partial_\rho^2 (\rho\epsilon) } = \pi\rho
   \label{vsff}
\ee where we noticed that, since in our notation $m=1$, the sound
velocity and Fermi velocity are the same for free fermions.

By introducing a complex field
\be
   w \equiv \pi \rho + \ii v,
   \label{w}
\ee we can rewrite both dynamical equations (\ref{hff}) as the
single complex Hopf equation\footnote{In real time formalism,
instead of $w$ and $\bar{w}$, one introduces right and left Fermi
momenta $k_{R,L}= \pi \rho \pm v$ which satisfy the Euler-Hopf
equations $\partial_{t} k + k \partial_{x} k = 0$, reflecting the
absence of interactions between fermions.}: \be
   \partial_{\tau} w - \ii w \partial_{x} w =0.
   \label{hopf}
\ee

The general solution of this equation can be written implicitly as
\be
   w = F (x + \ii w \tau) = F(z),
   \label{hsol}
\ee where $F(z)$ is an arbitrary analytic function of the complex
variable $z$ defined as
\be
   z \equiv  x + \ii w \tau .
   \label{zhopf}
\ee

We can determine the function $F(z)$ through the boundary
conditions. In \cite{abanovHydro} it was shown that by defining
\be
   F(z) \equiv \pi\bar\rho +\pi(\rho_{0}-\bar\rho) {z \over \sqrt{z^2 - R^2}}
   \label{Fff}
\ee equation (\ref{hsol}) satisfies the DFP problem and we can get
the EFP solution by setting $\bar{\rho} = 0$.

\begin{figure}
 \begin{center}
   \includegraphics[width=\columnwidth]{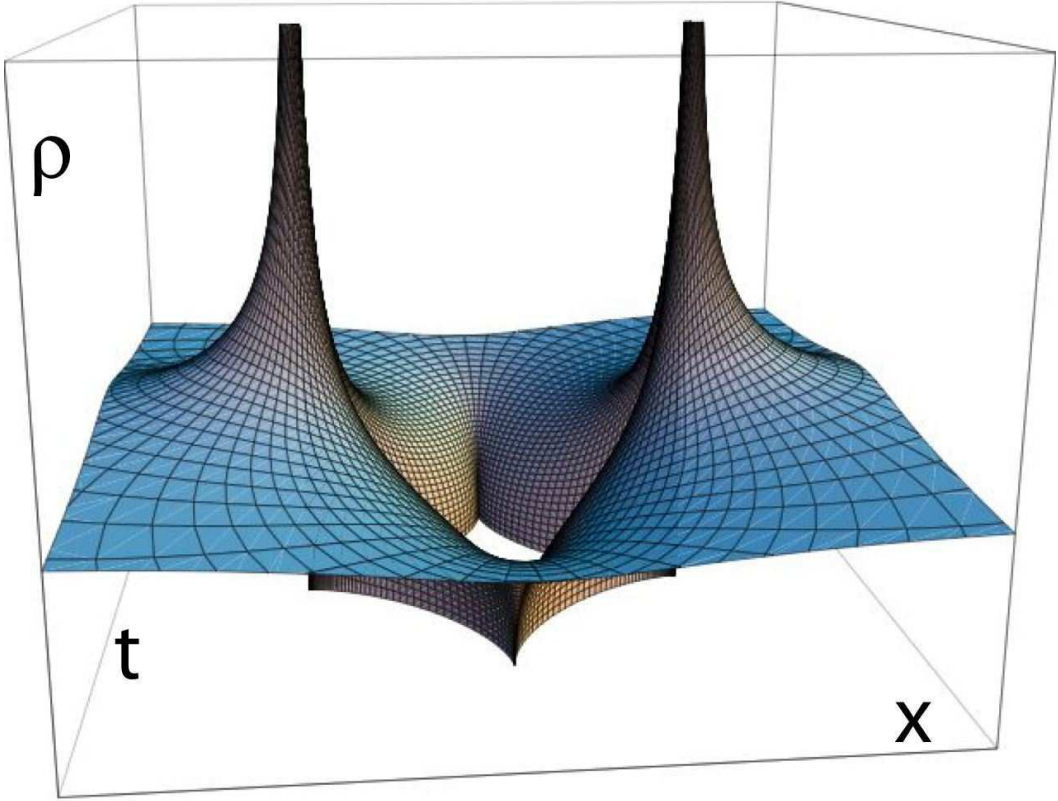}
 \end{center}
   \caption[Density profile for the EFP instanton for free fermions]
   {The density profile $\rho(x,\tau)$ is shown for the EFP instanton as implicitly given by (\ref{wffsol}).
   The density diverges at points $(x,\tau)=(\pm R,0)$.
   The shape of the ``Emptiness'' is shown in figure \ref{fig:droplet}. [From \cite{abanovHydro}]}
 \label{fig:density}
\end{figure}

\begin{figure}
 \begin{center}
   \includegraphics[width=\columnwidth]{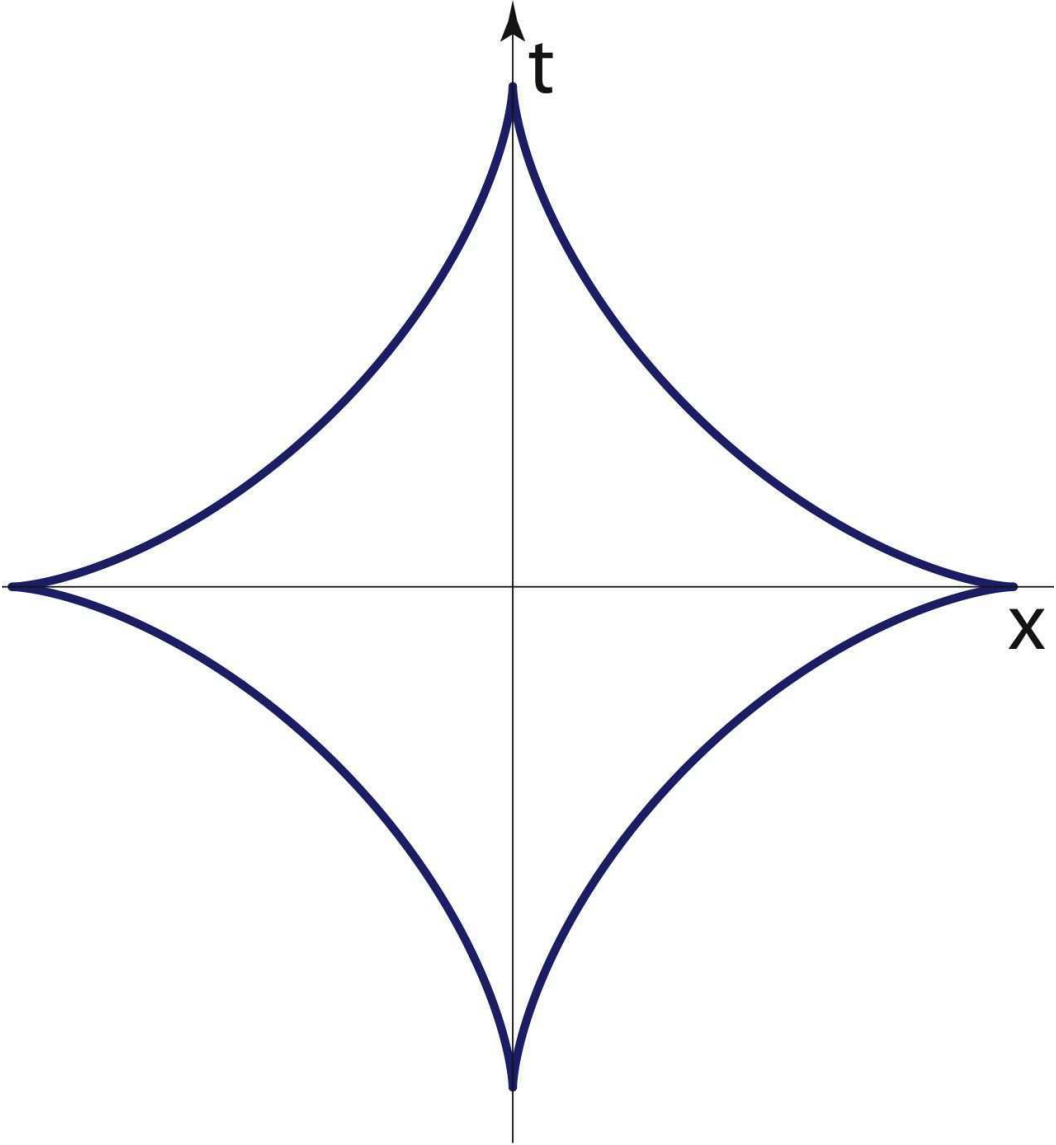}
 \end{center}
   \caption[Space-time region of vanishing density of the free fermion instanton solution]
   {The region of the $x-\tau$ plane in which $\rho(\tau,x)=0$
   for the EFP instanton solution for free fermions (\ref{wffsol} is shown.
   The boundary of the region can be found to be given by an astroid
   $x^{2/3}+(\pi \rho_0 \tau)^{2/3}=R^{2/3}$. [From \cite{abanovHydro}]}
 \label{fig:droplet}
\end{figure}

One obtains the density and velocity by taking the real and
imaginary part of $w$. To write down this instanton solution
explicitly is quite complicated, since one has to solve the
equation
\be
   w -  \pi \bar{\rho} = \pi (\rho_0 - \bar{\rho} ) {z \over \sqrt{z^2 - R^2} }.
   \label{wffsol}
\ee Abanov in \cite{abanovHydro} plotted a numerical solution that
we show in figures \ref{fig:density} and \ref{fig:droplet}).

However, as we argued in the previous section, to calculate the
leading behavior of the EFP (DFP), we do not need to know the full
solution of (\ref{wffsol}), but we just need the asymptotic
behavior of the instanton solution using (\ref{main}).

In the limit $x , \tau \to \infty$ we have $w \to \pi \rho_0$ and
$z= x +\ii w \tau \to x + \ii \pi \rho_0 \tau = z_0$. Therefore,
$w$ from (\ref{wffsol}) behaves asymptotically as
\be
  w - \pi \rho_0 \approx \pi (\rho_0 - \bar{\rho} ) {R^2 \over 2 z_0^2}
  \label{wasff}
\ee and taking, e.g., its real part we have
\be
  \rho - \rho_0 \approx {1 \over 4} (\rho_0 - \bar{\rho}) R^2
  \left( {1 \over z_0^2 } + {1 \over \bar{z}_0^2 } \right).
  \label{rhoffas}
\ee

Comparing (\ref{rhoffas}) with (\ref{asybeh}) we identify
\be
   \alpha = {1 \over 4} (\rho_0 - \bar{\rho}) R^2.
\ee Substituting this into (\ref{main}) we find
\be
   \partial_{\rho_0} {\cal S}_{opt} =
   \pi^2 (\rho_0 - \bar{\rho} ) R^2
\ee and, after integrating in $\rho_0$,
\be
  {\cal S}_{opt} = {1 \over 2} \left[ \pi ( \rho_0 - \bar{\rho} ) R \right]^2.
  \label{soptff}
\ee

We can conclude that DFP and EFP are respectively
\bea
    P_{{\rm DFP}} (R) & \sim & \exp \left\{
    - {1 \over 2} \left[ \pi ( \rho_0 - \bar{\rho}) \, R \right]^2 \right\},
    \label{dfpff} \\
    P_{{\rm EFP}} (R) & \sim & \exp \left\{
    - {1 \over 2} \left( \pi \rho_0 \, R \right)^2  \right\}.
    \label{efpff}
\eea

We can compare this result with the known results on the EFP for
free fermions \cite{dysmehta} that we recapitulated in Appendix
\ref{KnownRes} and confirm that (\ref{efpff}) gives the correct
first (Gaussian) term.

\section{EFP for the Calogero-Sutherland model}
\label{CSEFP}

Our final example for this chapter is the Calogero-Sutherland
model, an integrable model of one-dimensional particles
interacting with an inverse square potential. The Hamiltonian of
the model is
\bea
    H &=& -\frac{1}{2}\sum_{j=1}^{N}\frac{\partial^{2}}{\partial x_{j}^{2}}
    +\frac{1}{2} \sum_{1\leq j<k\leq N} \frac{\lambda(\lambda-1)}{(x_{j}-x_{k})^{2}}
 \la{CSh} \\
    &=& -\frac{1}{2}\sum_{i=1}^{N} \left(\frac{\partial}{\partial x_{i}}
    +\sum_{j=1, j\neq i}^{N}\frac{\lambda}{x_{i}-x_{j}} \right)
    \left(\frac{\partial}{\partial x_{i}}
    -\sum_{k=1, k\neq i}^{N}\frac{\lambda}{x_{i}-x_{k}} \right).
 \nonumber
\eea and we are going to consider it in the thermodynamic limit $N
\to \infty$ while keeping the density constant as we discussed in
the previous chapter (\ref{thermlimit})\footnote{Technically, in
order to go into the thermodynamic limit, we need either to add an
external harmonic potential to confine the particles or to
consider the model as defined on a closed ring; otherwise the
repulsive nature of the interaction would spread the particles and
we would not be able to keep the density fixed. We can neglect
these details here, since they do not change anything relevant for
us and, most of all, they both preserve the integrability of the
model.}.

This model is known to be integrable
\cite{Calogero-1969,Sutherland-1971} and we briefly analyzed it in
the appendix dedicated to the Bethe Ansatz (\ref{CSPot}). We can
write explicitly the ground state wavefunction of (\ref{CSh}) as
\be
   \Psi_{GS} = \prod_{j<k}(x_{j}-x_{k})^{\lambda}.
\ee As discussed in Appendix \ref{BetheInt}, this formula shows
that Calogero-Sutherland \hbox{particles} have intermediate
statistics interpolating between non-interacting bosons
($\lambda=0$) and non-interacting fermions ($\lambda=1$).

As a connection to the theory of Random matrices we mentioned in
\ref{EFPSystems}, we should point out that that the probability
distribution of particle coordinates
\be
    \left|\Psi_{GS}\right|^{2} = \prod_{j<k} |x_{j}-x_{k}|^{2\lambda}
\ee at some particular values of the coupling constant $\lambda
=1/2,\, 1,\, 2$ coincides with the joint probability of
eigenvalues for the orthogonal, unitary, and symplectic random
matrix ensembles respectively (see Eq. (\ref{jed})).

To calculate the leading behavior of the EFP for the
Calogero-Sutherland model we need to determine the equation of
motion of its hydrodynamic description (\ref{euler}), which means
that we need to know the internal energy function
$\epsilon(\rho)$. This can easily be found in
\cite{Sutherland-1971} or from a direct Bethe Ansatz calculation
and is
\be
   \epsilon(\rho) = {\pi^2 \over 6} \, \lambda^2 \, \rho^2.
   \label{ercs}
\ee

We immediately notice that (\ref{ercs}) differs from the free
fermion case ($\epsilon_{FF} (\rho) = {\pi^2 \over 6} \, \rho^2$)
just by a factor of $\lambda^2$ and coincides with the latter (as
expected) at $\lambda=1$.

Therefore, we can repeat the exact same analysis as in the last
section and derive the EFP (DFP) very easily. Introducing the
complex Riemann invariants
\be
    w \equiv \lambda \pi \rho + \ii v
\ee
and repeating the calculations of the previous section we obtain
\bea
    P_{{\rm DFP}} (R) & \sim & \exp \left\{
    - {1 \over 2} \lambda \left[ \pi ( \rho_0 - \bar{\rho} ) \, R \right]^2 \right\},
    \label{dfpcs} \\
    P_{{\rm EFP}} (R) & \sim & \exp \left\{
    - {1 \over 2} \lambda \left( \pi \rho_0 \, R \right)^2 \right\}.
    \label{efpcs}
\eea

Once again we can compare this result with what is already
known(\ref{CalogeroExact}) and confirming that, indeed,
(\ref{efpcs}) gives the exact leading asymptotics of the EFP for
the Calogero-Sutherland model. Subleading (in $1/R$) corrections
to (\ref{efpcs}) are due to gradient corrections to the
hydrodynamic action (\ref{ercs}) and to quantum fluctuations
around the found instanton which go beyond our accuracy.




\chapter{Hydrodynamics for a Spin-Charge System}
\label{Spin-ChargeHydro}
One of the most interesting predictions of the {\it Luttinger
Liquid} (LL) model is the effect known as the spin charge
separation, according to which, in a one-dimensional system, the
fundamental excitations are {\it ``holons''} and {\it ``spinons''}
and they carry independently the charge and spin degrees of
freedom of the electrons respectively.

These results of the LL model have been known for many years, but
only now are we reaching the technological advancements necessary
to test it experimentally. In recent years many laboratories have
attempted to confirm this prediction (see, for instance
\cite{GaAs}-\cite{coldfermigas}) and the data are definitely in
accordance with the theory. What is still missing is a conclusive
test that would firmly link the experimental results to the
theoretical prediction, in that the methods used so far are fairly
indirect and their interpretation is not unique.

While devices are developed right now to overcome this problem and
\hbox{confirm} with certainty the occurrence of spin-charge
separation in one dimension, we are interested in looking ahead
and investigating the correction to this prediction.

In fact, it is well known that this effect is valid only for low
energy excitations, in that its derivation assumes the spectrum to
be linear. It is also known that the curvature of the spectrum
will destroy perfect spin-charge separation for higher energy
excitations and introduce a coupling between the spin and charge
degrees of freedom.

Unfortunately, the field theory description underlying the LL
model ({\it bosonization}), becomes inconsistent if one were to
consider a non-linear spectrum: even a reasonable quadratic
spectrum generates a theory with no stable vacuum. Even if one
tries to perturbatively include corrections to the linear spectrum
approximation, the expectation values calculated with this theory
are divergent and to complete a reasonable bosonization
calculation beyond the linear spectrum approximation is still an
open challenge in many contexts.

To address the problem of determining the corrections to exact
spin-charge separation we propose the hydrodynamic approach. We
already discussed with the example of the EFP how the hydrodynamic
approach is a natural generalization of a bosonization description
to include the full spectrum of the microscopic theory. Therefore,
we expect to be able to generalize the bosonization description of
electrons in terms of non-interacting holons and spinons to
include curvature corrections that will couple the two degrees of
freedom.

This hydrodynamic description will involve two interacting fluids,
one describing the bulk motion of the system, i.e. the charge, and
one the internal degrees of freedom, i.e. the spin. In the limit
of linear spectrum the interactions between these two fluids will
vanish and one would recover traditional LL results, while
normally the interactions will mix spin and charge in the fluids.
Ideally, we might find that the gradient-less hydrodynamic
description is integrable in the sense of Appendix
\ref{Integrability}. Then, one could be able to derive the stable
excitations of the system, {\it ``solitons''}, i.e. one could
identify two stable quasi-particles that would carry a fraction of
the total charge and of the total spin of an electron each.

To derive the hydrodynamic description, we use the exactly
solvable model of electrons with contact interactions, which was
first solved using the Bethe Ansatz in \cite{yang67}. Another
interesting integrable model to consider would be the spin
Calogero-Sutherland model \cite{polychronakos}, which can be
solved using the asymptotic Bethe Ansatz. These two models are at
opposite limits, in that one assumes the shortest interaction
possible, i.e. contact interaction, while the other describes an
inverse quadratic long range potential. While these can be good
approximations to some physical systems, most systems lie in
between these two extremes. Nonetheless, studying the limiting
cases, where the theory of integrable models can help us, will
allow us to us gain some insights on the general structure of a
hydrodynamic description of spinful electrons.

In section \ref{QuadBos} we sketch how one would bosonize a theory
with quadratic spectrum and show that the resulting model has an
unstable vacuum and that a perturbative treatment of the curvature
of the spectrum generates divergences in the calculation of
physical observables. In section \ref{BetheDelta}, we describe the
Bethe solution for a gas of electrons with contact interaction and
we show how to implicitly derive its hydrodynamic description.


\section{Bosonization for a quadratic spectrum}
\label{QuadBos}

When one wants to construct the bosonization description of a
system, the first step is to linearize the spectrum. As we are
going to show, it is hard to make sense of the bosonization of a
non-linear spectrum. Therefore, often one refers to
``bosonization'' as the whole procedure of linearizing the
spectrum and then expressing the system in terms of its density
excitations. However, it is important to keep in mind that in this
section we are going to distinguish these two steps and call
``bosonization'' just the transformation from the microscopic
degrees of freedom to the collective ones. In doing so we are
aware that we are losing rigorousness and that the resulting model
has to be interpreted carefully.

With these remarks in mind, let us now sketch how one would
bosonize a quadratic theory. For simplicity, let us consider free
fermions:
\bea
   {\cal H} & =&  - \Psi^\dagger (x) \partial_x^2 \Psi (x)
   \nonumber \\
   & = & k^2 \Psi^\dagger (k) \Psi (k)
\eea where $\partial_x \equiv \partial / \partial x$ and the
second line shows the spectrum of the Hamiltonian in the Fourier
space representation.

The traditional bosonization approach linearizes the spectrum as
\bea
   {\cal H} & = & - \sum_{L,R}  \psi^\dagger_{L,R} \left( \partial_x \pm \ii k_F \right)^2 \psi_{L,R}
   \nonumber \\
   & \simeq & - k_F^2 \sum_{L,R} \psi^\dagger_{L,R} \psi_{L,R}
   \mp \ii k_F \sum_{L,R} \psi^\dagger_{L,R} \partial_x \psi_{L,R} + \ldots
   \label{linearspinH}
\eea where the first term is interpreted as a chemical potential
and can be absorbed in a redefinition of the ground energy, while
the second term expresses the linear spectrum of the excitations
around the Fermi points $\pm k_F$. Moreover, we defined left- and
right-moving fields $\psi_{L,R}$ as the fields obtained expanding
around the left/right Fermi Point:
\be
   \Psi (x) =  : \eu^{\ii k_F x} \psi_R (x) + \eu^{- \ii k_F x} \psi_L (x) : \; .
\ee

In bosonization one effectively describes the system with a single
collective field that captures the density fluctuation. In
practice, one makes the following transformation:
\be
    \psi_{L,R} (x) \equiv {1 \over \sqrt{2 \pi}} \eu^{\pm \ii \sqrt{4 \pi} \phi_{L,R} (x) },
    \label{BosTrans}
\ee where $\phi_{L,R}$ are the left and right moving bosonic
fields.

We use the transformation (\ref{BosTrans}) to calculate various
bilinears in the spin fields. For instance, one can consider a
quantity like
\bea
   : \psi_{L,R}^\dagger (x) \psi_{L,R} (x + \epsilon) : & = &
    {1 \over 2 \pi} : \eu^{\pm \ii \sqrt{4 \pi} \left( \phi_{L,R} (x + \epsilon) - \phi_{L,R} (x) \right) } :
   \eu^{\pm 4 \pi \langle \phi_{L,R} (x) \phi_{L,R} (x+\epsilon) \rangle} =
   \nonumber \\
   & = & { 1 \over 2 \ii \pi \epsilon }
   \left[ \eu^{\pm \ii \sqrt{4 \pi} \left( \phi_{L,R} (x + \epsilon) - \phi_{L,R} (x) \right) } - 1 \right]
   \label{bilinear}
\eea where we used the identity
\be
   \eu^A \eu^B = : \eu^{A+B} : \; \eu^{\langle AB + {A^2 + B^2 \over 2} \rangle}
\ee and the fact that
\be
   \langle \psi_{L,R} (0) \psi_{L,R} (x) - \psi_{L,R}^2 (0) \rangle =
   \lim_{\alpha \to 0} {1 \over 4 \pi} \ln {\alpha \over \alpha \pm \ii x}.
\ee The colons denote normal ordering\footnote{Normal ordering
means that in the evaluation we should put the creation operators
to the right of all the annihilation operators. This convention is
equivalent to subtracting un-physical zero energy contributions.}
and in the second line of (\ref{bilinear}) we used the fact the
normal ordering simply amounts to subtract $1$ from the
exponential.

We can expand (\ref{bilinear}) in powers of $\epsilon$ to find
\bea
   : \psi_{L,R}^\dagger (x) \psi_{L,R} (x + \epsilon) : & = &
   \sum_{n=0}^\infty {\epsilon^n \over n!} \psi_{L,R}^\dagger (x) \partial^n \psi_{L,R} (x)
   \nonumber \\
   & = &  { 1 \over 2 \ii \pi \epsilon }
   \left[ \eu^{\pm \ii \sqrt{4 \pi } \sum_{n=1}^\infty {\epsilon^n \over n!} \phi_{L,R}^{(n)} (x) } - 1 \right]
   \label{bilexp}
\eea which give the generating function of the fermionic currents
\be
   j^{L,R}_n (x) \equiv \psi_{L,R}^\dagger (x) \partial_{L,R}^n \psi (x)
\ee in terms of the bosonic field $\phi_{L,R}$.

By matching powers of $\epsilon$ in (\ref{bilexp}) we can write
down these expressions. The density of fermion is
\be
   \rho_{L,R} = j^{L,R}_0 = \psi_{L,R}^\dagger (x) \psi_{L,R} (x) =
   \pm {1 \over \sqrt{ \pi} } \partial_x \phi_{L,R} (x),
\ee the current density is
\be
   j^{L,R}_1 = \psi_{L,R}^\dagger (x) \partial_x \psi_{L,R} (x) =
   \ii \left( \partial_x \phi_{L,R} (x) \right)^2 \pm {1 \over \sqrt{4 \pi} } \partial_x^2 \phi_{L,R} \; .
\ee

The third term in the expansion can be identify with the original
quadratic Hamiltonian for the left/right movers
\bea
   {\cal H}_{L,R} & = & j^{L,R}_2 = \psi_{L,R}^\dagger (x) \partial^2_x \psi_{L,R} (x)
   \nonumber \\
   & = & \mp { \sqrt{ 4 \pi} \over 3} \left( \partial_x \phi_{L,R} (x) \right)^3
   + \ii \left( \partial_x \phi_{L,R} \right) \left( \partial_x^2 \phi_{L,R} \right)
   \pm {1 \over 3 \sqrt{ \pi} } \partial_x^3 \phi_{L,R} \; ,
   \label{cubicHam}
   \nonumber \\
\eea but is important to notice that while the first line is a
well defined Hamiltonian operator, the second line is not and can
be understood only as a perturbative interaction term.

The last terms in (\ref{cubicHam}) are total derivatives and can
therefore be neglected as boundary terms. Therefore a model of
free fermions with quadratic spectrum would be transformed into a
theory of bosonic fields with Hamiltonian
\be
   {\cal H} = \left( \partial_x \phi_R \right)^3 + \left( \partial_x \phi_L \right)^3 .
\ee This describes a cubic theory and therefore it cannot be
quantized, since the spectrum for the bosonic field has no lower
bound and the ground state of the theory is unstable and has an
infinite energy. This is the reason for which it does not make
sense to directly bosonize a non-linear theory.

If one were, instead, to consider the linearized version of the
fermion theory (\ref{linearspinH}), using the expressions found
above, the bosonized Hamiltonian would be
\be
   H \sim k_F \left( \partial_x \phi_R \right)^2 + k_F \left( \partial_x \phi_L \right)^2 + \ldots
   \label{bosHam}
\ee One can then include the additional terms like
(\ref{cubicHam}) neglected in (\ref{linearspinH}) in a
perturbative way and treat them as small correction.
Unfortunately, even this attempt is ill-fated, since the
calculations of observable quantities diverge and nobody has found
the correct way to resum the diagrams and cure these infinities.

To add spin degrees of freedom in the bosonization is very easy.
Essentially, the machinery we just outlined can be repeated for
two fields, corresponding to spin-up and spin-down
fermions/bosons. The fermionic version of the Hamiltonian will
therefore be
\be
   {\cal H}  = - \Psi_{\downarrow}^\dagger (x) \partial_x^2 \Psi_{\downarrow} (x)
   - \Psi_{\uparrow}^\dagger (x) \partial_x^2 \Psi_{\uparrow} (x)
\ee while the linearized bosonic one will look like
\be
   H \sim k_F \sum_{L,R} \left( \partial_x \phi^{\downarrow}_{L,R} \right)^2
   + k_F \sum_{L,R} \left( \partial_x \phi^{\uparrow}_{L,R} \right)^2 + \ldots \quad .
\ee

One of the most interesting properties of the bosonization
procedure is that all the interactions between fermions that are
relevant, i.e. that cannot renormalized to zero, can be expressed
in terms of quadratic interactions for the bosonic field. One can
show that the effect of the interactions can be absorbed in a
different value for the coefficients in front of the left and
right moving fields in (\ref{bosHam}).

One can define spin and charge bosonic fields as
\bea
   \phi_c & \equiv & \phi^{\uparrow} + \phi^{\downarrow}
   \nonumber \\
   \phi_s & \equiv & \phi^{\uparrow} - \phi^{\downarrow}
   \label{holonspinondef}
\eea which are often referred to as {\it ``holons''} and {\it
``spinons''} respectively. In terms of these fields, the
Hamiltonian is
\be
   H \sim v_c \left( \partial_x \phi_c \right)^2 + v_s \left( \partial_x \phi_s \right)^2 + \ldots,
   \label{spinchargeham}
\ee where we took into account that interaction can renormalize
the coefficients in front of the spin and charge degrees of
freedom differently, redefining the spin and charge (Fermi)
velocity.

The Hamiltonian being quadratic, transformation
(\ref{holonspinondef}) leaves it substantially unchanged, i.e.
(\ref{spinchargeham}) remains quadratic. But if we were to include
the additional cubic terms coming from the spectrum curvature,
transformation (\ref{holonspinondef}) would mix the spin and
charge degrees of freedom, effectively introducing interactions
between holons and spinons like
\be
   \Delta H = {\alpha \over k_F} \left( \partial_x \phi_c \right) \left( \partial_x \phi_s \right)^2
   + {\beta \over k_F} \left( \partial_x \phi_c \right)^2 \left( \partial_x \phi_s \right) .
\ee These terms are suppressed compared to the main ones in
(\ref{spinchargeham}) by a factor of $1 / k_F$ (throughout the
calculation, $k_F$ has been the ``big'' parameter in the theory).
One could therefore introduce them in the bosonization description
and use them to calculate the corrections to exact spin-charge
separation. Unfortunately, such calculations are again ill-fated
in that their results are divergent. This is quite normal in
bosonization calculations, but, as far as we know, nobody has been
able to devise a scheme to cure this divergences for this
problem\footnote{Let us remark that recently Pustilnik and
coauthors have successfully solved a similar problem in
\cite{glazman}.}.

That is the reason for which we think that a hydrodynamic approach
to the problem can be quite effective in finding the corrections
to spin-charge separation at low energies.

\section{Hydrodynamics for fermions with contact interaction}
\label{BetheDelta}

We now consider a system of $N$ spin-1/2 fermions interacting
through the Hamiltonian \be
   H = - \sum_{i=1}^N {\partial^2 \over \partial x_i^2 } + 4 c \sum_{i<j} \delta (x_i -x_j) ,
\ee where $c$ is the coupling determining the strength of the
contact interaction.

We want to consider the Bethe ansatz ground wavefunction for the
state with a fixed number of spin down particles $M$, with a fixed
total momentum $P$ and a fixed total spin momentum $P_s$ (defined,
in this case, as the momentum of the spin-down degree of freedom).
In a Grand Canonical approach we would write: \be
  {\cal{H}} = H + \mu_0 h_0(M) + \mu_1 h_1(2M-N) + \mu_2 h_2(P) + \mu_3 h_3(P_s) .
\ee These definitions are chosen so that $N$ and $P$, and $M$ and
$P_s$ are canonically conjugate hydrodynamics variables satisfying
a continuity equation for each pair.

The Bethe Ansatz construction is more complicated than the one
outlined in Appendix \ref{BetheInt}, because in this case we have
two species of particles coexisting, fermions with spin up and
fermions with spin down. Therefore, one needs to include an
additional degree of freedom in the Bethe Ansatz approach, a
degree of freedom corresponding to the spin of the particle that
is going to be parameterized by a spin quasi-momentum, or, to be
more precise, by a spin rapidity\footnote{Rapidities are variables
related to the quasi-momentum that allow to express the scattering
matrix in a simpler form.}.

The existence of this second sets of parameters for the spins
brings a second set of Bethe equations (\ref{betheeq}) that have
to be satisfied. In the thermodynamic limit this will produce two
coupled integral equations in the density of the quasi-momenta and
in the density of the spin rapidities. The Bethe Ansatz
construction for this model was firstly reported in \cite{yang67},
while its thermodynamic study was undertaken in
\cite{takahashi71}.

We assume a wavefunction of the form
\be
   \Psi (x_1 s_1, x_2 s_2, \ldots , x_N s_N) = \sum_j \Phi_j^M (x_1, x_2, \ldots , x_N)
   G_j^M
\ee where $s_i = \pm 1/2$ are the spin quantum numbers and $G_j^M$
is the spin part of the wavefunction. The spatial part of the
wavefunction is constructed as
\be
   \Phi_j^M = \sum_P [Q,P] \eu^{\ii \sum_{j=1}^N k_{pj} x_{Qj} }
\ee
with the coefficients given by
\be
   [Q,P] = \pm \sum_R A_R \prod_{j=1}^M F_P (\Lambda_{Rj},y_i)
\ee
where
\bea
   F_P (\Lambda,y) & = & \prod_{j=1}^{y-1} {k_{Pj} - \Lambda + \ii c \over k_{P(j+1)} - \Lambda - \ii c } \\
   A_R & = & \prod_{i<j; R_i > R_j} {\Lambda_{Rj} - \Lambda_{Ri} + 2 \ii c \over \Lambda_{Rj} - \Lambda_{Ri} - 2 \ii c }.
\eea
The $y_i$'s are the coordinates of the spin down particles (i.e. a subsets of the $x_i$'s);
$Q$, $P$ and $R$ are respectively permutations of the $x_i$'s, $k_j$'s and $y_i$'s.

By imposing periodic boundary conditions we find the following Bethe equations
\bea
   \eu^{\ii k_j L} & = & \prod_{\alpha=1}^M {k_j - \Lambda_{\alpha} + \ii c \over k_j - \Lambda_{\alpha} - \ii c } \\
   \prod_{j=1}^N {\Lambda_{\alpha} - k_j+ \ii c \over \Lambda_{\alpha} - k_j - \ii c } & = &
   \prod_{\beta \ne \alpha} {\Lambda_{\alpha} - \Lambda_{\beta} + 2 \ii c \over \Lambda_{\alpha} - \Lambda_{\beta} - 2 \ii c }
   \label{nestedbetheeq}
\eea
constituting a nested Bethe ansatz.
Taking the logarithm of these equations we obtain
\bea
    L k & = & 2 \pi I_k - 2 \sum_\Lambda \arctan \left( { k - \Lambda \over c } \right) \\
    0 & = & 2 \pi J_\Lambda - 2 \sum_\Lambda \arctan \left( { \Lambda - k \over c } \right)
     + 2 \sum_{\Lambda'} \arctan \left( {  \Lambda - \Lambda' \over 2 c } \right) .
\eea Due to the additional degrees of freedom, we now have two
sets of integers $I_k$ and $J_\lambda$ to define the state of the
system.

In the thermodynamic limit ($L \to \infty$, $N \to \infty$, $M \to
\infty$ with the condition that $ \rho = N / L$ and $ \rho_s = M
/ L$ are finite) the sums can be converted into integrals and the
densities of momenta $k$'s and rapidities $\Lambda$'s are
determined through integral equations:
\bea
    2 \pi \sigma (\Lambda) & = & - \int_{B_L}^{B_R} {4 c \sigma (\Lambda') \de \Lambda' \over
    4 c^2 + ( \Lambda - \Lambda')^2 } + \int_{Q_L}^{Q_R} {2 c \tau (k) \de k \over
    c^2 + ( \Lambda - k)^2 } \\
    2 \pi \tau (k) & = & 1 + \int_{B_L}^{B_R} {4 c \sigma (\Lambda) \de \Lambda \over
    c^2 + 4 ( k - \Lambda )^2 } \: .
\eea

These two integral equations define the densities $\tau (k)$ and
$\sigma (\Lambda)$ as a function of the parameters $Q_L, Q_R, B_L,
B_R$. To determine these parameters in terms of physical
observables we need to satisfy the following consistency
conditions: \bea
    \rho = \int_{Q_L}^{Q_R} \tau (k) \de k , \quad && \quad
    P =  \int_{Q_L}^{Q_R} k \; \tau (k) \de k ,
    \nonumber \\
    \rho_s = \int_{B_L}^{B_R} \sigma (\Lambda) \de \Lambda , \quad && \quad
    P_s = \int_{B_L}^{B_R} p(\Lambda) \sigma (\Lambda) \de \Lambda
    \label{hydroparameters}
\eea where $p (\Lambda)$ is the pseudo-momentum of the spin
degrees of freedom with rapidity $\Lambda$\footnote{These
definitions give $P = {2 \pi \over L} \sum_k I_k + {2 \pi \over L}
\sum_\Lambda J_\Lambda$ and $P_s = {2 \pi \over L} \sum_\Lambda
J_\Lambda$ in terms of the Bethe integers.}:
\be
   p (\Lambda) \equiv \ii M \int_{Q_L}^{Q_R} \ln \left(
   {\ii c + \Lambda - k \over \ii c - \Lambda +k }
   \right) \; \tau(k) \de k \; ,
\ee which comes from the definition
\be
   \eu^{\ii p(\Lambda_{\alpha}) M } \equiv
   \prod_{j=1}^N {\Lambda_{\alpha} - k_j+ \ii c \over \Lambda_{\alpha} - k_j - \ii c }.
\ee

Finally, the energy (Hamiltonian) of the system is given by
\be
   H (Q_L, Q_R, B_L, B_R) = \int_{Q_L}^{Q_R} k^2 \tau (k) \de k .
\ee We can now invert the system of equations
(\ref{hydroparameters}) to express the parameters $Q_L, Q_R, B_L,
B_R$ in terms of the hydrodynamic variables $\rho, J, \rho_s, J_s$
and to construct the hydrodynamic Hamiltonian
\be
  H (\rho, P, \rho_s, P_s) = H (Q_L, Q_R, B_L, B_R) .
\ee
The equations of motion can be easily derived using the fundamental commutation relations
\be
    [\rho_{\uparrow} (x), v_{\uparrow} (y) ] = [\rho_{\downarrow} (x), v_{\downarrow} (y) ] = - \ii \delta' (x-y)
\ee
where $v=j/\rho$ is the velocity and the spin/charge degrees degrees of freedom are connected
to the spin-up/down one by
\bea
   \rho = \rho_{\uparrow} + \rho_{\downarrow} \quad && \quad J = J_{\uparrow} + J_{\downarrow} \\
   \rho_s = \rho_{\downarrow} \quad && \quad J_s = J_{\downarrow} \: .
\eea





\chapter{Aharonov-Bohm effect with many vortices}
\label{ManyVortices}

We now turn to a problem that is very different from what we
considered so far and we study a two-dimensional configuration.
The Aharonov-Bohm effect is the prime example of a zero field
situation where a non-trivial vector potential acquires physical
significance, a typical quantum mechanical effect. We consider an
extension of the traditional A-B problem, by studying a
two-dimensional medium filled with many point-like vortices.
Systems like this might be present within a Type II
superconducting layer in the presence of a strong magnetic field
perpendicular to the layer. We are going to construct an explicit
solution for the wavefunction of a scalar particle moving within
one such layer when the vortices occupy the sites of a square
lattice. From this construction we infer some general
characteristics of the spectrum and imply that such a flux array
produces a repulsive barrier to an incident low-energy charged
particle, so that the penetration probability decays
exponentially.

\section{Introduction}

In classical mechanics it is said that the vector potential has no
physical meaning, and only the electromagnetic field has physical
(measurable) effects, since it is the only gauge invariant. In
quantum mechanics, however, the vector potential appears in gauge
invariant quantities that describe a new class of effects. In
these cases, corresponding to topologically non-trivial
configurations, we recognize the importance of the vector
potential, even when the electromagnetic field vanishes everywhere
in the regions accessible to a charged particle.

The standard example of this class of effects is recognized in the
Aharonov-Bohm effect \cite{ab}, in which a magnetic field is
confined to a region of space, and electrically charged particles
are only free to move outside this region. Although a particle
cannot experience the field strength directly, the covariant
momentum, i.e. the momentum derived through the minimal coupling
to the electromagnetic field, is
\be
  D_{\mu} = \partial_{\mu} - \ii e A_{\mu}
\ee and is affected by this overall configuration, because the
vector potential $A_{\mu}$ carries a {\it `memory'} of the
presence of the magnetic field even outside the region where the
field is localized.

In this way, the particle is influenced by the field, through a
shift in the phase of the wavefunction
\be
 {e \over \hbar} \oint {\bf A} \cdot \de {\bf x} =
 {e \over \hbar} \int {\bf H} \cdot \de {\bf s} =
 {e \over \hbar} \Phi,
 \label{covmomentum}
\ee where $\Phi$ is the total magnetic flux inside the circuit
(i.e. a closed path of the particle)\footnote{For a phase shift to
be measurable, we need to create an interference effect and
therefore the particle has to come back to the same point. It does
not make sense to speak of a phase shift for open paths.}. This
explains why the effect is called `topological': the behavior of
the particle is sensitive to the overall configuration of the
system, even though there is no classical magnetic force at any
point.

The extension of the A-B problem in the presence of many localized
fluxes cannot in general be tackled exactly. There exists a simple
argument \cite{Aharonov} due to Aharonov which shows, using the
Bloch theorem, that an infinite line of equispaced point-like
fluxes would constitute an impenetrable barrier to a particle of
sufficiently low energy.  The particle would not be able to pass
through such an array because it could not satisfy simultaneously
on both sides of the barrier the Bloch periodicity conditions on
its phase, in the light of the A-B effect.

We are interested in exploring a possibly more realistic set-up by
studying the propagation of a charged particle through a medium
filled with point-like fluxes.

Experimentally, one might find a situation similar to this inside
a Type II superconducting layer in the presence of a large
magnetic field perpendicular to the layer. Quasiparticles in the
layer would encounter numerous vortices, each containing a
superconductor flux quantum, and under some conditions might not
penetrate the vortices (see, for instance, \cite{typeII}).

Situations similar to this have been addressed by several authors
in recent years \cite{randomvortices}, especially in connection
with transport studies and with a special focus on the Hall
conductivity of two-dimensional electron gases on top of
superconducting material. These works have shown a depletion of
the density of states at the bottom of the spectrum.

Our aim is to consider such a 2-dimensional layer, punctured by
magnetic fluxes, and to study the wavefunction of a single scalar
particle entering this medium. For simplicity, we take the
vortices as point-like, so that the space available for the
particles is a punctured plane. A similar attempt was done by
Nambu in \cite{nambu}, and our aim is to extend his results.

We are going to show that a lattice of impenetrable magnetic
fluxes (vortices), such as the one described above, constitutes a
barrier to a low-energy charged particle trying to pass through
the medium. That is, the distribution of the vortices creates a
configuration whose topological constraints on the wavefunction
are comparable to an effective repulsive potential. Qualitatively,
there are a number of ways to see this:
\begin{itemize}

\item{
The presence of the fluxes generates a non-zero vector potential
inside the medium, raising the minimum energy (that is the square
of the covariant momentum, Eq. \ref{covmomentum}) required for an
electrically charged particle to exist in the medium,}

\item{
Particles are repelled by the vortices, as their wavefunctions
must vanish on the vortex sites. Therefore, the bigger the typical
amplitude of the wavefunction in the flux-containing region, the
bigger the energy due to the sharp spatial variation. This means
that for low-energy states the wavefunction will not be able to
reach a value appreciably different from zero in the presence of
fluxes,}

\item{
The analysis of Nambu indicates that the medium constitutes a
barrier even from the point of view of angular momentum. In
\cite{nambu}, he argues that the angular momentum of a particle
should be greater than the magnetic flux present in the medium if
the particle wavefunction is to satisfy the boundary conditions.
In other words, the lower angular momentum levels are missing and
are not part of the spectrum.}

\end{itemize}

This result is even more clear in light of the aforementioned
works \cite{randomvortices} showing a Lifshits tail in the density
of state at low energies for a random distribution of vortices.
From a physical point of view, it seems quite clear that a charged
particle approaching the medium with sufficiently low energy will
be repelled, that is, its penetration will be exponentially
damped. In the same way, if we localize a particle in its ground
state in a region without vortices, the particle will not be able
to escape outside that region through one containing vortices
except by tunneling, and we should be able to construct a bound
state of topological character (actually a very long-lived
resonance), even though there is no classical force. The fact that
a bound state can be topological in nature is new and was already
suggested by the work of Nambu \cite{nambu}.

\section{Mathematical preliminaries}

We concentrate on the case in which all the $N$ fluxes have equal
strength $\Phi = \Phi_0 / 2$, where $\Phi_0 = 2 \pi {\hbar \over
e}$ is the quantum unit of flux. In this case it can be shown
(see, for instance, \cite{aharogold}) that the problem is
invariant under time-reversal, and we can therefore choose the
wavefunctions to be real.

Indicating with $(x_i,y_i)$, $i=1..N$, the coordinates of the
vortices, we can write the vector potential in the standard
circular gauge as
\bea
 (A_x,A_y) & = & \Phi \left (\sum_{i=1}^N {y-y_i \over (x-x_i)^2 + (y-y_i)^2},
 - \sum_{i=1}^N {x-x_i \over (x-x_i)^2 + (y-y_i)^2} \right) \nonumber \\
 & = & \Phi {\bf \nabla} \sum_{i=1}^N \tan^{-1} \left( { y-y_i \over x-x_i },
 \right) = \nonumber \\
 & = & \ii \Phi {\bf \nabla} \sum_{j=1}^N \ln \left(
 { (x-x_j) + \ii (y-y_j) \over (x-x_j) - \ii (y-y_j) } \right) \\
 {\bf \nabla} \times {\bf A} & = & 2 \pi \Phi \sum_{i=1}^N \delta^2
 \left( x-x_i , y-y_i \right).
\eea

The equation of motion for a particle in this medium is given by
the Schr\"odinger equation (in units $\hbar = e =1$)
\bea
 {1 \over 2 m} \left( {\bf \nabla} - \ii {\bf A} \right)^2 \Psi +E \Psi =0,
\eea
and, in these units, integer $\Phi$'s correspond to a
quantized flux (in our case $\Phi = 1/2$).

Following the idea of Nambu \cite{nambu}, we implement a singular
gauge transformation $G$ to remove the vector potential:
\bea
 \Psi = G \psi \qquad G=\prod_{j=1}^N \left(
 { (x-x_j) - \ii (y-y_j) \over (x-x_j) + \ii (y-y_j) } \right)^{1/2}.
\eea
In this way, we reduce our problem to a free-field case
\bea
 -{1 \over 2 m} {\bf \nabla}^2 \psi = E \psi,
\eea with non-trivial (topological) boundary conditions on the
wavefunctions in the region surrounding each vortex.

In constructing our solutions, we must require that the
wavefunctions vanish on the vortex sites
\be
 \psi (x=x_i, y=y_i) = 0 \qquad i=1..N \ \ ,
\ee and that they acquire the A-B phase  $\eu^{2 \ii \pi \Phi} =
-1$ each time a particle completes a turn around a vortex. More
precisely stated, in this singular gauge the effect of the vector
potential is represented by a phase-matching condition on the
wavefunction
\be
 \psi (\theta) = - \psi (\theta + 2 \pi )
\ee
where $\theta$ is the azimuthal angle about the vortex.

We know from standard complex analysis that this condition implies
the existence in the 2-dimensional plane of a cut connecting two
distinguished Riemann sheets. For a real wavefunction this last
condition implies that there exists at least one line exiting each
vortex site on which the function has to vanish in order to change
its sign.

\section{Construction of the solutions}

\begin{figure}
  \dimen0=\textwidth
  \advance\dimen0 by -\columnsep
  \divide\dimen0 by 2
  \noindent\begin{minipage}[t]{\dimen0}
    \resizebox {\columnwidth}{!}{\includegraphics{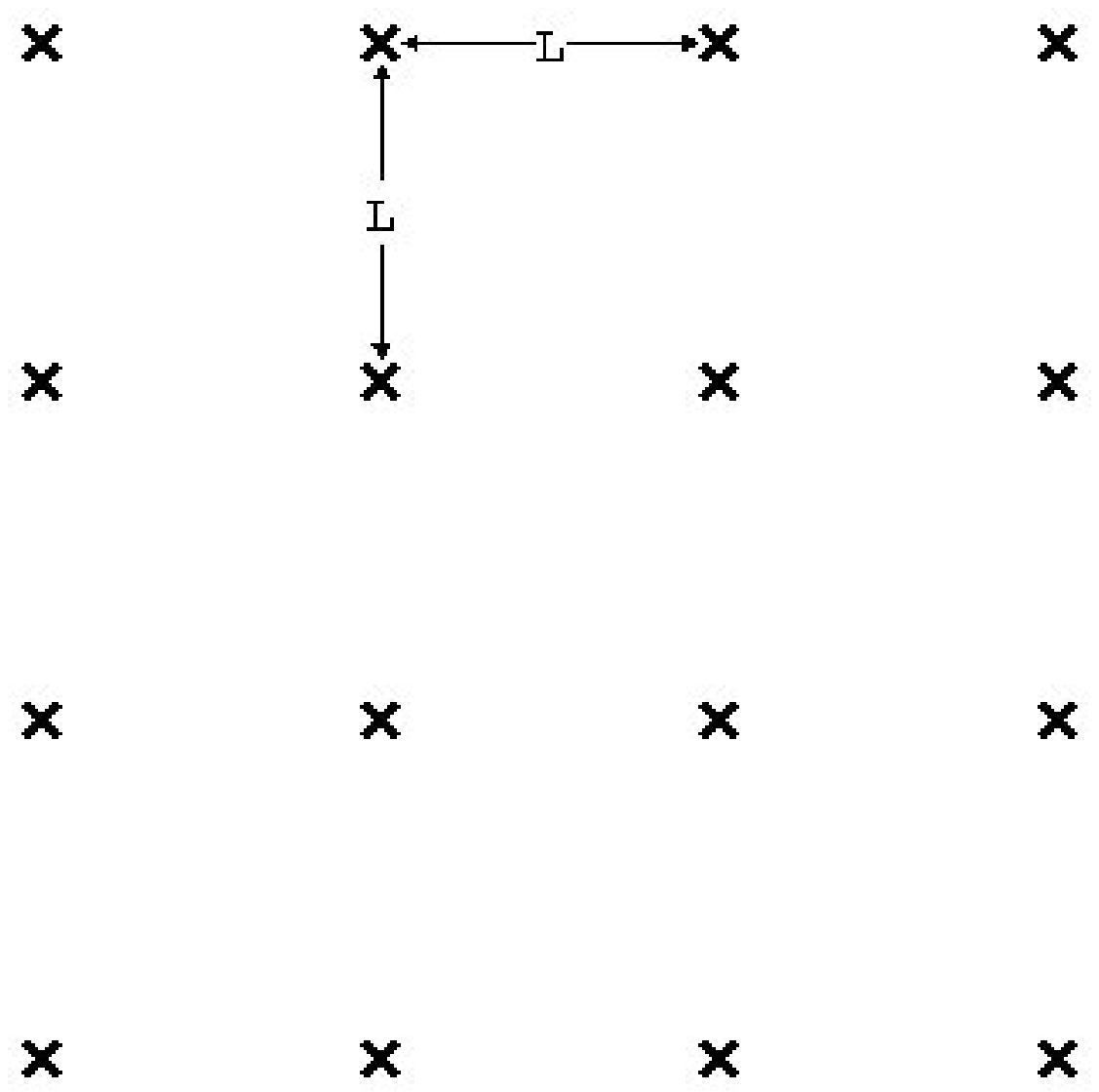}}
    \caption[The vortices are located on the sites of a square lattice]
    {The vortices are located on the sites of a square lattice.}
    \label{lattice}
  \end{minipage}
  \hfill
  \begin{minipage}[t]{\dimen0}
    \resizebox {\columnwidth}{!}{ \includegraphics{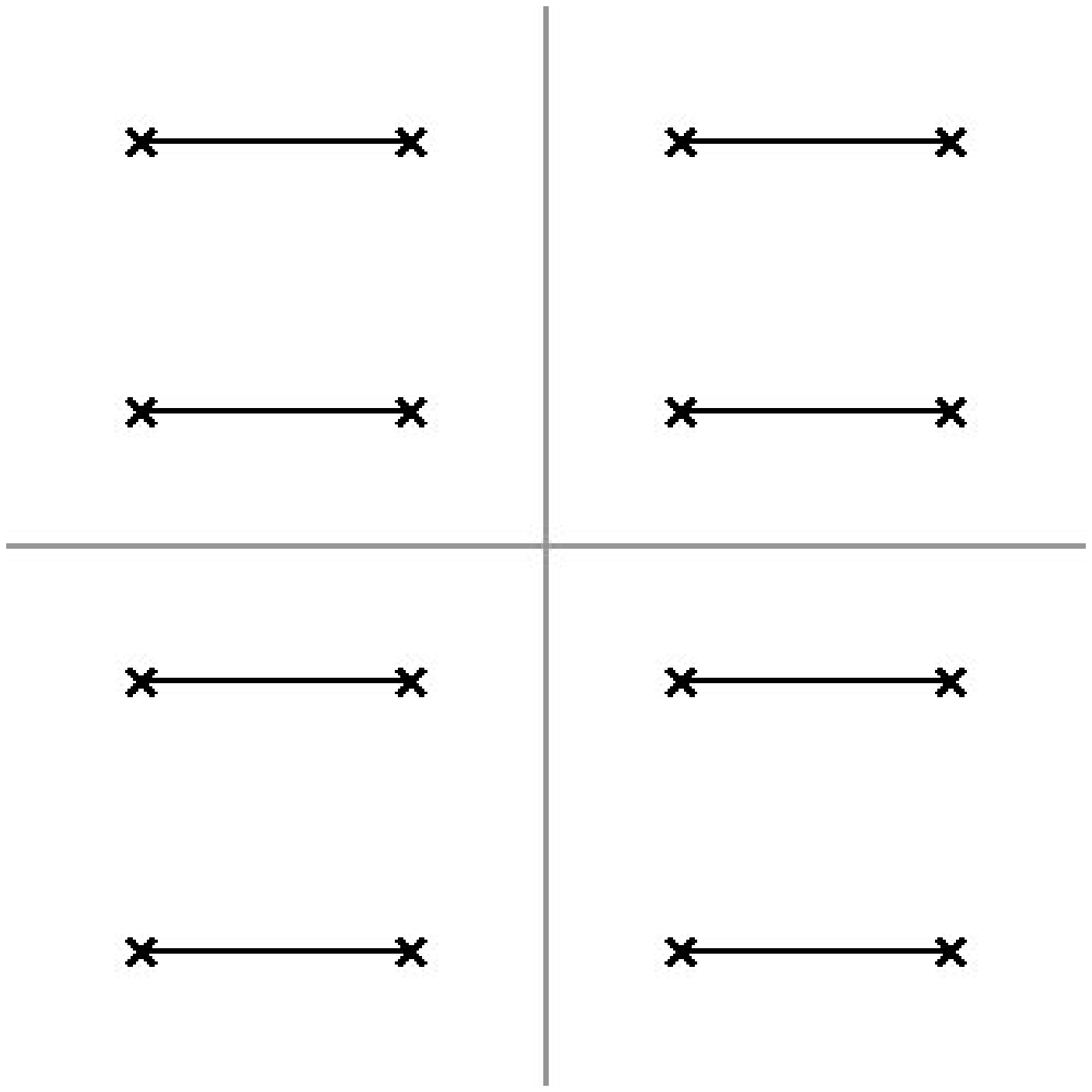} }
    \caption[Pairing of the vortices and fundamental lattice cell]
    {The vortices are paired and connected by segments on which the
    wavefunction has to vanish in order to satisfy the topological conditions.
    Grey lines indicate the real periodicity of the lattice and identify the
    fundamental region over which we will work.}
    \label{latticeline}
  \end{minipage}
\end{figure}

The construction of the solution on a general distribution of
fluxes is not easily attainable (as argued in \cite{nambu}). We do
not need to confront these complications in order to show our
point and so we shall simplify the problem by taking the vortices
as located on the vertices of a square lattice of lattice spacing
$L$ (Fig. \ref{lattice}), a case for which we shall be able to
give an explicit solution to the problem.

Inspired by a recent construction \cite{stodolsky}, we try to give
an estimate of the minimal energy required for a charged particle
to exist in the medium, and also to calculate the decay factor of
particles with zero energy in the lattice.

Before we construct the solution, it may be helpful to spend a few
more words on our boundary conditions. Since we can take the
wavefunction to be real, we translated its phase shift around each
vortex with the condition that the solution has to vanish along
one line, but we have not specified this line. This line is not
the familiar cut in a complex plane (which is, of course, a gauge
choice). In fact, we have some freedom in the choice of the line
along which the wavefunction vanishes, but this is not a gauge
freedom in that it has a measurable effect. It would be better to
say that the position of this line is a freedom of choice for the
wavefunction. Therefore, in order to impose it as a boundary
condition, we have to make this choice appropriately for the
problem we want to study (this consideration will be important
when we will consider the penetration of a zero-energy solution
inside the medium).

Let us consider for a moment just a pair of vortices. If we choose
the line on which the wavefunction has to vanish as the ray
exiting one vortex and pointing in the direction of the other one,
we can see that the boundary conditions become that the function
has to vanish only along the segment connecting the two fluxes;
this is certainly a very convenient choice, compared to other
solutions which would require the wavefunction to vanish on two
semi-infinite lines and therefore to develop higher gradients.

\begin{figure}
  \dimen0=\textwidth
  \advance\dimen0 by -\columnsep
  \divide\dimen0 by 2
  \noindent\begin{minipage}[t]{\dimen0}
    \vspace{0mm}
    \resizebox {\columnwidth}{!}{ \includegraphics{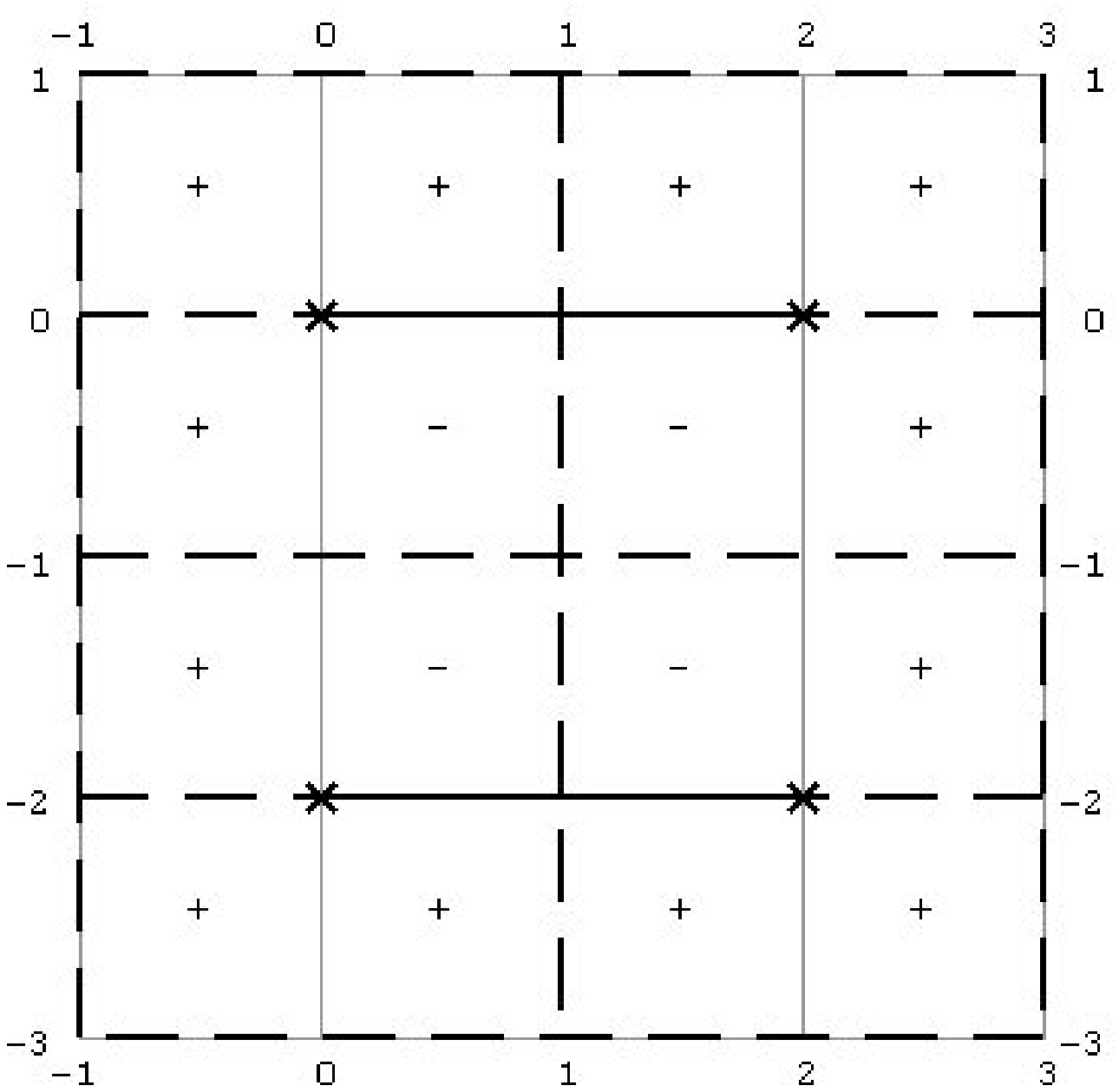} }
    \caption[Boundary conditions and parity of the wavefunction]
    {Boundary conditions and parity of the wavefunction: the black
    continuous lines represent Dirichlet boundary conditions, while
    the grey dashed lines indicate Neumann conditions.}
    \label{boundary}
  \end{minipage}
  \hfill
  \begin{minipage}[t]{\dimen0}
    \vspace{0mm}
    \resizebox {\columnwidth}{!}{ \includegraphics{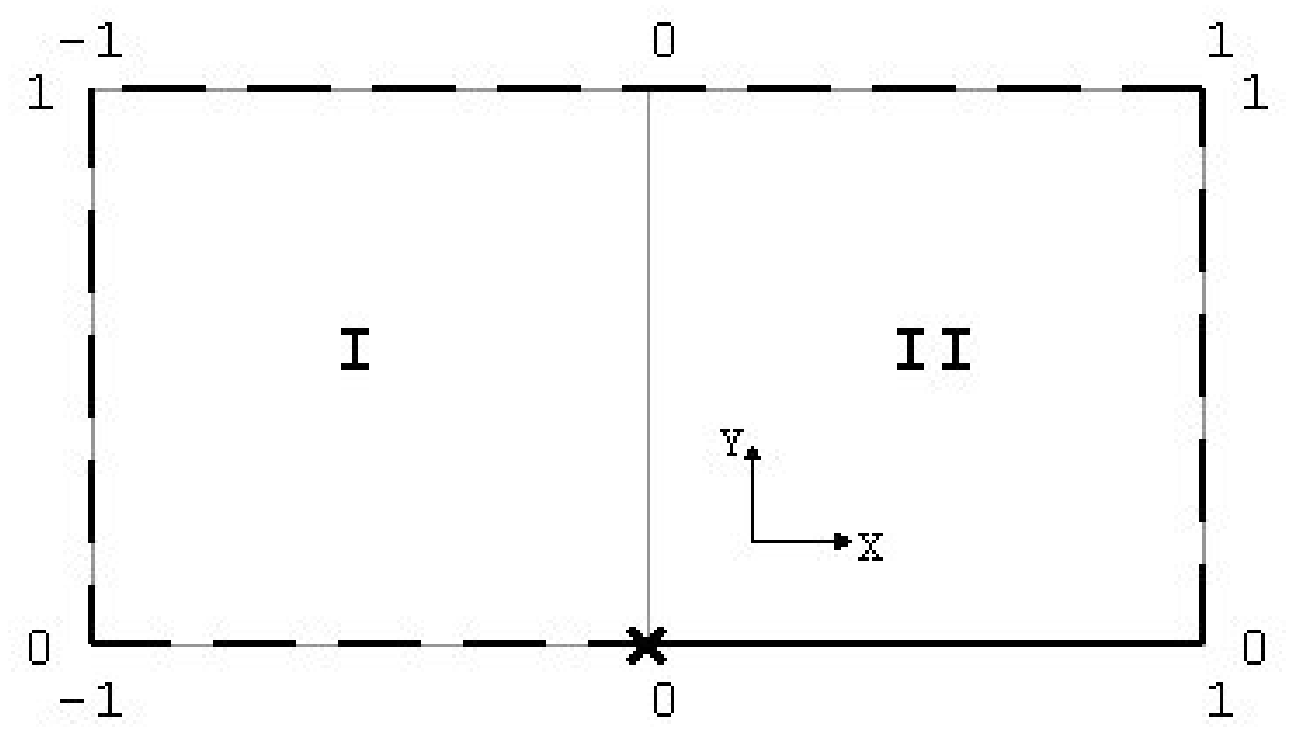} }
    \caption[Region over which we construct the fundamental solution and boundary conditions]
    {The region over which we construct the fundamental solution.
    The rest of the lattice can be covered starting from this basic tile.
    The continuous black line indicates where the wavefunction must vanish
    (Dirichlet condition) and the dashed ones where its derivative is zero
    (Neumann condition). We expand the solution on a basis in the region
    {\bf I} and on another basis in the region {\bf II} and we impose
    continuity of the function and derivative across the grey line.}
    \label{region1}
  \end{minipage}
\end{figure}

To construct the lowest energy solutions let us consider the
vortices in pairs, connecting nearest neighbors with line segments
along which the solution has to vanish. For definiteness, we
connect fluxes on the horizontal direction, requiring the
wavefunction to change sign when it crosses these segments (see
Fig. \ref{latticeline}).

Along these segments the wavefunction possesses odd parity. If we
are interested in the low energy modes, this means that along the
continuation of these segments, the function will be even and so
its derivative must vanish there. To conclude our analysis on the
boundary conditions, we notice that our system is clearly
periodic. To ensure periodicity of the wavefunction, we require
its derivative to vanish identically along the sides of each
square centered on a flux (see Fig. \ref{boundary}).

Bearing these considerations in mind, we now have to solve a
problem with mixed Dirichlet and Neumann boundary conditions. We
can further reduce the system under study and concentrate on two
of the quadrants around a flux site, because the rest of the
lattice can be covered by mirroring and flipping this unit (Fig.
\ref{region1} in reference to Fig. \ref{boundary}).

In summary, we now have to solve the problem of a free particle in
a rectangular box with sides of length $2$ and $1$ (in units of
half of a lattice spacing). We impose Neumann boundary conditions
everywhere, except on half of one of the long sides, where we
require the Dirichlet boundary condition.

\begin{figure}
 \resizebox {\columnwidth}{!}{ \rotatebox {270} {\includegraphics{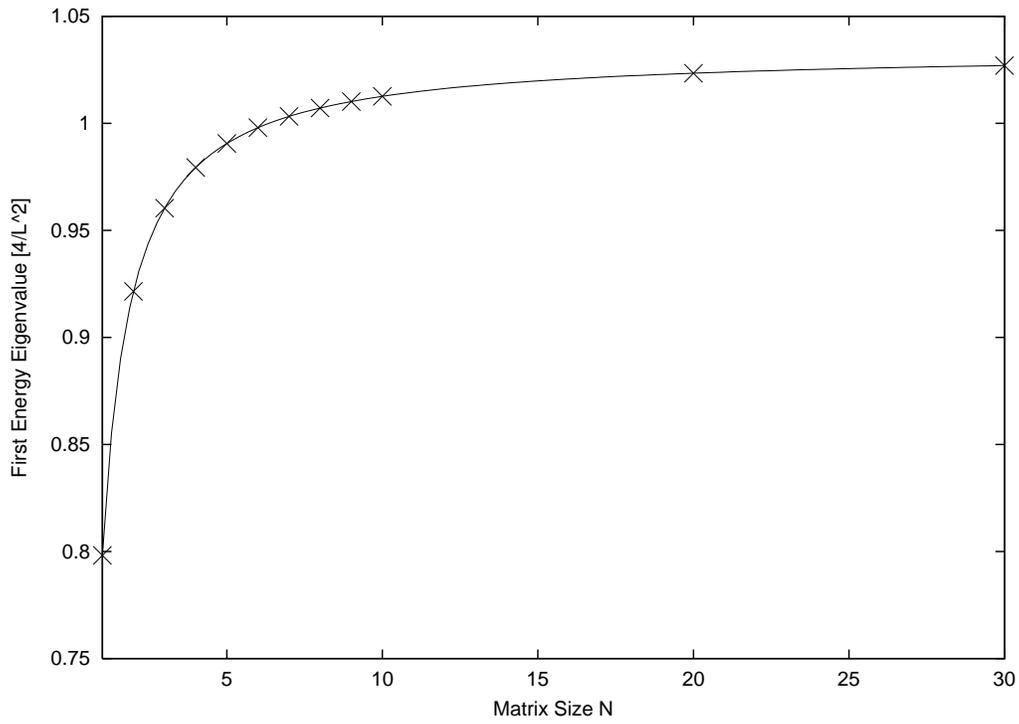} } }
 \caption[Plot of the Energy fit as a function of the matrix size $N$]
 {We truncate the infinite-dimensional matrix to a size $N$ and we
find the first energy eigenvalue $\varepsilon_0=2mE_0$
corresponding to this shorten system. This is the plot of $N$
versus $\varepsilon_0$ and its fit with a polynomial in inverse
powers of $N$ up to the third order (higher orders do not
contribute appreciably).}
  \label{energyfit}
\end{figure}

This is a non-standard problem; as we are not aware of any
previous study on a system with these boundary conditions, we
shall proceed in constructing the solution starting from a basis
compatible with the conditions. In region {\bf I} of Fig.
\ref{region1} we identify a convenient basis in the set
\be
   \left\{ \cosh \left[ k_n (1+x) \right] \cos (n \pi y)
   \right\}_{n=0}^{\infty} ,
   \nonumber
\ee while in region {\bf II} we expand the solution on
\be
   \left\{ \cosh \left[ K_n (1-x) \right] \sin \left[ (n+
   {1\over2})\pi y \right] \right\}_{n=0}^{\infty},
   \nonumber
\ee with the condition $n^2 \pi^2 - k_n^2 = (n+ {1\over2})^2\pi^2
- K_n^2 = 2mE$.

By matching the wavefunction and its derivative across the line
$x=0$, we may seek the values of $\varepsilon=2mE$ for which the
system admits a solution. In principle, this would involve the
calculation of the determinant of an infinite matrix. To obtain an
approximate solution, we truncated the system to a finite size,
and found the first energy eigenvalue $\varepsilon_0=2mE_0$ as a
function of the size of the matrix (see Fig. \ref{energyfit}).
Then, we plotted $\varepsilon_0$ versus the order $N$ of the
matrix and performed a fit with a polynomial in inverse powers of
$N$, taking the zeroth-order coefficient as the solution we would
have obtained by considering the whole infinite system.

In this way, we found the first energy eigenvalue for our solution
to be:
\be
 \varepsilon_0 = 2mE_0 = (1.0341 \pm 0.0002) \times {4 \over L^2},
\ee
that is
\be
 E_0 = (2.0682 \pm 0.0002) m^{-1} L^{-2}.
\ee

\begin{figure}
  \resizebox {\columnwidth}{!}{ \includegraphics{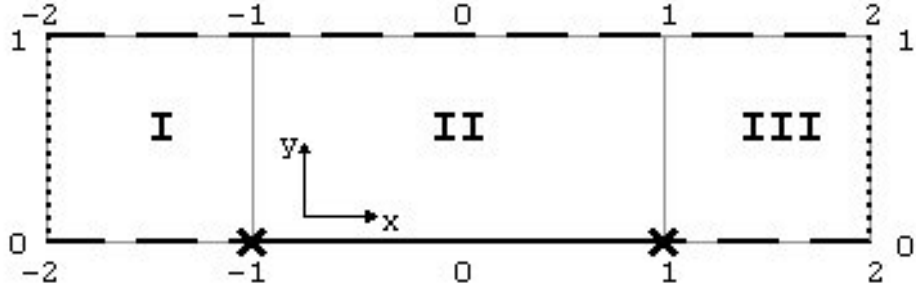} }
  \caption[Boundary conditions on the zero-energy decaying solution]
  {Decay of the zero-energy solution moving horizontally.
  We require periodicity on the vertical axis and exponential
  decay in the horizontal direction. The continuous black line
  indicates where the wavefunction must vanish (Dirichlet condition)
  and the dashed ones where its derivative is zero (Neumann condition).}
  \label{region2}
\end{figure}

Next, we are interested in estimating the decay factor of a
particle entering the medium with zero energy. This problem
depends on the direction in which the particle is traveling, in
that it is connected with the choice of the ray/segment over which
the solution has to vanish. Depending on the direction of motion,
the wavefunction may `choose' different configurations for these
segments.

We solve the problem for a particle moving along the $x$
direction. That is, we construct a solution which exhibits
periodic behavior in the $y$ direction and real decay in the $x$
(Fig. \ref{region2}). Again, we expand the wavefunction on
appropriate bases: in region {\bf I} and {\bf III} of Figure
\ref{region2} we use
\be
   \left\{ \eu^{n \pi x} \cos (n \pi y)
   \right\}_{n=0}^{\infty}
   \nonumber
\ee for right-moving and
\be
   \left\{ \eu^{-n \pi x} \cos (n \pi y)
   \right\}_{n=0}^{\infty}
   \nonumber
\ee for left-moving modes. In region {\bf II} we expand on
\be
   \left\{ \eu^{(n+{1\over2}) \pi x} \sin \left[ (n+ {1\over2})\pi y
   \right] \right\}_{n=0}^{\infty}
   \nonumber
\ee for right-moving and
\be
  \left\{ \eu^{-(n+{1\over2}) \pi x}
  \sin \left[ (n+ {1\over2})\pi y \right] \right\}_{n=0}^{\infty}
  \nonumber
\ee for left-moving modes.

\begin{figure}
  \resizebox {\columnwidth}{!}{ \rotatebox {270} {\includegraphics{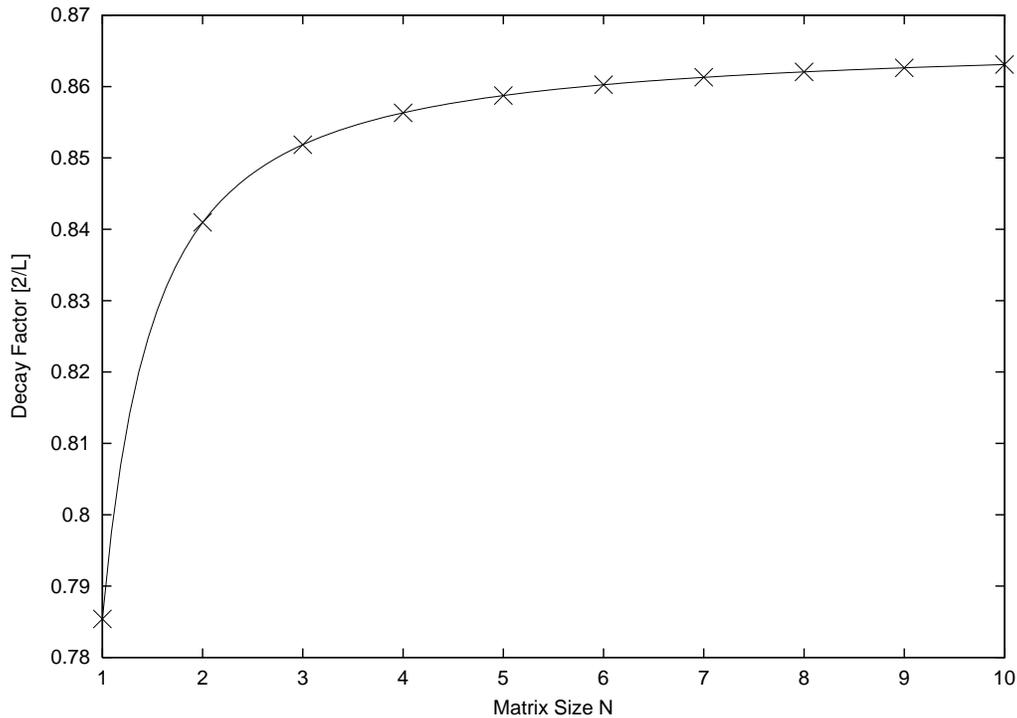} } }
   \caption[Plot of the decay factor fit as a function of the matrix size $N$]
   {We truncate the infinite-dimensional matrix to a size $N$
   and we find the lowest value for the decay factor $K$ corresponding
   to this finite system. This is the plot of $N$ versus $K$ and its
   fit with a polynomial in inverse powers of $N$ up to the third order
   (higher orders do not contribute appreciably).}
  \label{kappafit}
\end{figure}

We impose matching of the wavefunction and its derivative across
the line $x=-1$ and $x=1$ and we write the damping of the solution
by requiring an exponential suppression:
\be
 \psi (x=-2,y) = \eu^{4K} \psi (x=2,y) \qquad
 {\de \psi \over \de x} (x=-2,y) = \eu^{4K} {\de \psi \over \de x} (x=2,y).
\ee
We look for the values of $K$ for which the system admits solution.

As before, the system of equations is infinite-dimensional, so we
found the lowest value for $K$ as a function of the order $N$ of
the matrix and performed a fit with inverse powers of $N$ to
retain the zeroth order of the polynomial as the solution (see
Fig. \ref{kappafit}).

In this way, we find a decay factor for a particle moving along
the horizontal direction:
\be
 K = ( 0.88 \pm 0.01) \times {2 \over L} = ( 1.76 \pm 0.02 ) L^{-1} \ \ ,
\ee
and the same $K$ holds for a particle moving in the vertical
direction because we have the freedom to rotate the system by 90
degrees and rearrange the segments connecting the vortices in the
new direction.

\section{Conclusions}

Considering a lattice of point-like magnetic vortices, we showed
that the spectrum for a particle in such a medium is discrete and
that the lowest energy eigenvalue is greater than zero, by
explicitly constructing the wavefunction ($E_0 = (2.0682 \pm
0.0002) m^{-1} L^{-2}$).

This is quite in contrast with what was predicted by Y. Nambu in
\cite{nambu}. In this paper, the author argues that a solution of
the Schr\"odinger equation in our gauge would have to be either
holomorphic, or anti-holomorphic.

To see this, let us switch to complex coordinates to describe the plane.
The free particle equation now reads:
\be
 \partial_z \partial_{\bar{z}} \psi = E \psi
 \label{N}
\ee and therefore the solution for zero energy is either
analytical or anti-analytical. Nambu argues that, by continuity,
this property should persist at higher energies as well. In the
preceding section we constructed a non-zero energy solution which
clearly is neither holomorphic nor anti-holomorphic.

The analyticity or anti-analyticity of the solutions is an
important point of Nambu's construction that leads him eventually
to conclude that the states with lower angular momentum are not
admissible in the spectrum. This would imply that a particle
entering the medium with zero energy would undergo a suppression,
which is not merely exponential, but at least Gaussian. For that
reason, we argue that our approximation comes closer to the true
behavior, because by allowing more penetration it reduces
uncertainty-principle energy. This statement applies even for zero
energy, where Nambu's argument appears rigorous at first sight
from (\ref{N}). The loophole, we believe, is that in any case the
wavefunction is not analytic at the location of a vortex, and
therefore the factorization of the Laplace operator fails at that
set of points.

We computed the decay factor for a particle moving along one of
the lattice directions to be $K = ( 1.76 \pm 0.02 ) L^{-1}$, and
showed that this decay is purely exponential. The magnitude of
this suppression depends on the direction of travel. To compute
the decay factor in other directions it would be necessary to
modify ad hoc the boundary condition (the positioning of the ray
where the wavefunction vanishes). The condition we worked with is
the one that minimizes the extension of such rays and therefore
poses the minimal constraint on the solution. Any other choice
would have a greater impact on the shape of the wavefunction, and
this suggests it would shorten the decay length. The directional
dependence is easy to understand, because the coupling between
charge and vortex is strong, so that the lattice length scale and
the decay length are comparable: in the limit of vanishing lattice
constant the decay length also vanishes. A quantitative analysis
for generic directions would require a different formalism from
the one implemented here.

One can of course estimate the lowest allowed energy level with a
mean field approximation. A long wavelength particle ($k \simeq
0$) would see a uniform magnetic field
\be
   B = {\Phi \over L^2 } = {\pi \hbar \over e L^2}
\ee and it would occupy the lowest Landau level
\be
   E_0^{\it naive} = \hbar \omega_c {1 \over 2} =
   {\hbar^2 \over m c } \; {\pi \over 2 L^2} .
\ee This na\"ive calculation disagrees somewhat with the value we
found for the lowest energy level and indicates that is probably
possible to construct a better wavefunction:
\be
   {E_0 - E_0^{\it naive} \over E_0 } = 24 \%
\ee

Nonetheless, we believe that the order of magnitude of the effect
has been established, in that the energy eigenvalue and the decay
rate $K$ agree in this respect. This agreement implies that the
topological constraints imposed by the configuration of vortices
act as an effective repulsive potential of order unity (taking the
mass $m=1$), and that this potential is direction-dependent. This
effective potential could be used to trap a particle in a region,
just by surrounding that region with a medium of localized fluxes.
This may be a new form of trapping.

It would be interesting to investigate the more physical situation
in which the vortices are not forced to occupy the sites of a
square lattice, instead being randomly distributed, as in
\cite{randomvortices} and maybe dynamical objects themselves, but
this would require more powerful mathematical tools. It seems
plausible to us that the order of magnitude of the decay length
and the qualitative characteristics of the problem would not be
very different from the ones found with our model. Our reason for
saying this is that one could replace the random vortex
distribution with a random distribution of short line segments on
which the wavefunction vanishes, and this array surely would be
equivalent to a repulsive potential of characteristic magnitude,
leading to exponential, not Gaussian decay.


\chapter{Conclusions}
\label{TConclusions}
During my Ph.D. studies, we have devoted considerable efforts to
the study of the correlator known as {\it ``Emptiness Formation
Probability''} (EFP). This correlator measures the probability
that a one-dimensional system spontaneously develops a region of
space depleted of particles.

This correlator is a $n$-point correlator, where $n$ is the number
of particles that one has to move to empty the region of space, so
its complexity grows with $n$. Despite this apparent complexity,
the EFP was introduced as a very natural correlation function in
the development of a determinant representation for correlators of
integrable models using the Bethe Ansatz technique.

The calculation of correlation functions for exactly solvable
models is an arduous task and while some exact expressions have
been derived in terms of operator-valued determinants, they are
often of very limited practical use. The EFP has the simplest
expression in this representation and is therefore believed to be
the `fundamental' correlator for integrable theories.

In Chapter \ref{EFPinXY} we calculated the asymptotic behavior for
$n \to \infty$ of the EFP in the anisotropic spin 1/2 XY model. We
showed that for this system the EFP has an exact expression in
terms of the determinant of a complex matrix belonging to a family
of matrices known as {\it ``Toeplitz Matrices''}. We used some
results known in the theory of Toeplitz determinants to calculate
the asymptotic behavior of the EFP throughout the rich phase
diagram of the XY model. We found that the EFP decays
exponentially almost everywhere, except on the isotropic line,
where the decay is Gaussian. We also identified a power-law
prefactor with universal exponents on the critical lines of the
phase-diagram. Using a bosonization approach, We interpreted the
crossover from Gaussian to exponential behavior as an intermediate
asymptotics effect.

Beside its importance in the theory of integrable models, the EFP
is an interesting correlation function in that it measures a
substantial deviation from the equilibrium state of the system
that requires contributions even far from the Fermi points. For
this reason, the EFP is a perfect test field for a hydrodynamic
description of integrable theory.

Very often, calculations of correlators in one-dimensional
integrable systems are performed using the bosonization technique.
While being very powerful, the bosonization approach has a strong
limitation in the assumption of a linear spectrum. Because of
this, calculations done within this approximation are valid only
as long as only excitations close to the Fermi points are
involved. This assumption is clearly violated by the EFP.

We introduced the hydrodynamic description of integrable systems
and we showed its ability to correctly calculate the leading
asymptotic behavior of the EFP for a couple of Galilean Invariant
systems (Free Fermions and Calogero-Sutherland particles), as
firstly derived in \cite{abanovHydro}.

In \cite{abanovHydro}, the hydrodynamic description for free
fermions on a lattice was also considered and its EFP was
calculated to leading order. We have not included this project in
this thesis, but the problem of applying the hydrodynamic approach
to other lattice systems is definitely at hand. One should be able
to derive the EFP of the XY model and of the XXZ model from a
hydrodynamic point of view.

In Chapter \ref{Spin-ChargeHydro} we addressed the prediction of
exact spin-charge separation coming from the Luttinger Liquid
model. We argued that this result is a feature of the linear
spectrum approximation and that any curvature of the spectrum
would lead to a coupling between charge and spin degrees of
freedom. We also explained some difficulties encountered by the
bosonization method to account for these corrections.

A hydrodynamic description naturally takes into account the
interactions between spin and charge degrees of freedom. We
sketched the derivation of the hydrodynamic description of
fermions with contact repulsion. The hydrodynamic description is
given in an implicit form, leaving the analysis of the coupling
between the spin and charge degrees of freedom for further work.
Another interesting model to study in this respect is the spin
Calogero-Sutherland. Fermions with contact repulsion and this
latter model represent the two integrable extremes of short and
long range interaction, respectively, and therefore it is
important to understand how spin and charge couple in both
systems.

Another open question is the origin of the double infinite series
of conserved quantities for gradient-less, Galilean invariant,
hydrodynamic systems, as discussed in Appendix
\ref{Integrability}. The existence of these conserved currents was
discovered in the 1980's and we believe to be the first to
actually construct them explicitly. It is still unclear what
underlying symmetry is responsible for these conservation laws.

Finally, in Chapter \ref{ManyVortices} we looked at a
two-dimensional system in which a scalar particle moves into a
medium filled with point-like magnetic vortices pinned at the
sites of a square lattice. We studied the Aharonov-Bohm problem
connected with such a configuration when the magnetic fluxes are
all equal to a half of the quantum flux unit. We explicitly
constructed a wavefunction for this system, showing that the
spectrum is discrete, and considered the decay of a zero-energy
particle within this medium and showed that this decay is
exponential. We proposed that this medium acts effectively as a
repulsive potential of topological nature (since no real field is
present outside the vortices) and that one could use such a
configuration to confine (trap) electrons. This is a novel idea
and it would be interesting to extend this analysis to electrons
with spin, since they would couple to the vortices differently
depending on their spins.


\appendix

\chapter{Exact results for EFP in some integrable models}
\label{KnownRes}
In this appendix we are going to recapitulate some known results
obtained for the Emptiness Formation Probability $P(R)$ in various
integrable one-dimensional systems as they were reported in
\cite{abanovHydro}. We present the results as
\be
   S \equiv -\ln P(R) .
\ee for brevity, but also to facilitate the comparison with the
instanton action $S_{opt}$ we are going to introduce in the main
text. In this appendix we always set $m = \hbar =1$.

\section{Free continuous fermions}

For Free Fermions we can use the fact that the ground state
wavefunction (more precisely $|\Psi|^{2}$) coincides with the
joint eigenvalue distribution of the unitary random ensemble. For
the latter the probability of having no eigenvalues in the range
$2R$ of the spectrum was obtained in \cite{dysmehta} (see also
\cite{mehta}). The first few terms of the expansion in $1/R$ are
\be
    S = {1 \over 2} \, s^{2} + {1 \over 4} \ln s
    - \left( {1 \over 12} \ln 2 + 3 \zeta'(-1) \right) + \Ord (s^{-2}),
   \label{FreeFermionsExact}
\ee where we introduced the parameter
\be
  s \equiv \pi \rho_0 R.
  \label{s1}
\ee

\section{Calogero-Sutherland model}

The Calogero-Sutherland model \cite{Calogero-1969,
Sutherland-1971} with $N$-particles is defined as
\be
   {\cal H} = {1 \over 2}\sum_{j=1}^{N} p_j^2
   + {1 \over 2} \sum_{j,k=1; j\neq k}^{N}
   {\lambda(\lambda-1) \over (x_j-x_k)^2},
   \label{1}
\ee where $p_j=- \ii \partial/\partial x_j$ is the momentum
operator of $j$-th particle and $\lambda$ is a dimensionless
coupling constant (we will come back to the Calogero-Sutherland
model in section \ref{CSEFP}). For $\lambda=1$ we have free
fermions, at $\lambda=0$ free bosons, while in the case of general
$\lambda$ the model (\ref{1}) describes interacting particles with
fractional statistics.

The ground state wavefunction is
\be
  \Psi_G = \prod_{j<k}(x_j-x_k )^{\lambda} .
\ee Using the form of the ground state wavefunction and
thermodynamic arguments \cite{mehta} one obtains
\be
   S = {\lambda \over 2} \, (\pi \rho_0 R)^2  + (1-\lambda) \pi \rho_0 R + \Ord (\ln R).
   \label{CalogeroExact}
\ee or, by defining
\be
   s \equiv \sqrt{\lambda} \pi \rho_0 R
   \label{slambda}
\ee and
\be
   \alpha_0 \equiv {\lambda^{1/2}-\lambda^{-1/2} \over 2},
   \label{alpha0}
\ee we have
\be
   S = {1 \over 2} \, s^2 - 2 \alpha_0 s + \Ord (\ln s).
   \label{CalogeroExacta}
\ee

The notation $\alpha_0$ originates from conformal field theory,
since the theory with central charge $c=1-24\alpha_0^2$ is known
to be related to the Calogero-Sutherland model \cite{Awata}. We
are not aware of a determination of the coefficient in front of
the $\ln s$ term in the expansion for general $\lambda$. However,
$\lambda =1/2, 1, 2$ correspond to random matrix ensembles where
the full asymptotic expansion is known (see below). In those cases
the coefficient of $\ln s$ is $1/8, 1/4, 1/8$ respectively. The
natural guess for an interpolation is \cite{abanovHydro} \be
    S = {1 \over 2} \, s^2 - 2 \alpha_0 \, s
    +\left( {1 \over 4} - \alpha_0^2 \right) \ln s + \Ord(1).
    \label{CSguess}
\ee

\section{Random matrices}

For Random Matrix ensembles with $\beta = 2 \lambda =1,2,4$ the joint
eigenvalue distribution is proportional to
$\prod_{i<j}|z_{i}-z_{j}|^{\beta}$. The full asymptotic expansion
of the quantity $E_{\beta}(0,2R)$ corresponding to the EFP $P(R)$
was obtained using properties of Toeplitz determinants
\cite{dysmehta,mehta}. The first few terms of these expansions are
given by
\bea
    S_{\lambda=1/2} & = & {1 \over 4} \, s^2  + {1 \over 2} \, s
    + {1 \over 8} \ln s - {7 \over 24}\ln 2 - {3 \over 2} \zeta'(-1) + \Ord(s^{-1}),
    \nonumber \\
    S_{\lambda=1} & = & {1 \over 2} \, s^2
    +{1 \over 4} \ln s - {1 \over 12} \ln 2 - 3\zeta'(-1) +\Ord (s^{-2}),
    \nonumber \\
    S_{\lambda=2} & = & s^2 - s
    + {1 \over 8} \ln s + {4 \over 3} \ln 2 - {3 \over 2} \zeta'(-1) + \Ord(s^{-1})
    \label{rmexact}
\eea where
\be
  s \equiv \pi \rho_0 R.
\ee

Here we used $\lambda=\beta/2=1/2,1,2$ instead of $\beta$.
Redefining $s$ as in (\ref{slambda}) and using (\ref{alpha0}), we
can summarize the first three terms of (\ref{rmexact}) in a
compact form in (\ref{CSguess}).

\section{Free fermions on the lattice}

For non-interacting one-dimensional fermions on the lattice (and
the corresponding isotropic XY spin chain) the asymptotic behavior
of EFP was derived in \cite{shiroishi} using Widom's theorem on
the asymptotic behavior of Toeplitz determinants. Introducing the
Fermi momentum $k_{F}=\pi\rho_{0}$ and using units in which the
lattice spacing is $1$ we have
\be
    S = - 4 R^2 \ln\cos {k_F \over 2}
    + {1 \over 4} \ln \left[ 2 R \sin {k_F \over 2} \right]
    - \left( {1 \over 12} \ln 2 + 3\zeta'(-1) \right) + \Ord (R^{-2}).
    \label{exlf}
\ee In the continuous limit $k_{F}\to 0$ the (\ref{exlf}) goes to
its continuous version (\ref{FreeFermionsExact}).

\section{Bosons with delta repulsion}

The model of bosons with short range repulsion is described by
\be
   {\cal H} = {1 \over 2} \sum_{j=1}^{N} p_j^2
   +g \sum_{1\leq j< k\leq N} \delta (x_j-x_k),
   \label{2}
\ee where $g$ is the coupling constant. This model is integrable
by Bethe Ansatz \cite{LiebLiniger-1963}. It was derived
(conjectured) in \cite{ItsKorepinWaldron-1995} that the leading
term of the EFP is
\be
   S = {1 \over 2}(K R)^{2} \left[1+I(g/K)\right],
   \label{IKW}
\ee where $K$ is the Fermi momentum in the Lieb-Liniger solution
\cite{LiebLiniger-1963} and
\be
    I(x) = {1 \over 2 \pi^2} \int_{-1}^{1} {y \, \de y \over \sqrt{1-y^2}} \,
    \int_{-1}^{1} {z \, \de z \over \sqrt{1-z^2}} \,
    \log \left( {x^2 + (y+z)^2 \over x^2 + (y-z)^2} \right).
\ee

The limit $I(x\to \infty) = 0$ corresponds to the free fermion
result (\ref{FreeFermionsExact}) (Tonks-Girardeau gas), while the
limit $I(x\to 0) = 1$ is the result for the EFP of free bosons.

\section{XXZ model}

The Hamiltonian of the XXZ model is given by
\be
   H = J \sum_j \left[ \sigma_j^x \sigma_{j+1}^x
   + \sigma_j^y \sigma_{j+1}^y
   + \Delta \sigma_j^z \sigma_{j+1}^z \right],
\ee where the sum is taken over the sites $j$ of a one-dimensional
lattice and $\sigma^{x,y,z}$ are Pauli matrices.

Let us parametrize the anisotropy as
\be
   \Delta = \cos \pi \nu.
\ee The asymptotic behavior of the EFP as $n=2R\to \infty$ was
found in \cite{KMST_gen-2002,KLNS-2002}
\be
    P(n) \sim  A n^{-\gamma}C^{-n^{2}},
\ee where
\be
    C \equiv {\Gamma^2 (1/4) \over \pi \sqrt{2\pi}}
    \exp \left\{ -\int_0^\infty {\de t \over t} \,
    {\sinh^2 (t\nu) \eu^{-t} \over \cosh (2t\nu) \sinh (t)}\right\}
\ee and the exponent $\gamma$ was conjectured in \cite{KLNS-2002}
to be
\be
    \gamma = {1 \over 12} + {\nu^2 \over 3(1-\nu)}.
\ee

\chapter{Asymptotic behavior of Toeplitz Determinants}
\label{ToeplitzApp}
As we showed in Chapter \ref{EFPinXY}, the asymptotic behavior of
the EFP for (\ref{spinham}) at $n \to \infty$ is exactly related
to the asymptotic behavior of the determinant of the corresponding
Toeplitz matrix (\ref{Tg},\ref{genfunc},\ref{PnX}) and can be
extracted from known theorems and conjectures in the theory of
Toeplitz matrices. These types of calculations have been done
first in \cite{LSM-1961,mccoy} for spin-spin correlation
functions. It is well known that the asymptotic behavior of the
determinant of a Toeplitz matrix as the size of the matrix goes to
infinity strongly depends upon the zeros and singularities of the
generating function of the matrix.

A very good report on the subject has been recently compiled by T.
Ehrhardt \cite{Ehrhardt-2001}. Here we want to recapitulate what is known
about the determinant
\be
   D_n [\sigma] = \det ({\bf S_n}) =
   \det \left| s(j-k) \right|_{j,k=0}^n
\ee
of a $n+1 \times n+1$ Toeplitz matrix
\be
   {\bf S_n} = \pmatrix{  s(0) & s(-1) & s(-2) & \ldots & s(-n) \cr
                         s(1) & s(0)  & s(-1) & \ldots & s(1-N) \cr
                         s(2) & s(1) & s(0) & \ldots & s(2-N) \cr
                         \vdots & \vdots & \vdots & \ddots & \vdots \cr
                         s(n) & s(n-1) & s(n-2) &\ldots & s(0) \cr
                     }
\ee with entries generated by a function $\sigma(q)$:
\be
   s(l) \equiv \int_{-\pi}^{\pi} \sigma(q)
   \eu^{- \ii l q} {\de q \over 2 \pi} \: ,
\ee where the generating function $\sigma(q)$ is a periodic
(complex) function, i.e. $\sigma(q) = \sigma(2 \pi + q)$.

In this work we dealt only with generating functions with zero
winding number \be
   {\rm Ind } \, \sigma(q) \equiv
   \int_{-\pi}^{\pi} {\de q \over 2 \pi} \, {\de \over \de q}
   \log \sigma (q) = 0
\ee
and this brief review will be limited to this condition.
This was not the case in the study of Barouch et al. \cite{mccoy},
where the generating function (see footnote after (\ref{PnX}))
had non-zero winding number in some regions of the phase
diagram.

\section{The Strong Szeg\"o Theorem}

If $\sigma(q)$ is sufficiently smooth, \underline{non-zero} and
satisfies ${\rm Ind } \,\sigma(q) = 0$ (i.e., the winding number is $0$),
we can apply what is known as the {\it Strong Szeg\"o Limit Theorem}
(\cite{Hirschman}, \cite{mccoywu}), which states that the determinant
has a simple exponential asymptotic form
\be
   D_n [\sigma] \sim E[\sigma] G[\sigma]^n
   \qquad n \rightarrow \infty,
   \label{szego}
\ee
where $G[\sigma]$ and $E[\sigma]$ are defined by
\be
  G[\sigma] \equiv \exp{\hat{\sigma}_0}, \qquad
  E[\sigma] \equiv
  \exp{\sum_{k=1}^\infty k \hat{\sigma}_k \hat{\sigma}_{-k}}
  \label{szegoexp}
\ee
and $\hat{\sigma}_k$ are the Fourier coefficients of the expansion of
the logarithm of $\sigma(q)$:
\be
  \log \sigma(q) \equiv
  \sum_{k=-\infty}^\infty \hat{\sigma}_k \eu^{\ii k q}.
  \label{sigmak}
\ee

\section{The Fisher-Hartwig Conjecture}

Over the years, the Szeg\"o Theorem has been extended to consider
broader classes of generating functions by relaxing the continuity
conditions which define a ``smooth function'', but it remained limited to
never-vanishing functions.
Therefore, some extensions have been proposed to the Szeg\"o Theorem in
order to relax this latter hypothesis.
When the generating function has only pointwise singularities (or zeros), there
exists a conjecture known as the Fisher-Hartwig Conjecture (FH)
\cite{FisherHartwig-1968}.
\footnote{This conjecture is still not completely proven.
For details and status of the conjecture see Ref. \cite{basor}.}

When $\sigma(q)$ has $R$ singularities at $q= \theta_r$ ($r=1..R$), we
decompose it as follows:
\be
   \sigma(q) = \tau(q) \prod_{r=1}^R
   \eu^{\ii \kappa_r [(q - \theta_r) \mod 2 \pi - \pi]}
   \left( 2 - 2 \cos (q - \theta_r) \right)^{\lambda_r}
   \label{fishdec}
\ee
so that $\tau(q)$ is a smooth function satisfying the conditions
stated in the previous section.
Then according to FH the asymptotic formula for the determinant takes
the form
\be
   D_n [\sigma] \sim E \left[ \tau, \{ \kappa_a \}, \{ \lambda_a \}, \{ \theta_a \} \right]
   \; n^{\sum_r \left( \lambda_r^2 - \kappa_r^2 \right)}
   G[\tau]^n \qquad n \rightarrow \infty,
   \label{singexp}
\ee
where the constant prefactor is conjectured to be
\bea
  E \left[ \tau, \{ \kappa_a \}, \{ \lambda_a \}, \{ \theta_a \} \right]
  \equiv & E[\tau] & \prod_{r=1}^R
  \tau_- \left( \eu^{\ii \theta_r} \right)^{-\kappa_r - \lambda_r}
  \tau_+ \left( \eu^{- \ii \theta_r} \right)^{\kappa_r - \lambda_r}
  \nonumber \\  & \times &
  \prod_{1 \le r \ne s \le R} \left( 1 - \eu^{\ii (\theta_s - \theta_r)}
  \right)^{(\kappa_r + \lambda_r) (\kappa_s - \lambda_s)}
  \nonumber \\ & \times &
  \prod_{r=1}^R { {\rm G} (1 + \kappa_r + \lambda_r)
  {\rm G} (1 - \kappa_r + \lambda_r) \over {\rm G} (1 + 2 \lambda_r) }.
  \label{fisherhartwig}
\eea
$E[\tau]$ and $G[\tau]$ are defined as in (\ref{szegoexp}) and
$\tau_{\pm}$ are defined by decomposition
\be
   \tau(q) = \tau_- \left( \eu^{\ii q} \right)
   G[\tau] \tau_+ \left( \eu^{- \ii q} \right),
   \label{wienerhopf}
\ee
so that $\tau_+$ ($\tau_-$) are analytic and non-zero
inside (outside) the unit circle on which $\tau$ is defined and
satisfy the boundary conditions $\tau_+ (0) = \tau_- (\infty) = 1$.
${\rm G}$ is the {\it Barnes G-function}, an analytic entire
function defined as
\be
   {\rm G}(z + 1) \equiv (2 \pi)^{z/2} \eu^{-[z + (\gamma_E + 1) z^2]/2}
   \prod_{n=1}^\infty \left( 1 + {z \over n} \right)^k
   \eu^{-z + {z^2 \over 2n}},
   \label{BGfun}
\ee
where $\gamma_E \sim 0.57721 \ldots$ is the Euler-Mascheroni Constant.

This conjecture is actually proven  for some ranges of parameters
$\kappa_r$ and $\lambda_r$ or fully for the case of a single singularity
($R=1$), see \cite{widomsing,Ehrhardt-1997}.

In many simple cases it is possible to find  the
factorization of $\tau$ into the product of $\tau_+$ and $\tau_-$ by inspection.
More complicated examples like the ones presented in this work require
a special technique to obtain this factorization, which is known as the
{\it Wiener-Hopf decomposition}:
\bea
   \log \tau_+ (w) = \oint {\de z \over 2 \pi \ii}
   {\log \tau (z) \over z - w} & \qquad & |w| < 1,
   \nonumber \\
   \log \tau_- (w) = - \oint {\de z \over 2 \pi \ii}
   {\log \tau (z) \over z - w} & \qquad & |w| > 1,
   \label{wienint}
\eea
where the integral is taken over the unit circle.

In light of these formulas, it is useful to present the parametrization
(\ref{fishdec}) in a form which makes the analytical structure more
apparent.
Changing the variable dependence from $q$ to $z \equiv \eu^{\ii q}$, we
write
\be
   \sigma(z) = \tau(z) \prod_{r=1}^R
   \left( 1 - {z \over z_r} \right)^{\lambda_r + \kappa_r}
   \left( 1 - {z_r \over z_{} } \right)^{\lambda_r - \kappa_r},
\ee
where $z_r \equiv \eu^{\ii \theta_r}$.

\section{The Generalized Fisher-Hartwig Conjecture}
\label{gfhsec}

Despite the considerable success of the Fisher-Hartwig Conjecture, few
examples have been reported in the mathematical literature that do not
fit this result.
These examples share the characteristics that inequivalent
representations of the form (\ref{fishdec}) exist for the generating
function $\sigma(q)$.
Although no theorem has been proven concerning these cases, a
generalization of the Fisher-Hartwig Conjecture (gFH) has been suggested
by Basor and Tracy \cite{basor} that has no counter-examples yet.

If  more than one parametrization of the kind
(\ref{fishdec}) exists, we write them all as
\be
   \sigma(q) = \tau^i(q) \prod_{r=1}^R
   \eu^{\ii \kappa^i_r [(q - \theta_r) \mod 2 \pi - \pi]}
   \left( 2 - 2 \cos (q - \theta_r) \right)^{\lambda^i_r},
   \label{fishgendec}
\ee
where the index $i$ labels different parametrizations (for $R > 1$ there
can be only a countable number of different parametrizations
of this kind).
Then the asymptotic formula for the determinant is
\be
   D_n [\sigma] \sim \sum_{i \in \Upsilon}
   E \left[ \tau^i, \{ \kappa^i_a \}, \{ \lambda^i_a \}, \{ \theta_a \}
   \right] \; n^{\Omega(i)} G[\tau^i]^n \qquad n \rightarrow \infty,
   \label{singgenexp}
\ee
where
\bea
   \Omega(i) & \equiv &
   \sum_{r=1}^R \left( \left( \lambda^i_r \right)^2 -
   \left( \kappa^i_r \right)^2 \right),
 \\
   \Upsilon & = &
   \left\{ i \: \| \: \R[\Omega(i)] = \max_j \R[\Omega(j)] \right\}.
 \label{Leading}
\eea

The generalization essentially gives the asymptotics of the Toeplitz
determinant as a sum of (FH) asymptotics calculated separately for
different leading (see Eq.~(\ref{Leading})) representations
(\ref{fishgendec}).
In Sec.~\ref{ssOmegap} we used the sum of all (not necessarily leading)
representations and showed that it also correctly produces the first
subleading corrections to the asymptotics of our Toeplitz determinant.

\section{Widom's Theorem}

If $\sigma(q)$ is supported only in the interval $\alpha \le q
\le 2 \pi - \alpha$ as in our model for $\gamma=0$, singularities are
no longer pointwise and one should apply Widom's Theorem
\cite{widomsupp}.
It states that the asymptotic behavior of the determinant in this case
is
\be
   D_n [\sigma] \sim 2^{1/12} \eu^{3 \zeta'(-1)}
   \left( \sin {\alpha \over 2} \right)^{-1/4} E[\rho]^2 n^{-1/4} G[\rho]^n
   \left( \cos {\alpha \over 2} \right)^{n^2},
   \label{widomasympt}
\ee
where $E$ and $G$ are defined in (\ref{szegoexp}) and
\be
   \rho(q) = \sigma
   \left( 2 \cos^{-1} \left[\cos {\alpha \over 2} \cos q\right] \right)
\ee
with the convention $0 \le \cos^{-1} x \le \pi$.

For the case considered in Section \ref{gammazero}, the generating
function is constant, $E[\rho] = G[\rho] = 1$, and (\ref{widomasympt})
simplifies considerably giving
\be
   D_n [\sigma] \sim 2^{1/12} \eu^{3 \zeta'(-1)}
   \left( \sin {\alpha \over 2} \right)^{-1/4} n^{-1/4}
   \left( \cos {\alpha \over 2} \right)^{n^2}.
   \label{widomth}
\ee

\chapter{A brief introduction to the Bethe Ansatz}
\label{BetheInt}

The Bethe Ansatz technique is a very powerful tool to extract
information about an integrable system. Its main advantage is to
convert a quantum many body problem (the problem of solving a
system of coupled Schr\"odinger equations) into the solution of a
much simpler algebraic system. This solution specifies the energy
eigenfunctions of the system, although implicitly. Even more
importantly, it allows for the calculation of thermodynamic
quantities in the thermodynamic limit.

Integrable models are theories with a high degree of symmetry
which guarantee as many conserved quantities as the number of
degrees of freedom of the system. Most of the results obtained in
theoretical physics are derived as the consequence of some
approximation, since the whole theory is normally too complicated
to be solved directly. In one dimension all the systems are
strongly interacting because of the limited dimensionality and for
this reason is it hard to control perturbative calculations.

This is what makes integrable models so important in one
dimension. The Bethe Ansatz is one of the main tools in the theory
of quantum integrable models. It has a rich mathematical structure
and a wide range of applications. Clearly, we cannot recount its
full scope and cite the many important works in this subject in
the space we have. We refer the reader to the recent book by
Sutherland \cite{sutherland} for its clarity and for its ability
to introduce even the inexperienced reader to quite advanced
topics.

Here we summarize basic formulas of the Bethe Ansatz which are
needed in the hydrodynamic approach.

\section{The Bethe Wavefunction}

An integrable model is an exactly solvable theory. Beside this
intuitive definition, it is quite complicated to identify and
formalize the exact meaning of the notion of integrability. We are
not going to pursue the problem of identifying the integrability
of a model and we will assume that somehow we know we are dealing
with such a theory.

Let us consider a one-dimensional system of $N$ identical
particles interacting with a pair potential. The Hamiltonian for
the quantum problem can be written as:
\be
   H = -{\hbar^2 \over 2 m} \sum_{j=1}^N {\partial^2 \over \partial x_j^2}
   + \sum_{j>k=1}^N U \left( x_j - x_k \right) .
   \label{betheham}
\ee We assume that the potential is such that the model is
integrable and we will not specify it further at the moment.

The first ingredient in the Bethe Ansatz is to assume a plane
waves superposition form for the wavefunctions of the system:
\be
   \Psi (\{x_i\}) = \sum_P A (P) \eu^{\sum_i P_{k_i} x_i}
   \label{bethewavefunc}
\ee where the sum is carried over all the permutations $P$ of the
momenta of each particle in the system. This formula is valid when
the particles are in order, i.e. $x_i<x_{i+1}$ for every $i$, and
allows for an easy extension to the other configurations by
employing the symmetry or antisymmetry properties of the
wavefunction.

One of the fundamental conditions for a model to be integrable is
that there are no two-particle irreducible interactions, i.e. all
processes in the system can be viewed as a sequence of two
particle scattering, and the order of the scatterings does not
matter. This latter condition is formally expressed as the
Yang-Baxter equation.

The important quantity in a two-particle scattering is the phase
acquired by the particles after the interaction. In the Bethe
Ansatz construction the {\it ``scattering phase''} $\theta_\pm
(k)$ plays a very important role. We can define a different phase
for bosons $\theta_+ (k)$ and for fermions $\theta_- (k)$ to take
in account the different statistics under a particle exchange. One
can prove that $\theta_\pm (k)$ is a real odd function of $k$
($\theta_\pm^* (k)=\theta_\pm (k^*)$, $\theta_\pm (-k) = -
\theta_\pm (k)$), where $k=k_1 - k_2$ is the difference between
the momenta of the two interacting particles.

If the particles were distinguishable, we would describe the
scattering process through the transmission and reflection
amplitudes. These can be expressed in terms of the scattering
phase as
\bea
   T (k) & \equiv & {\eu^{-\ii \theta_- (k)} - \eu^{-\ii \theta_+( k)} \over 2} ,
   \nonumber \\
   R (k) & \equiv & - {\eu^{-\ii \theta_- (k)} + \eu^{-\ii \theta_+ (k)} \over 2} .
   \label{TRAmp}
\eea

It is straightforward to solve the one-dimensional Schr\"odinger
problem for a given potential and derive from it the scattering
phase. We skip these derivations here (see, for instance,
\cite{sutherland}) and just list the results for a couple of
integrable potential we will use in our hydrodynamic approach.

First, let us consider a contact interaction:
\be
   U(r) = c \delta(r).
\ee For fermions, the antisymmetry of the wavefunction prevents
the particles from interacting, so
\be
   \theta_- (k) = 0.
\ee For Bosons the solution of the scattering problem brings
\be
   \theta_+ (k) = -2 \arctan (k/c) .
\ee If the particles are distinguishable, one can then compute the
transmission and reflection amplitudes from (\ref{TRAmp}) as
\be
   T(k) = {k \over k + \ii c} \qquad
   R(k) = - {\ii c \over k + \ii c}.
\ee

Let us consider {\it ``Calogero-Sutherland''} particles, i.e.
particles interacting with the potential:
\be
   U(r) = {\lambda (\lambda -1) \over r^2} ,
   \label{CSPot}
\ee where $\lambda$ quantifies the interaction strength and the
statistics of the particles. In fact, the scattering phase is
\be
   \theta (k) = \pi (\lambda-1) \sgn (k).
\ee One sees that $\lambda =1$ corresponds to a free Hamiltonian
with antisymmetric wavefunction and therefore describes free
fermions, while for $\lambda = 0$ the potential vanishes again but
the wavefunction is symmetric and it describes free bosons. In
analogy to higher dimensional cases, Calogero-Sutherland particles
can be viewed as anyons with statistics given by
$\theta$.\footnote{In fact, in one dimension there is no way to
bring a particle pass another avoiding the interaction, so the
concept of fractional statistics, introduced in two-dimensional
physics, is of a different nature in one dimension.}

Once one knows the scattering phase $\theta (k)$, it is a just a
tedious application of algebra and combinatorics to write the
explicit form of the amplitudes $A(P)$ in ({\ref{bethewavefunc}).
We are not going to pursue the problem of writing down the
wavefunction explicitly here, especially since most of the
important physical results can be obtained already without this
knowledge. For a very detailed derivation and explanation of these
details, we refer the reader to \cite{takahashi}.

\section{Periodic boundary conditions}

We established that every scattering can be decomposed in a
sequence of two-particle scatterings and that the effect of such a
process on the wavefunction is just an additional phase that
depends on the difference between the momenta of the two
interacting particles.

We are now in position to establish the fundamental equations and
results for the Bethe Ansatz. Let us consider a system of $N$
particles in a box of size $L$. Eventually, we want to take the
thermodynamic limit for the size of the box $L$ and the number of
particles $N$ to go to infinity in such a way that the density
$N/L$ stays finite. So it doesn't matter the kind of boundary
conditions we impose and we choose periodic boundary condition on
the wavefunction (\ref{bethewavefunc}):
\be
   \Psi (x_1, x_2, \ldots, x_j + L, \ldots, x_N) =
   \Psi (x_1, x_2, \ldots, x_j, \ldots, x_N).
\ee

For a particle $j$ to wind around the box, it would have to
scatter through every other particle of the system acquiring this
way a phase
\be
   \prod_{i=1}^N \eu^{- \ii \theta (k_j-k_i) } ,
\ee where we remembered that $\theta(0) = 0$. Moreover, during the
motion the particle will acquire a dynamical phase \be
   \eu^{\ii L k_j / \hbar}.
\ee Putting both effects together, periodic boundary conditions
amount to impose
\be
   1 = \eu^{\ii L k_j / \hbar} \; \prod_{i=1}^N \eu^{- \ii \theta (k_j-k_i) }
   \label{betheeq1}
\ee for each of the N $k_j$ momenta in the system.

Taking the logarithm of (\ref{betheeq1}) we arrive at a system of
N coupled algebraic equations
\be
   2 \pi I_j = L k_j / \hbar + \sum_{i=1}^N \theta (k_j-k_i) ,
   \label{betheeq}
\ee which are called the {\it ``Bethe Equations''} or the {\it
``Fundamental Equations''}. The $N$ integers $I_j$ are the winding
numbers of the phase and are effectively the quantum numbers for
the state described by the wavefunction.

The importance of the equations (\ref{betheeq}) is paramount,
since they translate the problem of solving a complicated
differential equation into a system of algebraic equations. Their
solution gives the momenta $k_j$ for the system and this
information is already sufficient to calculate the total momentum
\be
   P = \sum_{j=1}^N k_j
   \label{PBethe}
\ee and energy
\be
  E = {1 \over 2 m} \sum_{j=1}^N k_j^2 .
  \label{EBethe}
\ee

It is important to note now that, although our derivation would
lead to the conclusions that the $k_j$ can be identified as the
actual momenta of the individual particles, this interpretation is
in general not correct. Not only is it incorrect to assign a
defined momentum to a particle in a region where the particle is
interacting, but, more fundamentally, more complicated models will
assign quasi-momenta labels $k_j$ to their degrees of freedom, but
these labels cannot be interpreted as momenta in a traditional
way.

In general, the quasi-momenta $k_j$ should be viewed as points in
an appropriate phase-space that describe the system. The integers
$I_j$ will specify the state of the system and it can be shown
that these $N$ integers have to be chosen all distinct (if this
was not the case, two particles would have the same momentum and
would not interact and this would make the system singular).

\section{Zero temperature thermodynamics}

The set of integers $I_j$ in (\ref{betheeq}) defines the state
described by the wavefunction. The ground state, the state with
the lowest energy, can be identified with the state with the set
of $I_j$ running from $-N/2$ to $N/2$, $N$ being the number of
particles in the system. Clearly, from (\ref{PBethe}) we see that
this state has zero momentum. By choosing a different set of
integers, one constructs the excited states of the model.

When we take the thermodynamic limit, we let the number of
particles $N$ and the length of the system $L$ go to infinity as
we keep their ratio, the density of particles, finite
\be
   N/L = \rho.
\ee

As we increase $N$, one can show that the distribution of the
momenta $k_j$ of a system grows more dense. In the thermodynamic
limit one can prove that the average distance between two
neighboring momenta scales like $1/L$ and we can define the
distribution function \be
   \tau (k_j) = \lim_{N \to \infty, L \to \infty }
   {1 \over L (k_{j+1} - k_j)} >0
\ee which defines the density of particle in this quasi-momentum
space.

We can write the Bethe Equations ({\ref{betheeq}) as
\be
   k_j + {1 \over L} \sum_{i=1}^N \theta (k_j-k_i) = y (k_j)
\ee where we defined the {\it ``counting function''} $y (k_j)
\equiv {2 \pi I_j \over L }$, which is a monotonically increasing
function that counts the integers as a function of the
quasi-momenta. By definition,
\be
   y (k_j) - y (k_l) = {2 \pi \over L} ( I_j - I_l)
\ee and one can show that
\be
   2 \pi \tau (k) = {\de y (k) \over \de k} \: ,
\ee establishing a direct connection between the distribution of
the integers and of the quasi-momenta.

In the thermodynamic limit, the system of algebraic equations
(\ref{betheeq}) can be written as an integral equation for the
counting function and the momentum distribution:
\be
   y (k) = k + \int_{k_{min}}^{k_{max}} \theta (k - k') \tau (k') \de k'
\ee and, by taking the derivative of this equation by $k$,
\bea
   \tau (k) & = &
   {1 \over 2 \pi} + {1 \over 2 \pi} \int_{k_{min}}^{k_{max}} \theta' (k - k') \tau (k') \de k'
   \nonumber \\
   & = & {1 \over 2 \pi} + \int_{k_{min}}^{k_{max}} K (k - k') \tau (k') \de k'
   \label{intbetheeq}
\eea where we introduced the kernel of the integral equation as
the derivative of the scattering phase:
\be
   K ( k ) \equiv {1 \over 2 \pi} {\de \theta (k) \over \de k} .
\ee

Equation (\ref{intbetheeq}) allows us to determine the
distribution of the quasi-momenta. This distribution depends on
the support of the kernel, in equation (\ref{intbetheeq}) the
limits of integration $k_{min}$ and $k_{max}$. The support is
determined by the choice of the integers in the original equations
(\ref{betheeq}). For the ground state, the limits of integration
are symmetric ($k_{min} = - k_{max} = k_F$). A direct way to
determine the limits of integration is to calculate the number of
particles per unit length:
\be
   N / L = \rho = \int_{-k_F}^{k_F} \tau (k) \de k
   \label{BetheN}
\ee and invert this equation to calculate $k_f$ in terms of $N$.

Finally, we can write (\ref{PBethe}) in the thermodynamic limit as
\be
   P / L = \int_{-k_F}^{k_F} k \; \tau(k) \de k = 0 ,
   \label{BetheP}
\ee where we have used the fact that $\tau(-k) = \tau(k)$, and we
rewrite (\ref{EBethe}) as
\be
   E / L = \int_{-k_F}^{k_F} {k^2 \over 2} \; \tau(k) \de k .
   \label{BetheE}
\ee

Equations (\ref{BetheN}) and (\ref{BetheE}), together with
(\ref{intbetheeq}), define the equation of state of the system:
\be
   \rho \; \epsilon (\rho) = {E \over L} .
\ee

\chapter{Integrability of Gradient-less Hydrodynamics}
\label{Integrability}
In this appendix we construct the conserved quantities of
gradient-less hydrodynamic theories. It is a little-known fact
that if we consider a Galilean invariant system and we neglect
terms containing spatial derivatives of the density or of the
velocity, we can construct an infinite series of conserved
quantities ({\it``integrals of motion}'').

Although this result has been derived in the 1980's in the study
of the mathematical structure of a class of differential equations
(defined as of {\it ``hydrodynamic type''} \cite{dubrovin}), it is
not well-known in the physics community. Mathematicians have
traced the source of these conserved quantities to a property
known as {\it ``multi-Hamiltonian structure''}
\cite{olver}-\cite{brunelli04}. This means that there exists more
than one set of Hamiltonian and symplectic structure, i.e. Poisson
brackets, that generates the dynamical equations of the model.

Besides the Hamiltonian (\ref{HydroH1}) which in connection with
the Poisson \hbox{brackets} (\ref{rhovcomm}) generates the
equations of motion (\ref{rhotHam},\ref{vtHam}), we can construct
a different Hamiltonian and Poisson brackets that would lead to
the same dynamical equations. This interesting property allows
integrals of motion to be translated between the different
Hamiltonian systems and this generates a ladder structure in which
one can generate new conserved quantities using the ones of the
other system and vice versa.

We are not going to describe this structure any further, since
what is missing in our opinion is a clear physical understanding
of the origin of all these conserved quantities. We know that in
general conserved charges are due to some symmetry of the model,
but we have not been able to identify what symmetry would
guarantee such an infinite series, actually, a double infinite
series\footnote{Certainly, any spatial symmetry like Galilean
invariance is not sufficient to guaranty such a high number of
conserved quantities, since it is not infinite-dimensional.}.

We know that the free fermions model possesses a $W_\infty$ (or
even $Gl(\infty)$) symmetry and this is the source of the
integrability of the system. We think that even interacting
hydrodynamic theories could possess the same vast symmetry and
that, in some sense, by neglecting gradient corrections we discard
most of the content of the system and end up with some essentially
free fermions in disguise.
It might be possible to establish an exact mapping between any two
hydrodynamic theories (and in particular to map an interacting
model into the free fermion model), but so far we failed in
constructing such transformation.

After introducing the model in section \ref{IntInt}, we discuss
the integrability of the system in section \ref{IntTheory} and
argue that the presence of an infinite number of conserved
densities does not guarantee the full integrability of the theory.
In section \ref{IntMot} we explicitly construct the double
infinite series of integrals of motion. As far as we know, this is
the first time that such an explicit construction has been
reported.
Finally, we are going to discuss these results in section
\ref{IntConc}.

\section{Hydrodynamic Hamiltonian and equations of motion}
\label{IntInt}

Let us consider a generic Hamiltonian of the Hydrodynamic type
like the ones consider in Chapter \ref{HydroApp}. We require the
theory to have Galilean invariance and neglect terms with
derivatives of the fluid density $\rho$ or of the fluid velocity
$v$: \be
   {\cal H} = {\rho v^2 \over 2} + \rho \epsilon (\rho) \; ,
   \label{hydham}
\ee where $\epsilon(\rho)$ is the internal energy of the fluid.
The theory is defined by specifying the Poisson brackets between
the density $\rho$ and the velocity $v$ as in (\ref{rhovcomm}):
\be
   \left\{ \rho(x), v(y) \right\} = - \ii \delta' (x-y),
\ee while the other Poisson brackets vanish identically.

From the Hamiltonian and the Poisson brackets we find the
equations of motion to be
\bea
    \rho_t & = & -\rho v_x -v \rho_x = - (\rho v)_x \; ,
    \label{continuityD} \\
    v_t & = & - v v_x - (\rho \epsilon)_{\rho \rho} \rho_x
    = - v v_x - {v_s^2 \over \rho} \rho_x \; ,
    \label{eulerD}
\eea
where
\be
   v_s \equiv \sqrt{ \rho (\rho \epsilon)_{\rho \rho} }
\ee is the {\it ``Sound Velocity''} of the fluid and where we
adopted a notation for which $\partial_A B = B_A$. Eq.
(\ref{continuityD}) expresses the continuity condition for the
fluid, while Eq. (\ref{eulerD}) is the Euler dynamical equation of
motion.

\section{Integrability of the Hydrodynamic theory}
\label{IntTheory}

Some integrals of motion of the theory are trivial: $\rho$, $\rho
v$ and $H$ are obviously conserved densities.

To find more conserved quantities let us start with
\be
   I = \int \de x f(\rho,v)
\ee and find under which conditions is the function $f(\rho,v)$ a
conserved density. To this end, we calculate the commutation
relation between $I$ and $H \equiv \int {\cal H} \de x$ and we
impose it to be zero:
\bea
   \{I,H\} & = & \int \de x \; \de y \left\{ \left. f(\rho,v)
   \right|_x , \left. {\rho v^2 \over 2} + \rho \epsilon(\rho)
   \right|_y \right\} \nonumber \\
   & = & \int \de x \Bigg( \left[ f_\rho v + f_v (\rho \epsilon)_{\rho\rho} \right]
   \rho_x + \left[ f_\rho \rho + f_v v \right] v_x \Bigg) \nonumber \\
   & = & \int \de x \; \partial_x g(\rho,v) \nonumber \\
   & = & \int \de x \bigg( g_\rho \rho_x + g_v v_x \bigg)= 0
\eea where $g(\rho,v)$ is some unknown function. Equating the
second and fourth line and imposing the condition $g_{\rho v} =
g_{v \rho}$, we can eliminate $g(\rho,v)$ from the equations and
find that $f(\rho,v)$ is a conserved density if
\be
   f_{\rho \rho} = {v_s^2 \over \rho^2} f_{v v} .
   \label{conservedcond}
\ee

Eq. (\ref{conservedcond}) allows us to find integrals of motion.
To determine how many of them can be simultaneously specified, let
us calculate the commutation between two conserved integrals
\be
   I_1 = \int \de x f(\rho,v) \qquad I_2 = \int \de x h
   (\rho,v)
\ee with conserved densities $f(\rho,v)$ and $h(\rho,v)$
satisfying Eq. (\ref{conservedcond}):
\be
   \{ I_1, I_2 \} = \int \de x \partial_x g(\rho,v) = 0.
\ee Solving this condition as before we find that the two
integrals of motion commute if
\be
   f_{vv} h_{\rho \rho} = f_{\rho \rho} h_{vv}
   \label{commutingcong}
\ee which is identically satisfied since both $f(\rho,v)$ and
$h(\rho,v)$ satisfy Eq. (\ref{conservedcond}).

Therefore, we see that any solution of (\ref{conservedcond}) is an
integral of motion and, since we can construct an infinite series
of solution, linearly independent from each others, the theory
admits an infinite series of mutually commuting integrals of
motion and their respective conserved densities. One would
conclude from this that the theory is integrable, but it is not
necessarily the case.

To be more precise, there are many definitions of integrability
for a system with an infinite number of degrees of freedom. If by
integrable we mean that every degree of freedom admits a
representation in terms of action-angle variables (integrability
according to Liouville), the existence of an infinite number of
integrals of motion (the ``actions'' conjugated to the ``angle''
variables) might not be enough to exhaust all the degrees of
freedom.

Therefore, whether gradient-less hydrodynamic theories are
integrable according to Liouville or only integrable in the lesser
meaning of possessing infinitely many conserved quantities is
still not clear.

\section{The integrals of motion}
\label{IntMot}

To understand better the structure of the integrals of motion, let
us start with the simplest ansatz for a solution of Eq.
(\ref{conservedcond}):
\be
    f(\rho,v) \equiv \eu^{- \kappa v} J^\kappa (\rho)
    \label{solansatz}
\ee so that Eq. (\ref{conservedcond}) becomes
\be
    J^k_{\rho \rho} = {v_s^2 \over \rho^2} \kappa^2 J^\kappa .
\ee

Expanding this solution in a Taylor series in powers of $\kappa$:
\be
    f(\rho,v) = \sum_n (-\kappa)^n f_n(\rho,v)
\ee we can calculate
\be
    \partial_v f(\rho,v) = \sum_n (-\kappa)^n \partial_v
    f_n(\rho,v).
\ee Using Eq. (\ref{solansatz}) we can calculate the same quantity
as
\bea
    \partial_v f(\rho,v) & = & - \kappa \eu^{-\kappa v} J^\kappa (\rho)
    \nonumber \\
    & = & \sum_n (-\kappa)^n f_{n-1} (\rho,v)
\eea and we find that
\be
    f_n (\rho,v) = \partial_v f_{n+1} (\rho,v).
    \label{densitygen}
\ee

Since $f(\rho,v)$ is a conserved density, so are the coefficients
of the Taylor expansion $f_n (\rho,v)$. We showed before that any
integral of motion commutes with any other; therefore, the
functions $f_n (\rho,v)$ are the conserved densities of a series
of mutually commuting integrals of motion.

The choice of the ansatz in Eq. (\ref{solansatz}) provides only
two linearly independent integrals of motion, but by expanding any
of these two solutions in a Taylor Series, we recover an infinite
series of conserved densities. It can be shown that there exist
exactly two infinite series of integrals of motion and each of
them can be generated from the two solutions of Eq.
(\ref{solansatz}). Moreover, we found a recurrence relation among
each series, Eq. (\ref{densitygen}). We can use this relation to
generate the entire series from the first element in the series
(the one with the lowest power in $v$) and successively
integrating in $v$; we only need to determine the integration
constant.

Let us calculate the elements of the first series. We already know
three of them:
\bea
    I_0^1 & = & \int \de x \rho \; , \\
    I_1^1 & = & \int \de x \rho v \; , \\
    I_2^1 & = & \int \de x \left[ {\rho v^2 \over 2} + \rho \epsilon (\rho) \right] \: .
\eea The general structure of this series is
\be
   I_n^1 = \int \de x j_n^1 (\rho,v)
   \label{In1}
\ee with
\be
    j_n^1 = {\rho v^n \over n} + \sum_{k=1}^{[n/2]} \phi_n^k (\rho)
    v^{n-2k} \; ,
\ee where $[n/2]$ means the highest integer smaller or equal to
$n/2$.

We can determine the coefficients $\phi_n^k (\rho)$ by requiring
$\partial_t j_n^1 = \partial_x g$ for some function $g(\rho,v)$.
Solving this condition brings:
\bea
    \phi_n^1 & = & (n-1) \rho \epsilon (\rho) \\
    (\phi_n^k)_{\rho \rho} & = & (n-2k+2)(n-2k+1) { (\rho \epsilon)_{\rho \rho}
    \over \rho} \phi_n^{k-1} \nonumber \\
    & = & (n-2k+2)(n-2k+1)
    { v_s^2 \over \rho^2} \phi_n^{k-1} \qquad \qquad k=2 \ldots [n/2]
    \nonumber \\
\eea and in this way we can recurrently generate the series. For
instance:
\be
   I_3^1 = \int \de x \left[ {\rho v^3 \over 3} + 2 \rho \epsilon (\rho) v \right] \: .
\ee

It is easy to show that the velocity $v$ is also a conserved
density and it constitutes the first element of the second series
of integrals of motion: \be
   I_n^2 = \int \de x j_n^2 (\rho,v)
   \label{In2}
\ee
with
\be
    j_n^2 = {v^n \over n} + \sum_{k=1}^{[n/2]} \varphi_n^k (\rho)
    v^{n-2k} .
\ee We can determine the coefficient $\varphi_n^k (\rho)$ as
before and find \bea
    (\varphi_n^1)_{\rho \rho} & = & (n-1) { (\rho \epsilon)_{\rho \rho} \over \rho}
    = (n-1) {v_s^2 \over \rho^2} \\
    (\varphi_n^k)_{\rho \rho} & = & (n-2k+2)(n-2k+1) { (\rho \epsilon)_{\rho \rho}
    \over \rho} \varphi_n^{k-1} \nonumber \\
    & = & (n-2k+2)(n-2k+1)
    { v_s^2 \over \rho^2} \varphi_n^{k-1} \qquad \qquad k=2 \ldots [n/2]
    \nonumber \\
\eea and generate the whole series:
\bea
    I_1^2 & = & \int \de x v \; , \\
    I_2^2 & = & \int \de x \left[ {v^2 \over 2} + \xi (\rho) \right] \; , \\
    I_3^2 & = & \int \de x \left[ {v^3 \over 3} + 2 \xi (\rho) v \right] \; , \\
    \ldots \nonumber
\eea where
\be
   \xi_{\rho\rho} = {v_s^2 \over \rho^2} \: .
\ee

It is interesting to notice that the conserved densities $j_n^l$,
with the given definition for the coefficients, naturally follow
the condition in Eq. (\ref{densitygen}), i.e.:
\be
    j_n^l (\rho,v) = \partial_v j_{n+1}^l (\rho,v).
\ee and it can be shown that they are the coefficients of the
Taylor expansion of the two solutions of Eq. (\ref{solansatz}).

\section{Conclusions and open questions}
\label{IntConc}

We have shown that a hydrodynamic theory having Galilean
invariance and without gradient corrections has two infinite
series of integrals of motion. We still do not understand
completely the nature of this integrability, i.e. the symmetry
group it comes from.

The hydrodynamic description of free fermions has the Hamiltonian:
\be
   H = {\rho v^2 \over 2} + {\pi^2 \over 6} \rho^3
   \label{freefermham}
\ee with sound velocity
\be
   v_s = \pi \rho.
\ee It is known to be integrable, its symmetry group being
$W_\infty$, and its integrals of motion being simply:
\be
    \int \de x {1 \over k} (v \pm \rho)^k
\ee which are linear combinations of the $I_n^1$ and $I_n^2$
(\ref{In1},\ref{In2}).

It is natural to think that the origin of the integrability of the
generic Hamiltonian in Eq. (\ref{hydham}) is the same as for the
free fermions, but this has not been confirmed yet.

It is worth noticing, however, that unlike other integrable
theories this system does not have solitons and that any wave
eventually develops singularities\footnote{Systems like these can
be viewed as {\it ``dispersionless limits''} of other integrable
theories and can therefore be considered singular, since more than
one integrable model can have the same dispersionless limit.}.
This is in sharp contrast with the intuitive concept of
integrability and is probably a sign that the double infinite
series of integrals of motion is not sufficient to completely
describe and constrain the dynamics of the system.

A lot of work has been done on equations of the hydrodynamic type
and a good understanding of the integrability of these theories in
terms of an underlying multi-Hamiltonian structure is established
(ref. \cite{olver}-\cite{brunelli04} for a very short and
non-comprehensive list), but we believe that the physical nature
of the integrability, i.e. the underlying symmetry group, is not
yet been explained. Reference \cite{dubrovin} has the most
symmetry-oriented approach.

It is also known that there exist integrable hydrodynamic theories
with gradient corrections, namely the ones describing the
Calogero-Sutherland model and the free bosons with delta repulsion
(Lieb-Liniger bosons). It would be interesting to understand what
preserves the integrability of this theories, which were
integrable even before the addition of the gradient corrections,
as we just showed.


\end{document}